\documentclass[preprint,sort&compress,12pt]{elsarticle}

\usepackage{mathptmx}
\usepackage[12pt]{moresize}

\usepackage{graphicx}
\usepackage{amsmath}
\usepackage{amssymb}
\usepackage{bm}
\usepackage[usenames]{color}
\usepackage[colorlinks=false,linktocpage=true]{hyperref}

\usepackage{booktabs}
\usepackage{multirow} 
\usepackage{multicol}
\usepackage{rotating} 

\usepackage[utf8]{inputenc}
\usepackage{amsmath}
\usepackage{amssymb}
\usepackage{multirow}
\usepackage[inline]{enumitem}

\usepackage{color}
\definecolor{blue}{RGB}{0,50,230}

\usepackage{accents}
\newcommand{\ubar}[1]{\underaccent{\bar}{#1}}

\usepackage{hyperref}
\hypersetup{
  colorlinks=true,
  linkcolor=blue,
  citecolor=blue,
  urlcolor=blue
}

\usepackage{algorithmicx}
\usepackage[ruled]{algorithm}
\usepackage[noend]{algpseudocode}
    
\usepackage{tikz}
\usetikzlibrary{shapes.geometric, arrows}
\usepackage{tikz}
\usetikzlibrary{shapes}
\usetikzlibrary{positioning,arrows}
\usepackage{scalefnt}
    
\usepackage[percent]{overpic}
    
\journal{Computer Physics Communications}

\def\be{\begin{equation}}
\def\ee{\end{equation}}
\def\bea{\begin{eqnarray}}
\def\eea{\end{eqnarray}}

\newcommand{\q}{{\bf q}}

\newcommand{\V}{n}
\newcommand{\n}{{\cal N}}
\newcommand{\ed}{{\cal E}}
\newcommand{\p}{{\cal P}}
\newcommand{\peq}{{\cal P}_{0}}

\begin{document}

\begin{frontmatter}

\title{Massively parallel simulations of relativistic fluid dynamics on graphics processing units with CUDA}

\author[a]{Dennis Bazow\corref{author}}
\author[a]{Ulrich Heinz}
\author[b]{Michael Strickland}

\cortext[author] {Corresponding author.\\\textit{E-mail address:} bazow.1@osu.edu (D. Bazow)
}
\address[a]{Department of Physics, The Ohio State University,
  Columbus, OH 43210 United States}
\address[b]{Department of Physics, Kent State University, 
Kent, OH 44242 United States}

%
\begin{abstract}
Relativistic fluid dynamics is a major component in dynamical simulations of the quark-gluon plasma created in relativistic heavy-ion collisions. Simulations of the full three-dimensional dissipative dynamics of the quark-gluon plasma with fluctuating initial conditions are computationally expensive and typically require some degree of parallelization. In this paper, we present a GPU implementation of the Kurganov-Tadmor algorithm which solves the 3+1d relativistic viscous hydrodynamics equations including the effects of both bulk and shear viscosities. We demonstrate that the resulting CUDA-based GPU code is approximately two orders of magnitude faster than the corresponding serial implementation of the Kurganov-Tadmor algorithm. We validate the code using (semi-)analytic tests such as the relativistic shock-tube and Gubser flow.
\end{abstract}
%

\begin{keyword}
Relativistic fluid dynamics \sep Quark-gluon plasma \sep GPU \sep CUDA \sep Parallel computing

\end{keyword}
\end{frontmatter}

%
{\bf PROGRAM SUMMARY}

\begin{small}
\noindent
{\em Manuscript Title:} Massively parallel simulations of relativistic fluid dynamics on graphics processing units with CUDA                                       \\
{\em Authors:} Dennis Bazow, Ulrich Heinz, Michael Strickland                                               \\
{\em Program Title:} GPU-VH                                        \\
{\em Journal Reference:}                                      \\
{\em Catalogue identifier:}                                   \\
{\em Licensing provisions: none}                                   \\
{\em Programming language: CUDA C}                                   \\
{\em Computer: Any machine with an Nvidia graphics processing unit}                                               \\
{\em Operating system:} GNU/Linux distributions                                     \\
{\em Global memory usage:} 0.5 GB (for a $128^3$ grid)                                            \\
{\em Keywords:} Relativistic fluid dynamics, Quark-gluon plasma, GPU, CUDA, Parallel computing  \\
{\em Classification:} 12 Gases and Fluids, 17 Nuclear Physics                                       \\
{\em External routines/libraries:} Google Test, GNU Scientific Library (GSL)                           \\
{\em Nature of problem:}\\
  Dynamical evolution of the fluid dynamic stage of the quark-gluon plasma produced in nuclear collisions. 
   \\
{\em Solution method:}\\
  Kurganov-Tadmor algorithm
   \\
{\em Running time:}
Typical running time on an Nvidia GeForce GTX 980 Ti graphics card for fluid dynamic simulations that includes a nonconformal equation of state with bulk and shear viscosities on a $128^3$ grid is 38.4 sec/time step.
\end{small}
\section{Introduction}
\label{sec:intro}
Relativistic fluid dynamics is a major workhorse in simulations of the quark-gluon plasma (QGP) created in relativistic heavy-ion collisions~\cite{Gale:2013da}. The fluid dynamic equations (conservation laws plus the convective-diffusion equations for the dissipative currents) form a coupled set of hyperbolic partial differential equations. In general, computationally expensive high-resolution schemes are needed when the solutions exhibit shock waves, discontinuities, or large gradients. Although the QGP is short lived (the fluid dynamic stage lasting ${\cal O}(10)\,\mathrm{fm/c}$ at the highest energies), simulations still take on the order of days. However, the computational cost is even more demanding; no two nuclei are alike, so one must average over thousands of randomly generated nucleus-nucleus collisions. To deal with this, parallel computing resources such as Beowulf clusters or supercomputers are typically taken advantage of to run these simulations. Even with the use of these resources, it remains intractable to systematically determine the best values of the parameters that enter into a $(3{+}1)$-dimensional simulation of the QGP using Bayesian statistics~\cite{Novak:2013bqa,Bernhard:2016tnd}.

All of these simulations relied on distributing the data over multiple central processing units (CPUs). However, for the algorithms employed to solve the fluid dynamic equations, this is not optimal; they are inherently parallel algorithms where the interactions between grid points is local rather than long range and which would benefit from concurrent execution of a large number of threads. In fact, the CPU is optimized for serial code performance (i.e. designed to minimize the execution latency of a single thread) by making use of large last-level on-chip cache memories to reduce the instruction and data access latencies as well as sophisticated control logic that allows a single thread to execute in parallel (while still maintaining the appearance of sequential execution). The control logic and cache memories, neither of which contribute to the peak calculation speed, consume hardware resources (such as chip area and power) that could otherwise be used to provide more arithmetic execution units and memory access channels. 

In contrast, the graphics processing unit (GPU) has a highly efficient, high throughput computation architecture designed to optimize the performance of a simultaneous execution of many threads. Giving up the sophisticated control logic and large cache memories results in long-latency pipelined memory channels and arithmetic operations, but actually increases the total execution throughput by reducing the chip area and power of the memory access hardware and arithmetic units, allowing more of them. This leads to a large peak-performance gap (measured by floating-point operations per second or FLOPS) between CPUs and GPUs shown in Table~\ref{XeonPhiTitanZComparision} which compares the state of the art Intel Xeon Phi coprocessor 7120 and the GeForce GTX Titan Z graphics card. The Titan Z performs about $3$ and $2$ times more single and double-precision FLOPS than the Intel Xeon Phi, respectively. Graphics cards also have a significant advantage in terms of memory bandwidth - the rate at which data can be accessed from memory - which operates about two times faster than the CPU. In this paper, we will use GPUs to accelerate our relativistic fluid dynamic code.  
\begin{table}[htbp]
\small
\begin{center}
\begin{tabular}{|c|c|c|c|}
\hline
Model & {} & Intel Xeon Phi & GeForce GTX Titan Z
\\ \hline
Processor cores
& \multirow{1}{2.5cm}{}
& 61 & 5760
\\ \hline
\multirow{2}{2.5cm}{\centering Clock speed (MHz)} & 
\multicolumn{1}{c|}{\centering Core} 
& 1238 & 705 \\
\cline{2-4} & \multicolumn{1}{c|}{Turbo/Boost}  
& 1333 & 876
\\ \hline
\multirow{2}{2.5cm}{\centering Memory Configuration} & 
\multicolumn{1}{c|}{\centering Size (GB)} 
& 16 & 12.288 \\
\cline{2-4} & \multicolumn{1}{c|}{\centering Bandwidth (GB/s)}  
& 352 & 672
\\ \hline
\multirow{2}{3.15cm}{\centering Processing power (GFLOPS)} & 
\multicolumn{1}{c|}{\centering Single precision} 
& 2416.58 & 8121.6 \\
\cline{2-4} & \multicolumn{1}{c|}{\centering Double precision} 
& 1208.29 & 2707.2
\\ \hline
TDP (W) 
& \multirow{1}{2.5cm}{} 
& 300 & 375
\\ \hline
\end{tabular}
\end{center}
\caption{Comparision between the Intel Xeon Phi coprocessor 7120P/D/X/A and GeForce GTX Titan Z graphics card based on official Intel and Nvidia specifications. (The thermal design power (TDP) is the maximum amount of heat generated by the CPU/GPU.)}\label{XeonPhiTitanZComparision}
\end{table}

This paper is organized as follows. In section \ref{sec:fluid_dynamics}, we write down the fluid dynamic equations that will be solved for simulations of nuclear collisions. The numerical scheme which we have adopted and its implementation onto graphics processing units is presented in Secs.~\ref{sec:num_scheme} and~\ref{sec:gpu_imp}. In Section~\ref{sec:num_tests} we describe the validation of our code against various test problems. We then show simulations of the $(3{+}1)$-dimensional QGP in a single nuclear collision,
and idendify the regions where fluid dynamics is the correct effective description for the problem. In Sec.~\ref{sec:performance} the performance benchmarks of our code against a highly optimized serial CPU implementation are presented. Our conclusions are summarized in Sec.~\ref{sec:conc}. 
\section{Relativistic fluid dynamics}
\label{sec:fluid_dynamics}
\subsection{Conservation laws and relaxation equations for dissipative flows}
\label{sec:conLaws}
Relativistic fluid dynamics is described by the macroscopic conservation laws of {\it any} conserved current $J^{\mu}_{a}$ (in this case the net baryon current $N^\mu$) and the energy-momentum tensor $T^{\mu\nu}$. In general coordinates these five equations are:
\begin{eqnarray}
d_{\mu}N^{\mu}\equiv N^{\mu}_{;\mu}&\equiv&\frac{1}{\sqrt{g}}\partial_{\mu}(\sqrt{g}N^{\mu})=0\;,
\nonumber \\
d_{\mu}T^{\mu\nu}\equiv T^{\mu\nu}_{;\mu}&\equiv&\frac{1}{\sqrt{g}}\partial_{\mu}(\sqrt{g}T^{\mu\nu})+\Gamma^{\nu}_{\mu\lambda}T^{\mu\lambda}=0\;,
\label{hydro_eqs}
\end{eqnarray}
where $d_\mu$ stands for the covariant derivative and is denoted compactly by a semicolon, $g\equiv-\det{(g_{\mu\nu})}$ is the negative determinant of the metric tensor $g^{\mu\nu}$ with mostly minus convention $(+,-,-,-)$, $\partial_\mu\equiv\partial/\partial x^\mu$ denotes the four-derivative, and $\Gamma^{\mu}_{\alpha\beta}\equiv\Gamma^{\mu}_{\beta\alpha}=
\frac{1}{2}g^{\mu\nu}(\partial_{\beta}g_{\alpha\nu}
+\partial_{\alpha}g_{\nu\beta}-\partial_{\nu}g_{\alpha\beta})$ are the affine connections (Christoffel symbols). 

The macroscopic fields can be decomposed with respect to the fluid velocity four-vector $u^\mu$ (defined in the Landau frame as the timelike eigenvector of $T^{\mu\nu}$, $T^{\mu\nu}u_{\nu}=\ed u^{\mu}$, where the energy density $\ed$ is its eigenvalue) as:
\begin{align}
N^{\mu}&=\n u^{\mu}+\V^{\mu}\;, \label{N}\\
T^{\mu\nu}&= \ed u^{\mu}u^{\nu}-(\peq+\Pi)\Delta^{\mu\nu}+\pi^{\mu\nu}\;.
\label{T}
\end{align}
By identifying the number density $\n$ and energy density $\ed$ with their equilibrium form via the Landau matching conditions we can characterize the fluid by a local temperature $T$ and a chemical potential $\mu$, and the thermal pressure $\peq$ can be obtained from the equilibrium equation of state (EoS) $\peq(\ed,\n)$.~The quantity $\V^\mu$ is the particle-diffusion current, the total isotropic pressure is obtained from the sum of the thermodynamic pressure $\peq$ and the bulk viscous pressure $\Pi$, and $\pi^{\mu\nu}\equiv T^{\langle\mu\nu\rangle}$ is the shear-stress tensor. The transverse projection tensor $\Delta^{\mu\nu}\equiv g^{\mu\nu}-u^{\mu}u^{\nu}$ is used to project four-vectors and tensors into the space orthogonal to $u^\mu$ (i.e. onto 3-space in the local rest frame (LRF)). Angular brackets $\langle\rangle$ around any two Lorentz indices indicates 
the result of applying the transverse projector $\Delta^{\mu\nu}$ to $u^\mu$ and traceless projector 
$\Delta^{\mu\nu}_{\alpha\beta}\equiv\frac{1}{2}(\Delta^{\mu}_{\alpha}\Delta^{\nu}_{\beta}
+\Delta^{\nu}_{\alpha}\Delta^{\mu}_{\beta})
-\frac{1}{3}\Delta^{\mu\nu}\Delta_{\alpha\beta}$ to a rank-2 tensor. By construction, the dissipative terms satisfy the constraints $u_{\mu}\V^{\mu}=u_{\mu}\pi^{\mu\nu}=\pi^{\mu\nu}u_{\nu}=\pi^{\mu}_{\mu}\equiv 0$.

The fluid dynamic equations (\ref{hydro_eqs}) are exact, but not closed due to the appearance of dissipative terms appearing in Eqs.~(\ref{N}) and (\ref{T}). In order to close this system, nine additional evolution or ``transport" equations for the independent components of the shear-stress tensor $\pi^{\mu\nu}$, bulk viscous pressure $\Pi$, and diffusion four-current $\V^\mu$ must be provided, along with the equation of state, from the underlying microscopic theory. These can obtained from the relativistic Boltzmann equation as~\cite{Denicol:2012cn}
\begin{align}
\tau _{\Pi }\dot{\Pi}+\Pi & =-\zeta \theta +\mathcal{J}+\mathcal{K}+\mathcal{%
R}\;,  \notag \\
\tau _{n}\dot{n}^{\left\langle \mu \right\rangle }+n^{\mu }& =\kappa
I^{\mu }+\mathcal{J}^{\mu }+\mathcal{K}^{\mu }+\mathcal{R}^{\mu }\text{ }%
,  \notag \\
\tau _{\pi }\dot{\pi}^{\left\langle \mu \nu \right\rangle }+\pi ^{\mu \nu }&
=2\eta \sigma ^{\mu \nu }+\mathcal{J}^{\mu \nu }+\mathcal{K}^{\mu \nu }+%
\mathcal{R}^{\mu \nu }\;.  
\label{relaxationEquations}
\end{align}
The explicit form of the terms on the r.h.s.~is discussed below.
Along with the conservation laws (\ref{hydro_eqs}), the above equations (\ref{relaxationEquations}) define a
resummed transient relativistic fluid-dynamical theory. The overdot denotes the covariant time derivative $D\equiv u^{\mu}d_{\mu}$, with the notation $\dot{\V}^{\langle\mu\rangle}\equiv\Delta^{\mu}_{\nu}\dot{\V}^{\nu}$ and 
$\dot{\pi}^{\langle\mu\nu\rangle}\equiv\Delta^{\mu\nu}_{\alpha\beta}
\dot{\pi}^{\alpha\beta}$.
Microscopic physics occurs on time scales equal to or faster than $\tau_{\Pi}$, $\tau_{n}$, and $\tau_{\pi}$ which describe how fast $\Pi$, $\V^\mu$, and $\pi^{\mu\nu}$ exponentially decay towards their respective Navier-Stokes values $-\zeta\theta$, $\kappa I^\mu$, and $2\eta\sigma^{\mu\nu}$~\cite{Denicol:2011fa}.
Here, $\zeta$ is the bulk viscosity coefficient, $\kappa$ is the particle-diffusion coefficient, $\eta$ is the shear viscosity, $\theta\equiv d_{\mu}u^{\mu}=\nabla{\cdot}u$ (where $\nabla^{\mu}\equiv\Delta^{\mu\nu}d_{\nu}$ is the spatial gradient in the LRF) is the expansion scalar, $I^{\mu}\equiv\nabla^{\mu}(\mu/T)$ is the gradient of the chemical potential to temperature ratio, and $\sigma^{\mu\nu}\equiv\nabla^{\langle\mu}u^{\nu\rangle}$ is the velocity shear tensor.

The r.h.s.~of Eqs.~(\ref{relaxationEquations}) are organized as an algebraic series in powers of the Knudsen number $\mathrm{Kn}$ (ratio between a characteristic microscopic and macroscopic time/length scale of the fluid \cite{Denicol:2012cn}) and the inverse Reynolds number $\mathrm{R}^{-1}$ (ratio of dissipative quantities and local thermodynamic equilibrium values).~Terms of order $\mathcal{O}(\mathrm{Kn}^{3})$, $%
\mathcal{O}(\mathrm{R}_{i}^{-1}\mathrm{R}_{j}^{-1}\mathrm{R}_{k}^{-1})$, $%
\mathcal{O}(\mathrm{Kn}^{2}\mathrm{R}_{i}^{-1})$, $\mathcal{O}(\mathrm{Kn}%
\,\mathrm{R}_{i}^{-1}\mathrm{R}_{j}^{-1})$ and higher are omitted in the effective theory. The tensors $\mathcal{J}$, $\mathcal{J}^{\mu }$, and $\mathcal{%
J}^{\mu \nu }$ contain all terms made up from products of terms of first order in Knudsen and inverse
Reynolds numbers,
\begin{align}
\mathcal{J}& =-\ell _{\Pi n}\nabla \cdot n-\tau _{\Pi n}n\cdot F-\delta
_{\Pi \Pi }\Pi \theta -\lambda _{\Pi n}n\cdot I+\lambda _{\Pi \pi }\pi ^{\mu
\nu }\sigma _{\mu \nu }\;,  \notag \\
\mathcal{J}^{\mu }& =-n_{\nu }\omega ^{\nu \mu }-\delta _{nn}n^{\mu }\theta
-\ell _{n\Pi }\nabla ^{\mu }\Pi +\ell _{n\pi }\Delta ^{\mu \nu }\nabla
_{\lambda }\pi _{\nu }^{\lambda }+\tau _{n\Pi }\Pi F^{\mu }-\tau _{n\pi }\pi
^{\mu \nu }F_{\nu }  \notag \\
& \hspace{5mm} -\lambda _{nn}n_{\nu }\sigma ^{\mu \nu }+\lambda _{n\Pi }\Pi I^{\mu
}-\lambda _{n\pi }\pi ^{\mu \nu }I_{\nu }\;,  \notag \\
\mathcal{J}^{\mu \nu }& =2\tau_{\pi}\pi _{\lambda }^{\left\langle \mu \right. }\omega
^{\left. \nu \right\rangle \lambda }-\delta _{\pi \pi }\pi ^{\mu \nu }\theta
-\tau _{\pi \pi }\pi ^{\lambda \left\langle \mu \right. }\sigma _{\lambda
}^{\left. \nu \right\rangle }+\lambda _{\pi \Pi }\Pi \sigma ^{\mu \nu
}\;-\tau _{\pi n}n^{\left\langle \mu \right. }F^{\left. \nu \right\rangle } 
\notag \\
& \hspace{5mm} +\ell _{\pi n}\nabla ^{\left\langle \mu \right. }n^{\left. \nu
\right\rangle }+\lambda _{\pi n}n^{\left\langle \mu \right. }I^{\left. \nu
\right\rangle }\;,  \label{14_moment_terms}
\end{align}
where we defined $F^{\mu }\equiv\nabla ^{\mu }\peq$, and $\omega^{\mu\nu}\equiv\frac{1}{2}(\nabla^{\mu}u^{\nu}-\nabla^{\nu}u^{\mu})$ is the vorticity tensor. 
The tensors $\mathcal{K}$, $\mathcal{K}^{\mu }$, and $\mathcal{K}%
^{\mu \nu }$ contain all terms of second order in Knudsen number,
\begin{align}
\mathcal{K}& =\zeta _{1}\,\omega _{\mu \nu }\omega ^{\mu \nu }+\zeta
_{2}\,\sigma _{\mu \nu }\sigma ^{\mu \nu }+\zeta _{3}\,\theta ^{2}+\zeta
_{4}\,I\cdot I +\zeta _{5}\,F\cdot F \notag \\
& \hspace{5mm} +\zeta _{6}\,I\cdot F+\zeta _{7}\,\nabla
\cdot I+\zeta _{8}\,\nabla \cdot F,  \notag \\
\mathcal{K}^{\mu }& =\kappa _{1}\sigma ^{\mu \nu }I_{\nu }+\kappa _{2}\sigma
^{\mu \nu }F_{\nu }+\kappa _{3}I^{\mu }\theta +\kappa _{4}F^{\mu }\theta \notag \\
& \hspace{5mm} +\kappa _{5}\omega ^{\mu \nu }I_{\nu }+\kappa _{6}\Delta _{\lambda }^{\mu
}\partial _{\nu }\sigma ^{\lambda \nu }+\kappa _{7}\nabla ^{\mu }\theta , 
\notag \\
\mathcal{K}^{\mu \nu }& =\eta _{1}\omega _{\lambda }^{\left. {}\right.
\left\langle \mu \right. }\omega ^{\left. \nu \right\rangle \lambda }+\eta
_{2}\theta \sigma ^{\mu \nu }+\eta _{3}\sigma ^{\lambda \left\langle \mu
\right. }\sigma _{\lambda }^{\left. \nu \right\rangle }+\eta _{4}\sigma
_{\lambda }^{\left\langle \mu \right. }\omega ^{\left. \nu \right\rangle
\lambda }  \notag \\
& \hspace{5mm} +\eta _{5}I^{\left\langle \mu \right. }I^{\left. \nu \right\rangle }+\eta
_{6}F^{\left\langle \mu \right. }F^{\left. \nu \right\rangle }+\eta
_{7}I^{\left\langle \mu \right. }F^{\left. \nu \right\rangle }+\eta
_{8}\nabla ^{\left\langle \mu \right. }I^{\left. \nu \right\rangle }+\eta
_{9}\nabla ^{\left\langle \mu \right. }F^{\left. \nu \right\rangle }.
\label{secondOrderKn}
\end{align}
The tensors $\mathcal{R}$, $\mathcal{R}^{\mu }$, and $\mathcal{R}%
^{\mu \nu }$ contain all terms of second order in inverse Reynolds number,
\begin{align}
\mathcal{R}& =\text{ }\varphi _{1}\Pi ^{2}+\varphi _{2}n\cdot n+\varphi
_{3}\pi _{\mu \nu }\pi ^{\mu \nu },  \notag \\
\mathcal{R}^{\mu }& =\varphi _{4}n_{\nu }\pi ^{\mu \nu }+\varphi _{5}\Pi
n^{\mu },  \notag \\
\mathcal{R}^{\mu \nu }& =\varphi _{6}\Pi \pi ^{\mu \nu }+\varphi _{7}\pi
^{\lambda \left\langle \mu \right. }\pi _{\lambda }^{\left. \nu
\right\rangle }+\varphi _{8}n^{\left\langle \mu \right. }n^{\left. \nu
\right\rangle }.
\label{secondOrderR}
\end{align}
These last terms arise from non-linear terms in the collision kernel~\cite{Denicol:2012cn}.

The conservation laws (\ref{hydro_eqs}) together with (\ref{relaxationEquations}) and the equilibrium equation of state $\peq(\ed,\n)$ define the fluid dynamic equations for a relativistic system. We remark that although the above equations were derived from the weakly coupled Boltzmann equation, their structure is generic and remains valid at strong coupling where the Boltzmann equation breaks down. Only the transport coefficients appearing in Eqs.~(\ref{relaxationEquations})-(\ref{secondOrderR}) and an appropriate EoS would need to be calculated self-consistently from a different microscopic theory.  

\subsection{Fluid dynamics for nuclear collisions}
\label{sec:fluidDynamicsNuclearCollisions}
The primary application that we have in mind is the simulation of the (3+1)-dimensional matter created in ultrarelativistic heavy-ion collisions. Here, the conditions are rather extreme: the fluid has the largest space-time gradients of the fields and the smallest space-time volume ever encountered. Under these extreme conditions it is not known {\it a priori} if terms contained in ${\cal K}^{\mu_1\cdots\mu_\ell}$ can be neglected--they might be of the same order as ${\cal J}^{\mu_1\cdots\mu_\ell}$. The problem is that the terms of second (and higher) order in the Knudsen number, e.g. $\nabla^{\mu}\theta\subset{\cal K}^{\mu}$, have second-order gradients resulting in acasual equations of motion. It was shown in Ref.~\cite{Denicol:2012vq} how to include such terms in a casual way by promoting additional dissipative currents other than those appearing in Eq.~(\ref{hydro_eqs}) to dynamical variables.~(Another way to cure this problem is to include non-hydrodynamic modes in the leading-order single-particle distribution function~\cite{Bazow:2015zca}.)~This is beyond the scope of this work; instead we use the 14-moment approximation where all terms in Eq.~(\ref{secondOrderKn}) vanish identically. This will at least be an acceptable approximation for asymptotically late times when the dissipative quantities approach their respective Navier-Stokes limit. Furthermore, this saves us the laborious task of determining all of their transport coefficients; although they have been formally written down in Ref.~\cite{Denicol:2012cn}, they have not yet been put into a convenient form to use in simulations of heavy ion collisions. Nonlinear effects of the collisional kernel seem to not play a major role in the transient regime~\cite{Bazow:2016oky}. It was explicitly shown in Ref.~\cite{Molnar:2013lta} that for a massless Boltzmann gas with constant scattering cross section in the 14-moment approximation, the contributions from the $\varphi_7$ term is an order of magnitude smaller compared to the other terms and can then safely be set to zero. We therefore neglect all terms ${\cal O}(\mathrm{R}^{-1}_{i}\mathrm{R}^{-1}_{j})$, i.e. all terms in Eq.~(\ref{secondOrderR}), in the equations of motion.~We further restrict ourselves to systems with vanishing chemical potential where energy and momentum are the only conserved currents and only the dissipative currents $\Pi$ and $\pi^{\mu\nu}$ are needed. They are governed by the following simplified relaxation equations:
\begin{align}
\tau_{\Pi}D\Pi+\Pi &=
-\zeta\theta -\delta_{\Pi\Pi}\Pi\theta
+\lambda_{\Pi\pi}\pi^{\mu\nu}\sigma_{\mu\nu}\;,
\\
\tau _{\pi }D\pi^{\mu \nu}+\pi ^{\mu \nu }&
=2\eta \sigma ^{\mu \nu }
+2\tau_{\pi}\pi^{\langle\mu}_{\lambda}\omega^{\nu\rangle\lambda}
-\delta_{\pi\pi}\pi^{\mu\nu}\theta
-\tau_{\pi\pi}\pi^{\lambda\langle\mu}\sigma^{\nu\rangle}_{\lambda}
+\lambda_{\pi\Pi}\Pi\sigma^{\mu\nu}
\nonumber \\
&-\tau_{\pi}( \pi ^{\lambda \mu }u^{\nu }+\pi ^{\lambda \nu
}u^{\mu }) Du_{\lambda }
\;. \label{pimunuSimplified}
\end{align}
The transport coefficients are unknown for the quark-gluon plasma; at this moment in time it is as good an approximation as any to obtain them from kinetic theory. We take formulas for the transport coefficients derived near the conformal limit (i.e. expanded to a particular order in $z\equiv m/T\ll 1$ that gives the lowest nonzero expression written entirely in terms of the energy density or equilibrium quantities $\peq$ and $c^2_s$ directly accessible from lattice QCD)~\cite{Denicol:2014vaa}. For the bulk viscous pressure, these transport coefficients are
\begin{align}
\frac{\zeta}{\tau_\Pi}&=15\left(\frac{1}{3}-c^2_s\right)^{2}(\ed+\peq)
+{\cal O}(z^5)\;, \label{betaPi}
\\
\frac{\delta_{\Pi\Pi}}{\tau_\Pi}&=\frac{2}{3}+{\cal O}(z^2\log z)\;,
\\
\frac{\lambda_{\Pi\pi}}{\tau_\Pi}&=\frac{8}{5}\left(\frac{1}{3}-c^2_s\right)
+{\cal O}(z^4)\;, 
\end{align}
and for the shear-stress tensor
\begin{align}
\frac{\eta}{\tau_\pi}&=\frac{\ed+\peq}{5}+{\cal O}(z^2)\;, 
\\
\frac{\delta_{\pi\pi}}{\tau_\pi}&=\frac{4}{3}+{\cal O}(z^2)\;, 
\\
\frac{\tau_{\pi\pi}}{\tau_\pi}&=\frac{10}{7}+{\cal O}(z^2)\;, 
\\
\frac{\lambda_{\pi\Pi}}{\tau_\pi}&=\frac{6}{5}+{\cal O}(z^2\log z)\;. 
\label{lambdapiPi}
\end{align}
%

\begin{figure*}[h!]
  \centering
  \includegraphics[width=0.75\linewidth]{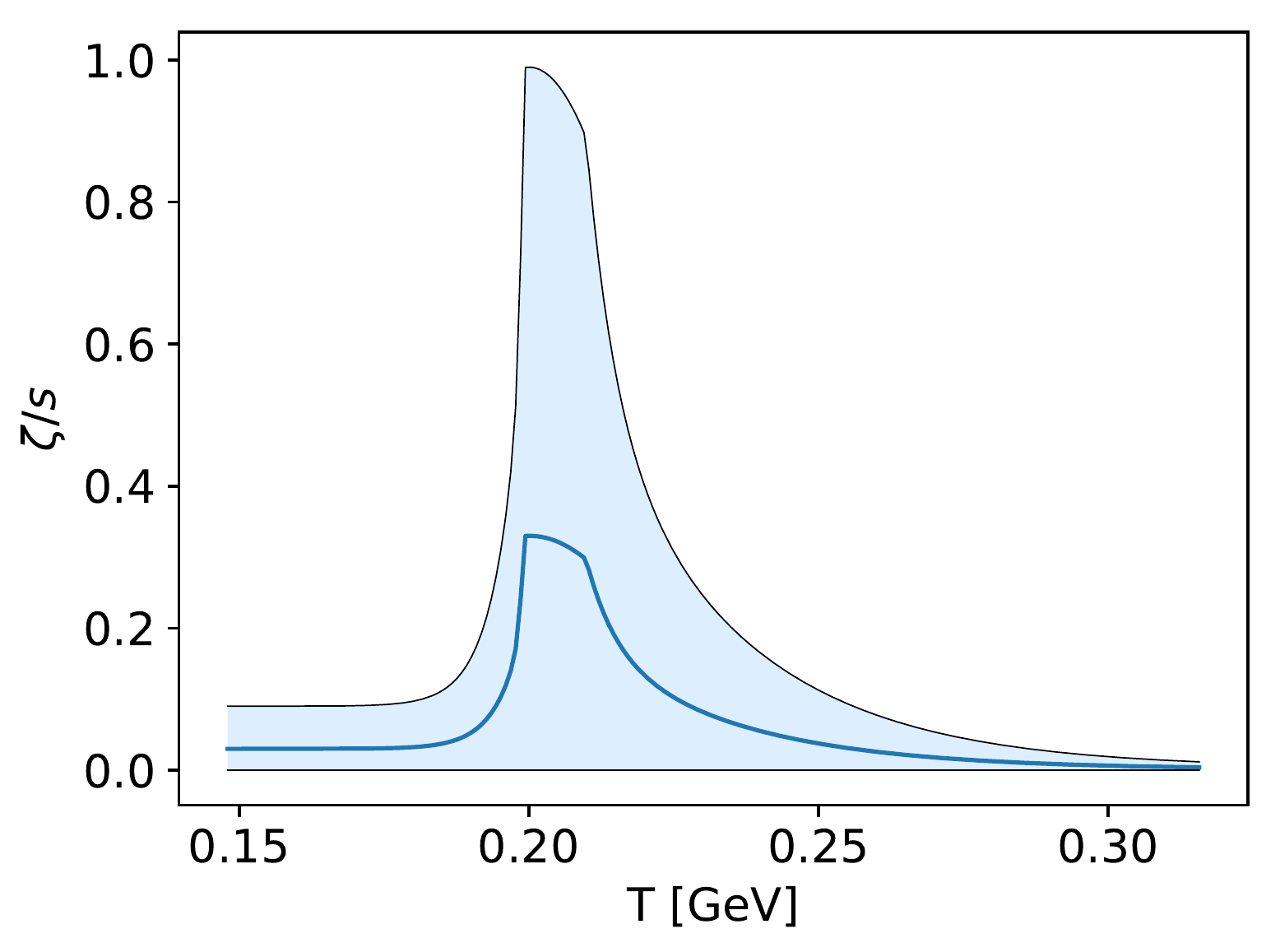}
  \caption{The specific bulk viscosity $\zeta/s$ as a function of temperature. The shaded light blue region indicates the estimated allowed range for $\zeta/s$~\cite{Heinz:2015arc,Bernhard:2016tnd}, while the darker blue line indicates the parametrization in Eq.~(\ref{eq:zetas}).}
  \label{specificBulkViscosityFig}
\end{figure*}
For the bulk viscosity, we use a parametrization that interpolates between data from lattice QCD for the QGP phase and results obtained from the hadron resonance gas model for the hadronic phase, connected quadratically around $T_c=200\,\mathrm{MeV}$~\cite{Denicol:2009am}:%
\footnote{In Ref~\cite{Denicol:2009am} $T_c=200\,\mathrm{MeV}$ was used, and we follow this choice here. Subsequent users have often used the value $T_c=180\,\mathrm{MeV}$ instead, see e.g.~\cite{Denicol:2015bpa,Heinz:2015arc,Bernhard:2016tnd}.}
\begin{equation}
{\small
  \frac{\zeta}{s} =
  \begin{cases}
  A_0 + A_1 x + A_2 x^2 &0.995T_c \ge T \ge 1.05T_c \\
    \lambda_1 \exp [-(x-1)/\sigma_1] + \lambda_2 \exp [-(x-1)/\sigma_2]+0.001
    &T > 1.05T_c \\
  \lambda_3 \exp [(x-1)/\sigma_3] + \lambda_4 \exp [ (x-1)/\sigma_4]+0.03
    &T < 0.995T_c 
  \end{cases},
  \label{eq:zetas}
}
\end{equation}
with $x = T/T_c$ and fitted parameters 
\begin{align*}
  &A_0=-13.45,\quad A_1=27.55,\quad A_2=-13.77, \\
  &\lambda_1=0.9,\quad \lambda_2=0.25,\quad\lambda_3=0.9,\quad \lambda_4=0.22,\\
  &\sigma_1=0.025,\quad \sigma_2=0.13, \quad\sigma_3=0.0025,\quad \sigma_4=0.022.
  \end{align*}
Fig.~\ref{specificBulkViscosityFig} shows an estimated range of the magnitude of the bulk viscosity by multiplying the parametrization (\ref{eq:zetas}) by an arbitrary normalization factor $0\leq(\zeta/s)_\mathrm{norm}\leq{3}$. A normalization constant $(\zeta/s)_\mathrm{norm}=1$ corresponds to Eq.~(\ref{eq:zetas}) indicated by the dark blue line in Fig.~\ref{specificBulkViscosityFig} and $(\zeta/s)_\mathrm{norm}=3$ has a peak value of $1$ similar to the two parametrization studied in Ref.~\cite{Denicol:2015bpa}. Simulations using Bayesian statistics to estimate the best parameter values from the experimental data find that $(\zeta/s)_\mathrm{norm}\sim 1.2$~\cite{Bernhard:2016tnd}. For simplicity, in all simulations we will use $(\zeta/s)_\mathrm{norm}=1$, since we are not worried here about precise comparison to data, but only interested that all qualitative features expected in nuclear collisions are implemented. With a peak value around $1$ (corresponding here to a $(\zeta/s)_\mathrm{norm}=3$) it has been shown that cavitation occurs~\cite{Denicol:2015bpa}, but for peak values lower than this, cavitation does not occur.
\subsection{Equation of state for nuclear collisions}
\label{sec:eos}
\begin{figure}[t!]
\begin{center}
  \begin{tabular}{cc}
  \includegraphics[width=0.45\linewidth]{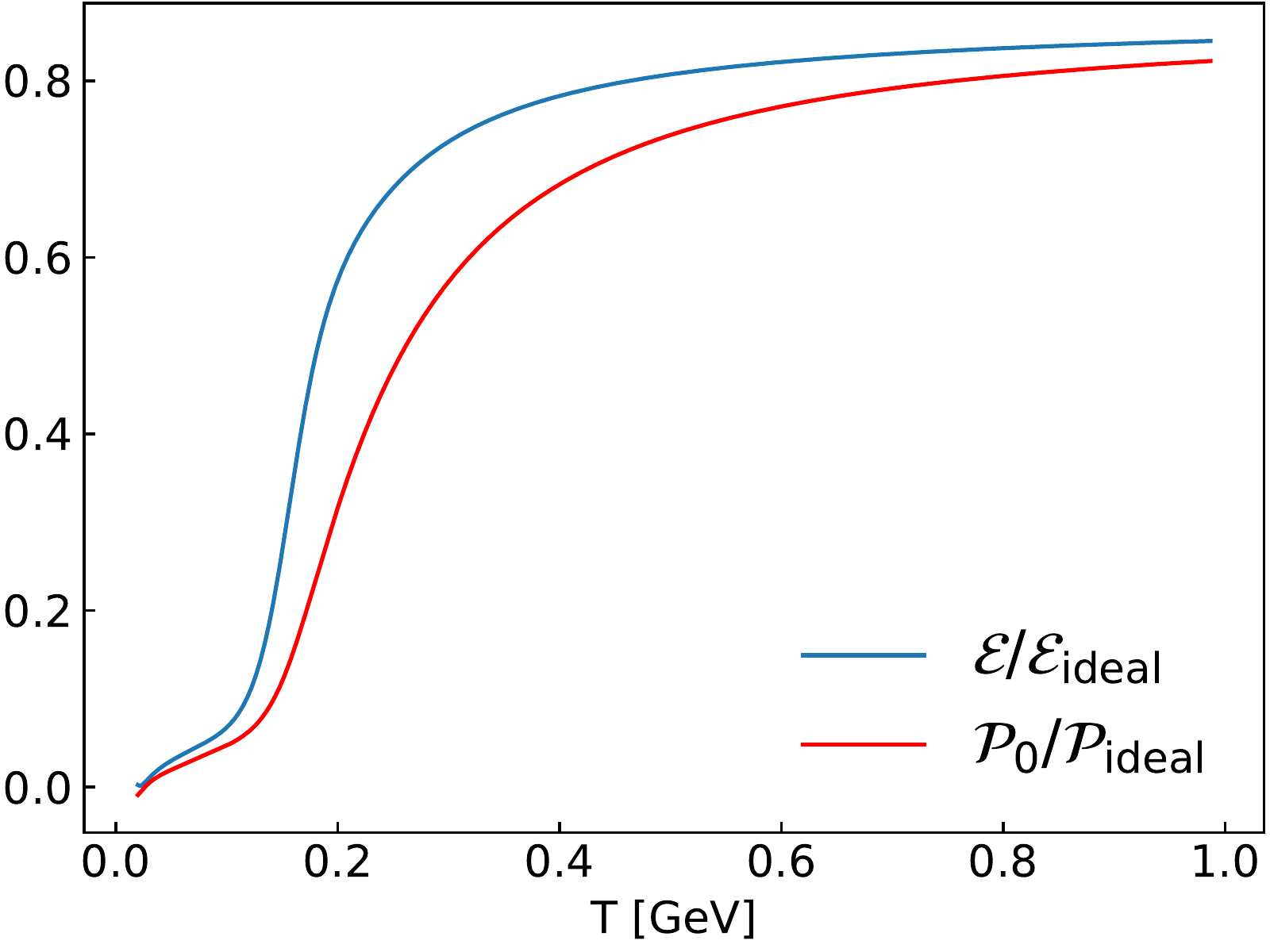} &
  \includegraphics[width=0.45\linewidth]{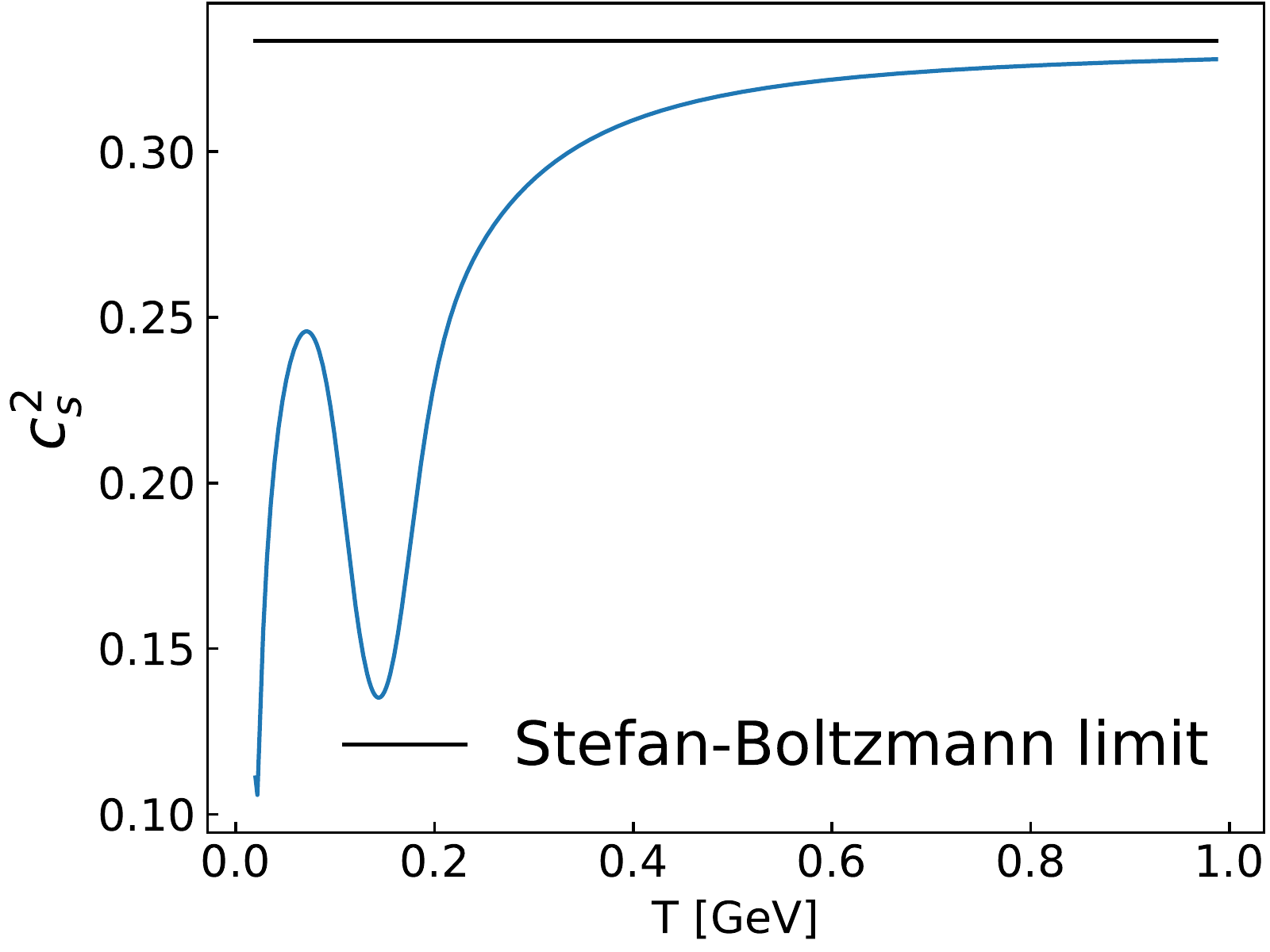} 
  \end{tabular}
\end{center}
\vspace{-7mm}
\caption{Pressure and energy density as a function of temperature normalized by their Stefan-Boltzmann limits (left panel) and the speed of sound (right panel) using the parametrization from the Wuppertal-Budapest collaboration~\cite{Borsanyi:2010cj}.}
\label{fig:eos}
\end{figure}
For an ideal gas of massless quarks and gluons (Stefan-Boltzmann limit)
\begin{align}
\ed_\mathrm{ideal}=3\left[
2(N^{2}_{c}-1)+\frac{7}{2}N_{c}N_{f}
\right]
\frac{\pi^{2}}{90}T^{4}\;,
\end{align}
where $N_{c}=3$ and $N_{f}=2+1$ are the numbers of colors and dynamical massless (2) and massive (1) quarks, respectively. The conformal equation of state gives $\ed_\mathrm{ideal}=3\p_\mathrm{ideal}$. However, when simulating the dynamics of the quark-gluon plasma, interactions and the running of the strong coupling parameter break conformal invariance. The corrections to the ideal equation of state can be determined from lattice QCD. Here, we use the analytic parametrization of the QCD trace anomaly $I(T)=\ed_\mathrm{eq}-3\p_\mathrm{eq}$ determined from lattice QCD by  the Wuppertal-Budapest collaboration~\cite{Borsanyi:2010cj}:
\begin{equation}
\frac{I(T)}{T^4}=\left[\frac{h_0}{1+\alpha t^2}+\frac{f_0\big[\tanh(f_1t+f_2)+1\big]}{1+g_1t+g_2t^2}\right]\exp\!\left(\frac{h_1}{t}-\frac{h_2}{t^2}\right)\;.
\label{traceAnomaly}
\end{equation}
Here, $t\equiv T/(0.2 \; \rm GeV)$, and for $N_f\equiv 2+1$ the fitting parameters are $h_0=0.1396$; $h_1=-0.1800$; $h_2=0.0350$; $f_0=2.76$; $f_1=6.79$; $f_2=-5.29$; $g_1=-0.47$; $g_2=1.04$; and $\alpha=0.01$.\footnote{We follow Ref.~\cite{Nopoush:2015yga} where a small $\alpha>0$ was introduced in order to guarantee that the pressure approaches the Stefan-Boltzmann limit in the high-temperature limit without affecting the parametrization near the phase transition.} The normalized pressure can be determined from the definite integral of the trace anomaly:
\begin{equation}
\frac{\p_{\rm eq}(T)}{T^4}=\int_0^T \frac{dT}{T}\frac{I(T)}{T^4} \, \label{peqDividedByT}.
\end{equation}
It is then straightforward, starting from the definition $\ed_\mathrm{eq}=3\p_\mathrm{eq}+I(T)$, to construct the inverse function $T(\ed_\mathrm{eq})$ and the speed of sound squared $c^2_s\equiv \partial \p_\mathrm{eq}/\partial \ed_\mathrm{eq}$. The energy density and pressure determined from the lattice and normalized by their Stefan-Boltzmann limits are plotted in the left panel of Fig.~\ref{fig:eos}. The right panel of Fig.~\ref{fig:eos} shows how $c^2_s$ approaches the Stefan-Boltzmann limit of $1/3$ as a function of temperature.
\subsection{Final equations in hyperbolic coordinates}
\label{sec:hyperbolicCoordinates}
Before proceeding, let us recall the decomposition of the covariant derivative into parts parallel and orthogonal to the fluid four-velocity for an arbitrary rank tensor:
\begin{equation}
A^{\mu_1\cdots\mu_\ell}_{;\alpha}\equiv
u_{\alpha}DA^{\mu_1\cdots\mu_\ell}+\nabla_{\alpha}A^{\mu_1\cdots\mu_\ell}\;,
\end{equation}
where 
\begin{align}
DA^{\mu_1\cdots\mu_\ell}&\equiv u^{\beta}A^{\mu_1\cdots\mu_\ell}_{;\beta}\;,
\\
\nabla_{\alpha}A^{\mu_1\cdots\mu_\ell}&\equiv
\Delta^{\beta}_{\alpha}A^{\mu_1\cdots\mu_\ell}_{;\beta}\;,
\end{align}
are the expressions for the  time derivative and the spatial gradient operators in the LRF, respectively.
Additionally, the convective time derivative is defined as 
$dA^{\mu_1\cdots\mu_\ell}\equiv u^{\beta}\partial_{\beta}A^{\mu_1\cdots\mu_\ell}$.
The covariant derivative of scalar quantities, contravariant four-vectors, and rank-two tensors are:
\begin{align}
(A^{\mu}A_{\mu})_{;\alpha}&\equiv\partial_{\alpha}(A^{\mu}A_{\mu})\;,
\\
A^{\mu}_{;\alpha}&\equiv\partial_{\alpha}A^{\mu}
+\Gamma^{\mu}_{\alpha\beta}A^{\beta}\;,
\\
A^{\mu\nu}_{;\alpha}&\equiv\partial_{\alpha}A^{\mu\nu}
+\Gamma^{\mu}_{\alpha\beta}A^{\beta\nu}
+\Gamma^{\nu}_{\alpha\beta}A^{\beta\mu}\;.
\end{align}
When there is a strong, approximately boost-invariant longitudinal flow (as is the case in relativistic nuclear collisions), the Milne coordinate system $x^{\mu}=(\tau,x,y,\eta_s)$ is the most natural one.
The longitudinal proper time is given by $\tau=(t^2-z^2)^{1/2}$ and the space-time rapidity is defined as $\eta_{s}=\frac{1}{2}\ln{[(t+z)/(t-z)]}$. Minkowski coordinates $t$ and $z$ are the laboratory time and the longitudinal (or beam) direction, respectively. The metric tensor in this coordinate system is
\begin{align}
g^{\mu\nu}&=\mathrm{diag}(1,-1,-1,-\tau^{-2})\;,\nonumber\\
g_{\mu\nu}&=\mathrm{diag}(1,-1,-1,-\tau^{2})\;,
\end{align}
which leads to $g=\tau^2$. The contravariant flow velocity is parametrized as
\begin{equation}
u^\mu\equiv(u^0,u^1,u^2,u^3)=(u_\tau,u_x,u_y,u_\eta)\;,
\label{fluidVelocity}
\end{equation}
and $u_{\mu}\equiv g_{\mu\nu}u^{\nu}=(u_\tau,-u_x,-u_y,-\tau^{2}u_\eta)$. The number of independent components of $u$ is fixed by the normalization condition $u^{\mu}u_{\mu}=1$, leading to $u_\tau=(1+u_x^2+u_y^2+\tau^2 u_\eta^2)^{1/2}$. For later convenience, we also introduce the scaled flow velocities (3-velocities of the fluid) $v_{i}\equiv u_{i}/u_{\tau}$.
The four-derivative is $\partial_{\mu}=(\partial_\tau,\partial_x,\partial_y,\partial_\eta)$ and $\partial^{\mu}\equiv g^{\mu\nu}\partial_{\nu}=(\partial_\tau,-\partial_x,-\partial_y,   -\tau^{-2}\partial_\eta)$. The only nonvanishing components of $\Gamma^{\mu}_{\alpha\beta}$ are
\begin{equation}
\Gamma^{\eta}_{\tau\eta}=\Gamma^{\eta}_{\eta\tau}=\frac{1}{\tau}\;,
\;\;\;
\Gamma^{\tau}_{\eta\eta}=\tau\;.
\end{equation}
As a result, the conservation laws become
%
%
\begin{eqnarray}
{\partial }_{\mu }T^{\mu\tau}& =&-\frac{1}{\tau}(T^{\tau\tau}+\tau^{2}T^{\eta\eta})\ , \label{dT001}\\
{\partial }_{\mu }T^{\mu x}& =&-\frac{1}{\tau}T^{\tau x}\ , \quad 
{\partial }_{\mu }T^{\mu y}=-\frac{1}{\tau}T^{\tau y}\ , \quad 
{\partial }_{\mu }T^{\mu \eta }=-\frac{3}{\tau}T^{\tau \eta}\ . \qquad
\label{dT0i1}
\end{eqnarray}
These equations can be decoupled by using the constituent equations (\ref{T}). Using $\p$ as a shorthand for $\peq+\Pi$, one has
\begin{align}
T^{\tau\tau}&\equiv(\ed+\p)u^{\tau}u^{\tau}-\p+\pi^{\tau\tau}\;,
\label{Ttt} \\
T^{\tau i}&\equiv(\ed+\p)u^{\tau}u^{i}+\pi^{\tau i}
=v^{i}T^{\tau\tau}+\p v^{i}-v^{i}\pi^{\tau\tau}+\pi^{\tau i}\;,
\label{Tti} \\
T^{ij}&\equiv(\ed+\p)u^{i}u^{j}-\p g^{ij}+\pi^{ij}
=v^{i}T^{\tau j}-\p g^{ij}-v^{i}\pi^{\tau i}+\pi^{ij}\;,
\label{Tij}
\end{align}
leading to the energy conservation equation
{\small
\begin{align}
\partial _{\tau }T^{\tau \tau }&+\partial _{x}(v_{x}T^{\tau \tau
})+\partial _{y}(v_{y}T^{\tau \tau })+\partial _{\eta }(v_{\eta }T^{\tau
\tau }) 
=-\frac{1}{\tau }\left( T^{\tau \tau }+\tau ^{2}T^{\eta \eta
}\right) \notag \\
 & \hspace{-5mm} -\partial _{x}\left( v_{x}\p-v_{x}\pi ^{\tau \tau }+\pi ^{\tau x}\right)
-\partial _{y}\left( v_{y}\p-v_{y}\pi ^{\tau \tau }+\pi ^{\tau y}\right) 
 -\partial _{\eta }\left( v_{\eta }\p-v_{\eta }\pi ^{\tau \tau }+\pi ^{\tau
\eta }\right) \,,
\label{dtT00}
\end{align}
}
and the momentum conservation equations
\begin{align}
\partial _{\tau }T^{\tau x}&+\partial _{x}(v_{x}T^{\tau x})+\partial
_{y}(v_{y}T^{\tau x})+\partial _{\eta }(v_{\eta }T^{\tau x})
=-\frac{1}{\tau }T^{\tau x}  \\
& -\partial _{x}\left( \p-v_{x}\pi ^{\tau x}+\pi ^{xx}\right) -\partial
_{y}\left( -v_{y}\pi ^{\tau x}+\pi ^{xy}\right)  
 -\partial _{\eta}\left( -v_{\eta }\pi ^{\tau x}+\pi ^{x\eta }\right) \,, \notag \\
\partial _{\tau }T^{\tau y}&+\partial _{x}(v_{x}T^{\tau y})+\partial
_{y}(v_{y}T^{\tau y})+\partial _{\eta }(v_{\eta }T^{\tau y})  
=-\frac{1}{\tau }T^{\tau y}  \\
& -\partial _{x}\left( -v_{x}\pi ^{\tau y}+\pi ^{xy}\right) -\partial
_{y}\left( \p-v_{y}\pi ^{\tau y}+\pi ^{yy}\right) 
 -\partial _{\eta }\left( -v_{\eta }\pi ^{\tau y}+\pi ^{y\eta }\right) \,, 
 \notag \\
\partial _{\tau }T^{\tau \eta }&+\partial _{x}(v_{x}T^{\tau \eta
})+\partial _{y}(v_{y}T^{\tau \eta })+\partial _{\eta }(v_{\eta }T^{\tau
\eta })
=-\frac{3}{\tau }T^{\tau \eta } \label{dtT03}\\
&-\partial _{x}\left( -v_{x}\pi ^{\tau \eta }+\pi ^{x\eta }\right)
-\partial _{y}\left( -v_{y}\pi ^{\tau \eta }+\pi ^{y\eta }\right)  
 -\partial _{\eta }\left( \frac{\p}{\tau ^{2}}-v_{\eta }\pi ^{\tau \eta
}+\pi ^{\eta \eta }\right) \,. \notag
\end{align}
We remark that the substitutions (\ref{Ttt})-(\ref{Tij}) manifest themselves as additional source terms containing spatial derivatives in the second lines of Eqs.~(\ref{dtT00})-(\ref{dtT03}). The relaxation equations~(\ref{pimunuSimplified}) are
\begin{align}
d\Pi&=-\frac{\zeta}{\tau_\Pi}\theta-\frac{\Pi}{\tau_\Pi}-I_\Pi \label{relEqs_Pi}
\;,\\
d\pi^{\mu\nu}&=2\frac{\eta}{\tau_{\pi}}\sigma^{\mu\nu}-\frac{\pi^{\mu\nu}}{\tau_{\pi}}-I^{\mu\nu}_{\pi}-G^{\mu\nu}_{\pi}\;,
\label{relEqs}
\end{align}
where we introduced the shorthand notation $G^{\mu\nu}_{\pi}\equiv u^{\alpha}\Gamma^{\mu}_{\alpha\beta}\pi^{\beta\nu}
+u^{\alpha}\Gamma^{\nu}_{\alpha\beta}\pi^{\beta\mu}$ which contains the geometrical source terms from writing the covariant derivative of $\pi^{\mu\nu}$ in terms of the convective derivative, and 
\begin{align}
I_{\Pi}&\equiv \frac{\delta_{\Pi\Pi}}{\tau_\Pi}\Pi\theta-\frac{\lambda_{\Pi\pi}}{\tau_\Pi}\pi^{\mu\nu}\sigma_{\mu\nu}\;,\label{dPi}
\\
I^{\mu\nu}_{\pi}&\equiv I^{\mu\nu}_{1}
+\frac{\delta_{\pi\pi}}{\tau_\pi}I^{\mu\nu}_{2}
-I^{\mu\nu}_{3}
+\frac{\tau_{\pi\pi}}{\tau_\pi}I^{\mu\nu}_{4}
-\frac{\lambda_{\pi\Pi}}{\tau_\pi}\Pi\sigma^{\mu\nu}\;,
\end{align}
which contains all terms in the theory that are of second order in the Knudsen number:
\begin{align}
I_{1}^{\mu \nu }& =\left( \pi ^{\lambda \mu }u^{\nu }+\pi ^{\lambda \nu
}u^{\mu }\right) Du_{\lambda }\ , \label{I1} \\
I_{2}^{\mu \nu }& =\theta \pi ^{\mu \nu }\ , \\
I_{3}^{\mu \nu }& =\pi ^{\mu \lambda }\omega _{\left. {}\right. \lambda
}^{\nu }+\pi ^{\nu \lambda }\omega _{\left. {}\right. \lambda }^{\mu }\ , \\
I_{4}^{\mu \nu }& =\frac{1}{2}g_{\lambda \kappa }\left( \pi ^{\mu \kappa
}\sigma ^{\nu \lambda }+\pi ^{\nu \kappa }\sigma ^{\mu \lambda }\right) -%
\frac{1}{3}\Delta ^{\mu \nu }\pi _{\beta }^{\alpha }\sigma _{\alpha }^{\beta
}\ .\label{I4}
\end{align}
We have not written out the convective derivative in ($\tau$-$\eta_{s}$)-coordinates, $d\equiv u_{\tau}\partial_{\tau}+u_{x}\partial_{x}
+u_{y}\partial_{y}+u_{\eta}\partial_{\eta}$, nor the individual components of the other tensors in Eqs.~(\ref{I1})-(\ref{I4}); they are listed in~\ref{sec:milne_eqns}.
Eq.~(\ref{relEqs}) contains 10 components of the symmetric shear stress tensor. By using the tracelessness and orthogonality of $\pi^{\mu\nu}$ to $u$ we can further reduce the number of unknowns to five. The choice is arbitrary as to which five components of $\pi^{\mu\nu}$ are chosen as dynamical quantities -- for example, one could choose $\pi^{xx}$, $\pi^{xy}$, $\pi^{x\eta}$, $\pi^{yy}$, and $\pi^{y\eta}$ -- and then the remaining components are determined algebraically. Four equations follow from $\pi^{\mu\nu}u_\nu=0$:
\begin{eqnarray}
\pi ^{\tau \tau } &=&\pi ^{\tau x}v_{x}+\pi ^{\tau y}v_{y}+\tau ^{2}\pi
^{\tau \eta }v_{\eta }\,,  \label{pi_tautau} \\
\pi ^{\tau x} &=&\pi ^{xx}v_{x}+\pi ^{xy}v_{y}+\tau ^{2}\pi ^{x\eta }v_{\eta
}\,, \\
\pi ^{\tau y} &=&\pi ^{xy}v_{x}+\pi ^{yy}v_{y}+\tau ^{2}\pi ^{y\eta }v_{\eta
}\,, \\
\pi ^{\tau \eta } &=&\pi ^{x\eta }v_{x}+\pi ^{y\eta }v_{y}+\tau ^{2}\pi
^{\eta \eta }v_{\eta }\,,
\end{eqnarray}
and the final component is obtained by using $\pi^{\mu\nu}g_{\mu\nu}=0$:
\begin{align}
\pi ^{\eta \eta }& \equiv \tau ^{-2}\left( \pi ^{\tau \tau }-\pi ^{xx}-\pi
^{yy}\right)  \notag \\
& =\tau ^{-2}\left[ \pi ^{xx}\left( v_{x}^{2}-1\right) +\pi ^{yy}\left(
v_{y}^{2}-1\right) +2\pi ^{xy}v_{x}v_{y}\right.  \notag \\
& \left. +2\tau ^{2}\left( \pi ^{x\eta }v_{x}v_{\eta }+\pi ^{y\eta
}v_{y}v_{\eta }\right) \right] /\left( 1-\tau ^{2}v_{\eta }^{2}\right) \ .
\label{pi_etaeta}
\end{align}
We will actually explicitly propagate all ten components of the shear-stress tensor and use the above algebraic equations as a check that our code preserves the traceless and orthogonality conditions numerically to a given precision. 

The solution to the final fluid dynamic equations of motion (\ref{dT001})-(\ref{relEqs}) gives us dynamical information about $T^{\tau\nu}$ and $\pi^{\mu\nu}$ -- this information then allows us to determine the local fields $\ed$ and $u^{\mu}$. To do this, we first define the known quantities $E\equiv T^{\tau\tau}-\pi^{\tau\tau}=T_{\tau\tau}-\pi_{\tau\tau}$, $\bar{M}_{i}\equiv T^{\tau i}-\pi^{\tau i}$, and $\ubar{M}_{i}\equiv T_{\tau i}-\pi_{\tau i}$. Then, from Eqs.~(\ref{Ttt}) and~(\ref{Tti}), the local rest frame energy density is solved iteratively via
\begin{equation}
\ed=E-\frac{M^2}{E+\p(\ed)}\;,
\label{eInferred}
\end{equation}
where $M^2\equiv \bar{M}_{i}\ubar{M}_{i}=\bar{M}^{2}_{x}+\bar{M}^{2}_{y}+\tau^{2}\bar{M}^{2}_{\eta}$.~[For a conformal system with $\Pi\equiv 0$ and equation of state $\ed=3\peq$, $\ed$ is determined algebraically as $\ed=\sqrt{4E^2-3M^2}-E$.] Knowing $\ed$, we can calculate the equation of state $\peq=\peq(\ed)$ as well as $\p(\ed)=\peq(\ed)+\Pi$ and then fully construct the components of the fluid velocity:
\begin{align}
u_{\tau}&=\sqrt{\frac{E+\p}{\ed+\p}}\;,
\label{utInferred}
\\
u_{i}&=\frac{\bar{M}^{i}}{(\ed+\p)u_{\tau}}\;.
\label{uiInferred}
\end{align}
The numerical scheme used to solve Eqs.~(\ref{dT001})-(\ref{relEqs}) is described in Sec.~\ref{sec:num_scheme}, followed in Sec.~\ref{sec:gpu_imp} by its implementation on graphics processing units. We then validate our code against various numerical tests in Sec.~\ref{sec:num_tests}.
\section{Numerical scheme}
\label{sec:num_scheme}
\subsection{The KT-RK algorithm}
\label{sec:ktrkalg}
The relativistic dissipative fluid dynamic equations (the conservation laws of the physical currents together with the relaxation equations) can be cast in first-order flux conservative vector form,
\begin{equation}
\frac{\partial {\bf q}}{\partial{\tau}}+\frac{\partial {\bf F}^{i}[{\bf q}]}{\partial{x}^{i}}
={\bf J}\;.
\label{numericalEquations}
\end{equation}
In the above equation, ${\bf q}$ is the conserved state vector, ${\bf F}^{i}$ is the flux functional of ${\bf q}$ (where Latin indices run from 1 to 3), and ${\bf J}$ is the source term. [See~\ref{sec:cons_form} for details concerning how to write Eqs.~(\ref{relEqs_Pi}) and (\ref{relEqs}) in conservative form.]
We will now explain our notation: Let $N_{cons}$ be the number of physical conserved currents in (\ref{hydro_eqs}) and $N_{diss}$ be the number of dissipative currents, respectively. We define quantities with an over-hat, i.e.~$\hat{a}$, as $N_{cons}$-dimensional vectors while quantities with an over-vector symbol, i.e~$\vec{a}$, are $N_{diss}$-dimensional vectors. The boldfaced notation denotes a $N$-dimensional vector; $N\equiv N_{cons}+N_{diss}$ is the sum of the number of variables from the physical conservation laws and the number of dissipative quantities promoted to conserved variables for our numerical scheme. For example, we define  
\begin{eqnarray}
&&\hat{T}\equiv 
 \begin{pmatrix}
  T^{\tau\tau} & T^{\tau x} & T^{\tau y} & T^{\tau\eta} 
 \end{pmatrix}^T \,, \\
&&\vec{\pi}\equiv 
 \begin{pmatrix}
  \pi^{\tau \tau} & \pi^{\tau x} & \pi^{\tau y} & \pi^{\tau \eta} &
  \pi^{xx} & \pi^{xy} & \pi^{x\eta} &
  \pi^{yy} & \pi^{y\eta} & \pi^{\eta\eta} & \Pi  
 \end{pmatrix}^T \,,\quad
\end{eqnarray}
so that the vector of (numerical) conservative variables is the concatenation of the two of them, given by 
\setcounter{MaxMatrixCols}{20}
\begin{eqnarray}
\label{qComponents}
&&\bm{q}\equiv 
 \begin{pmatrix}
  \hat{T} & \vec{\pi}
 \end{pmatrix}^T \,,
\end{eqnarray}
and the flux vectors are ${\bf F}_{x}\equiv v_{x}{\bf q}$, ${\bf F}_{y}\equiv v_{y}{\bf q}$, ${\bf F}_{\eta}\equiv v_{\eta}{\bf q}$.
The source vector is given by ${\bf J}=(\partial_{i}v_{i}+\hat{S}_{c}+\hat{G}_{c},\partial_{i}v_{i}+\vec{S}_\pi+\vec{G}_\pi)$ 
with the individual terms specified in Eqs.~(\ref{B13}). 

We now briefly describe the numerical scheme used for the integration of the system of hyperbolic equations (\ref{numericalEquations}).~Godunov-type schemes belong to a class of finite volume methods that guarantee the conservation of the primary variables (or vary correctly based on the true physical solution). The two main kinds of high resolution Godunov-type methods are the upwind and central schemes. In upwind schemes, the fluxes at the cell boundaries are computed based on the exact analytical or approximate solutions to the Riemann problem.~Central schemes bypass the need to employ costly (approximate) Riemann solvers -- there are no known analytic or approximate solutions in the case of dissipative fluid dynamics -- and in contrast to the former, this makes them simpler, more efficient, and universal. The Lax-Friedrichs scheme (LxF)~\cite{Lax,Friedrichs} is a first-order central scheme based on piecewise constant approximate solutions and forms the basis for all higher-order schemes. Nessyahu and Tadmor (NT) introduced a generalization of the staggered LxF scheme by replacing the first order piecewise constant solutions with a van-Leer's MUSCL-type piecewise linear second order approximation~\cite{Nessyahu}. Higher order schemes have also been developed (see e.g. Refs.~\cite{Liu,Bianco}). These staggered central schemes have numerical viscosity which is of order ${\cal O}[(\Delta x)^{2r}/\Delta \tau]$, where $r$ is the formal order of the scheme.

The numerical viscosity in the central schemes becomes problematic when the time step $\Delta\tau$ is enforced to be small due to stability restrictions or in order to sufficiently resolve all relevant time scales in the problem. For fluid dynamics, the latter requires that $\Delta \tau\ll\tau_\mathrm{micro}$, where $\tau_\mathrm{micro}$ is the largest microscopic time scale. For heavy-ion collisions considered herein, the slowest microscopic time scales are thought to be of the same order as the macroscopic scales. However, faster varying time scales would need to be included to handle problems with strong initial gradients in pressure or number density (see the discussion above in Sec.~\ref{sec:fluidDynamicsNuclearCollisions} or Ref.~\cite{Denicol:2012vq}). Additionally, important microscopic physics that might develop on time scales much shorter than the macroscopic ones (provided such time scales are included in the effective description of the underlying microscopic theory~\cite{Denicol:2012cn,Bazow:2015zca}) would require a sufficiently small $\Delta\tau$ (in the infinite resolution limit, $\Delta\tau\to 0$) in order to numerically resolve these fine details~\cite{PhysRevD.66.124013,PhysRevD.67.124013,
PhysRevLett.102.211601,Chesler2014,PhysRevLett.113.261601,Bazow:2015zca}. Moreover, the NT scheme (and the higher order generalizations thereof) do not permit a semi-discrete form.

Kurganov and Tadmor (KT) cured the aforementioned problems by introducing a new family of high-order Godunov-type central schemes with a much smaller numerical viscosity of order ${\cal O}[(\Delta x)^{2r-1}]$ (independent of $\Delta\tau$)~\cite{Kurganov}. The main ingredient was to use more precise characteristic information of the local speed of propagation. In the KT algorithm, the high resolution of the upwind Godunov-type schemes is retained without the need to solve the actual Riemann problem. By letting $\Delta \tau\to 0$ it permits a simple semi-discrete update equation for Eq.~(\ref{numericalEquations}):
{\small
\begin{align}
\frac{d}{d\tau}{\bf q}_{ijk}&=
-\frac{{\bf H}^{x}_{i+1/2,j,k}-{\bf H}^{x}_{i-1/2,j,k}}{\Delta x}
-\frac{{\bf H}^{y}_{i,j+1/2,k}-{\bf H}^{y}_{i,j-1/2,k}}{\Delta y}
-\frac{{\bf H}^{\eta}_{i,j,k+1/2}-{\bf H}^{\eta}_{i,j,k-1/2}}{\Delta \eta}
\notag \\
&\hspace{5mm} +{\bf J}[{\bf q}_{ijk}] \label{semiDiscreteKT}
\\
&\equiv {\bf C}[\q]\;,
\label{semiDiscreteKT_C}
\end{align}
}
where the $ijk$ indices on $\q$ are the integer labels for the $x$, $y$, and $\eta_{s}$ coordinates of the grid point, and $\Delta x$, $\Delta y$ and $\Delta \eta$ are the numerical resolution in the $x$, $y$, and $\eta_{s}$ coordinates. The numerical fluxes are given by
\begin{align}
{\bf H}^{x}_{i\pm 1/2,j,k}\equiv
\frac{{\bf F}^{x}[{\bf q}^{+}_{i\pm 1/2,j,k}]
+{\bf F}^{x}[{\bf q}^{-}_{i\pm 1/2,j,k}]}{2}
-a^{x}_{i\pm 1/2,j,k}
\frac{{\bf q}^{+}_{i\pm 1/2,j,k}-{\bf q}^{-}_{i\pm 1/2,j,k}}{2}\;,
\label{Hflux}
\end{align}
and similarly in the $y$ and $\eta_s$ directions. The KT algorithm makes use of the local propagation speed at each cell interface, defined as
\begin{align}
a^{x}_{i\pm 1/2,j,k}\equiv
\mathrm{max}\left\{
\rho\left(\frac{\partial {\bf F}^{x}}{\partial {\bf q}}
[{\bf q}^{+}_{i\pm 1/2,j,k}]\right),\,
\rho\left(\frac{\partial {\bf F}^{x}}{\partial {\bf q}}
[{\bf q}^{-}_{i\pm 1/2,j,k}]\right)
\right\}\;,
\label{ax}
\end{align}
where $\rho(A)\equiv\mathrm{max}_{i}(|\lambda_{i}(A)|)$ is the spectral radius and $\lambda_{i}(A)$ are the eigenvalues of the Jacobian matrix $A\equiv \partial {\bf F}/\partial {\bf q}$. For our situation with ${\bf F}^{i}=v^{i}\q$, the local propagation speeds become quite simple: $a^{i}\equiv |v_{i}|=|u_{i}/u_{\tau}|$. Additionally, the superscripts $+$ or $-$ stand for the reconstructed values of ${\bf q}$ at the left and right sides of the corresponding numerical cells $i+1/2$ and $i-1/2$, respectively. These intermediate values are given by
\begin{align}
{\bf q}^{+}_{i+1/2,j,k}&\equiv
{\bf q}_{i+1,j,k}-\frac{\Delta x}{2}({\bf q}_{x})_{i+1,j,k}\;,
\label{rightHalfCellExtrapolationForward}
\\
{\bf q}^{-}_{i+1/2,j,k}&\equiv
{\bf q}_{i,j,k}+\frac{\Delta x}{2}({\bf q}_{x})_{i,j,k}\;,
\label{leftHalfCellExtrapolationForward}
\\
{\bf q}^{+}_{i-1/2,j,k}&\equiv
{\bf q}_{i,j,k}-\frac{\Delta x}{2}({\bf q}_{x})_{i,j,k}\;,
\label{rightHalfCellExtrapolationBackwards}
\\
{\bf q}^{-}_{i-1/2,j,k}&\equiv
{\bf q}_{i-1,j,k}+\frac{\Delta x}{2}({\bf q}_{x})_{i-1,j,k}\;,
\label{leftHalfCellExtrapolationBackwards}
\end{align}
where the approximate spatial derivatives $(\q_x)_{ijk}$ satisfy the so-called total variation diminishing (TVD) property \cite{leveque2002finite}, thereby achieving flux limitation. An appropriate flux limiter function does not introduce spurious oscillations and interpolates between low and high resolution schemes when there are sharp (or zero) gradients and smooth solutions, respectively. The use of the generalized minmod limiter for the numerical derivatives, 
\begin{align}
({\bf q}_{x})_{i,j,k}\equiv
\mathrm{\texttt{minmod}}\left(
\theta\frac{{\bf q}_{i,j,k}{-}{\bf q}_{i-1,j,k}}{\Delta x},\,
\frac{{\bf q}_{i+1,j,k}{-}{\bf q}_{i-1,j,k}}{2\Delta x},\,
\theta\frac{{\bf q}_{i+1,j,k}{-}{\bf q}_{i,j,k}}{\Delta x}
\right)\;,
\end{align}
guarantees the TVD non-oscillatory property in the sense of satisfying a local scalar maximum principle (see Thm.~1 in Ref.~\cite{Jiang}). The parameter $\theta\in[1,2]$; $\theta=1$ corresponds to the most dissipative limiter while $\theta=2$ the least dissipative limiter. The multivariate minmod function is defined as
\begin{align}
\mathrm{\texttt{minmod}(x,y,z)}\equiv \mathrm{\texttt{minmod}(x,\mathrm{\texttt{minmod}}(y,z))}\;,
\end{align}
where $\mathrm{\texttt{minmod}(x,y)}\equiv[\mathrm{\texttt{sgn}(x)}+\mathrm{\texttt{sgn}(y)]}\cdot
\mathrm{\texttt{min}(|x|,|y|)/2}$ and $\mathrm{\texttt{sgn(x)}}\equiv\texttt{|x|/x}$. The numerical flux functions ${\bf H}^{y}$ and ${\bf H}^{\eta}$ with their corresponding local propagation speeds, intermediate values, and numerical derivatives can be defined accordingly, by permuting the triad $(i,j,k)$ and the spatial resolutions.  Finally, we note that it was shown in Ref.~\cite{Naidoo} how to deal with time-dependent nonlinear source terms in the KT scheme.

The importance of the semi-discrete formulation (\ref{semiDiscreteKT}) is that  we can use the method of lines, resulting in a system of ordinary differential equations to solve.~In the fully discrete formulation of the KT (or NT) algorithm one is constrained to multi-level time differencing. Instead, we couple the spatial discretization of the KT scheme with an efficient Runge-Kutta ODE solver for their time integration. We prefer to use a high-order explicit Runge-Kutta algorithm built up from convex combinations of the forward Euler step, ${\bf E}[w]\equiv w+\Delta\tau {\bf C}[w]$, where ${\bf C}[w]$ is the spatial recipe of the KT scheme as defined in Eq.~(\ref{semiDiscreteKT_C}).
A one-parameter family of RK schemes is defined by~\cite{Shu,ShuOsher}:
\begin{align}
\q^{(1)}&={\bf E}[\q^{n}]\;,
\label{step1}
\\
\q^{(\ell+1)}&=\beta_{\ell}\q^{n}+(1-\beta_{\ell}){\bf E}[\q^{(\ell)}]\;,\;\;\;\;
\ell=1,\dots,s-1\;,
\\
\q^{n+1}&\equiv \q^{(s)}\;.
\label{finalStep}
\end{align}
Here, ${\bf q}^{n}_{ijk}$ is an approximate value of 
${\bf q}(\tau=\tau^{n},x=x_{i},y=y_{j},\eta_{s}=\eta_k)$ at the four-dimensional grid point $(\tau^{n}\equiv n\Delta\tau,\,x_{i}\equiv i\Delta x,\,y_{j}\equiv j\Delta y,\,\eta_{k}\equiv k\Delta\eta)$. For the two-step modified Euler algorithm used in our code ($s=2$), the only coefficient needed is $\beta_{1}=1/2$.

The final piece of the algorithm is that the local propagation speeds in Eq.~(\ref{ax}) and the source terms ${\bf J}$ depend on the primary variables $\ed$ and $u^{\mu}$ calculated via Eqs.~(\ref{eInferred})-(\ref{uiInferred}). When calculating the numerical fluxes, the primary variables must be calculated from the intermediate values ${\bf q}^{\pm}_{\pm 1/2}$ so that $a^{x}_{i\pm 1/2,j,k}=|v_{x}|$ etc. can be determined. To deal with the source terms, the primary variables must be calculated before every step listed in Eqs.~(\ref{step1})-(\ref{finalStep}). Whenever the source terms depend on spatial derivatives of the conserved or primary variables, they are calculated using second-order central differences,\footnote{See Sec.~\ref{sec:compLattice} for how we treat the boundary conditions.} $\partial_{x}A_{i}=(A_{i+1}-A_{i-1})/(2\Delta x)$. The time derivatives in the source terms are calculated using first-order forward differences, $\partial_{\tau}A^{n}_{i}=(A^{n}_{i}-A^{n-1}_{i})/\Delta\tau$.

\subsection{Regularization of the dissipative currents}
\label{sec:reg}
In practice, to solve Eq.~(\ref{eInferred}) we write
\begin{equation}
f(\ed)\equiv\ed-E+\frac{M^2}{E+\p(\ed)}\;,
\end{equation}
and solve for the roots $f(\ed)=0$. The root solving algorithm (for which we use the Newton-Raphson method) will not be able to find a solution if $\pi^{\mu\nu}$ becomes too large compared to the local equilibrium energy-momentum tensor $T^{\mu\nu}_{0}\equiv\ed u^{\mu}u^{\nu}-\peq\Delta^{\mu\mu}$. This can happen at early times and in the cold/dilute regions of plasma. To ensure that $\pi^{\mu\nu}$ remains smaller than $T^{\mu\nu}_{0}$ we regulate it by~\cite{Shen:2014vra}
\begin{equation}
\pi^{\mu\nu}\to\frac{\tanh\rho}{\rho}\pi^{\mu\nu}\;,
\label{regPimunu}
\end{equation}
where
\begin{equation}
\rho\equiv\mathrm{max}\left[
\frac{\sqrt{\Pi^{\mu\nu}\Pi_{\mu\nu}}}{\rho_\mathrm{max}\sqrt{\ed^2+3\p^2}}\;,
\frac{g_{\mu\nu}\pi^{\mu\nu}}{\xi_{0}\rho_\mathrm{max}\sqrt{\Pi^{\mu\nu}\Pi_{\mu\nu}}}\;,
\frac{\pi^{\lambda\mu}u_{\mu}}{\xi_{0}\rho_\mathrm{max}\sqrt{\Pi^{\mu\nu}\Pi_{\mu\nu}}}
\right]\;, \;\forall\;\lambda\;.
\label{rho}
\end{equation}
Above, we used the shorthand notation $\Pi^{\mu\nu}\equiv\Pi\Delta^{\mu\nu}+\pi^{\mu\nu}$. We adopted the values $\xi_0\equiv 0.1$ and $\rho_\mathrm{max}\equiv 1$ used in Ref.~\cite{Shen:2014vra} in order to make sure that the root finding algorithm was able to converge on a solution. 
\section{GPU implementation: GPU-VH}
\label{sec:gpu_imp}
The KT algorithm described above is readily adaptable for parallelism. It is a single-instruction, multiple data (SIMD) algorithm which makes its implementation onto GPUs straightforward. The most easily parellizable problems are those where every grid point decouples from all others, so-called ``embarrassingly parallel problems". The KT algorithm is classified as a distributed parallel problem because it needs communication between various processes and communication of intermediate results. However, because this grid point coupling is local rather than long range, there is not a significant communication bottleneck between different parallel processes. We will now describe how we implement this problem on a single GPU. We call this code GPU-VH.  
\subsection{Lattice}
\label{sec:compLattice}
\begin{figure}[t!]
\begin{center}
  \begin{tabular}{cc}
    \includegraphics[width=0.45\linewidth]{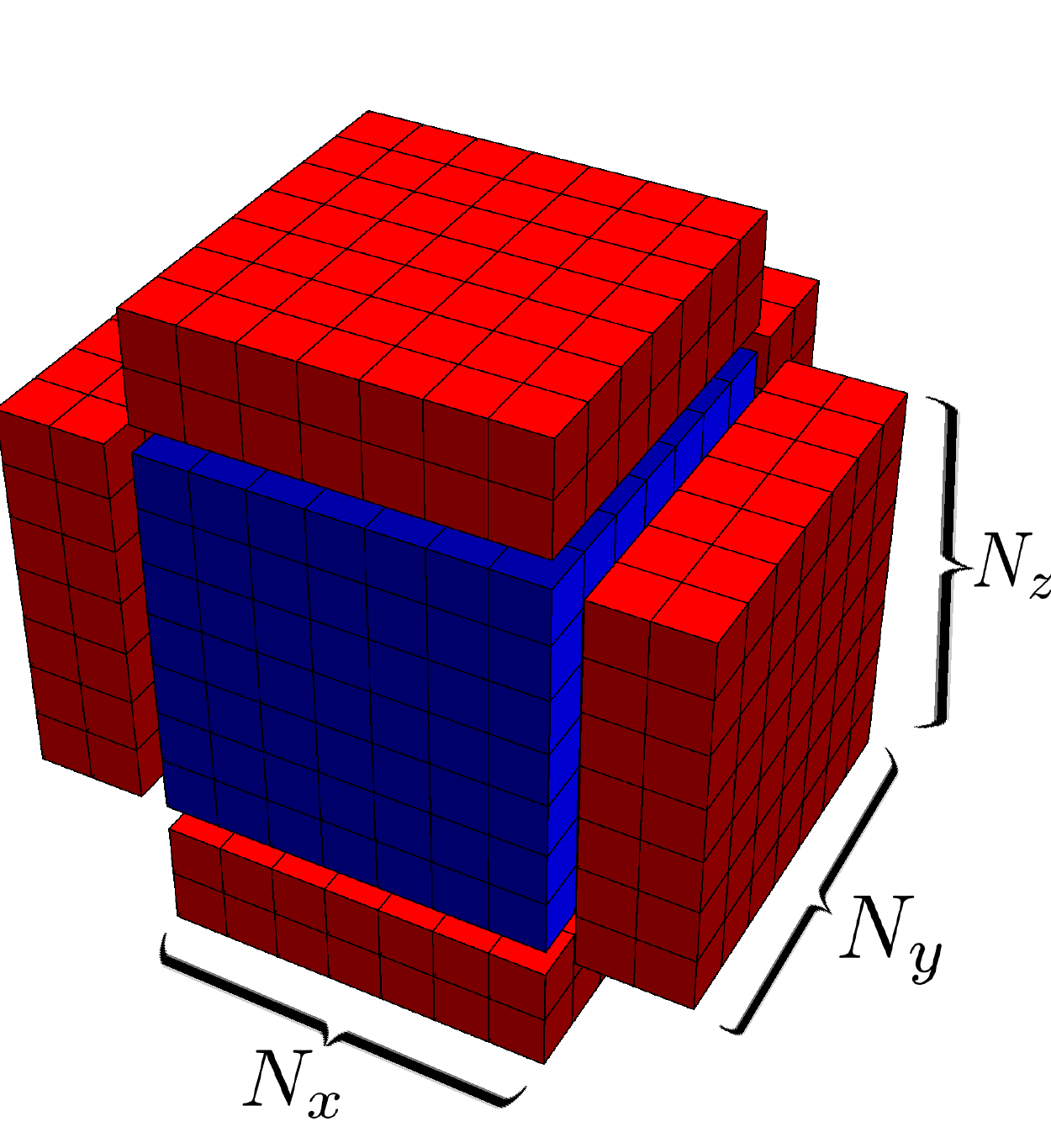} &
  	\includegraphics[width=0.45\linewidth]{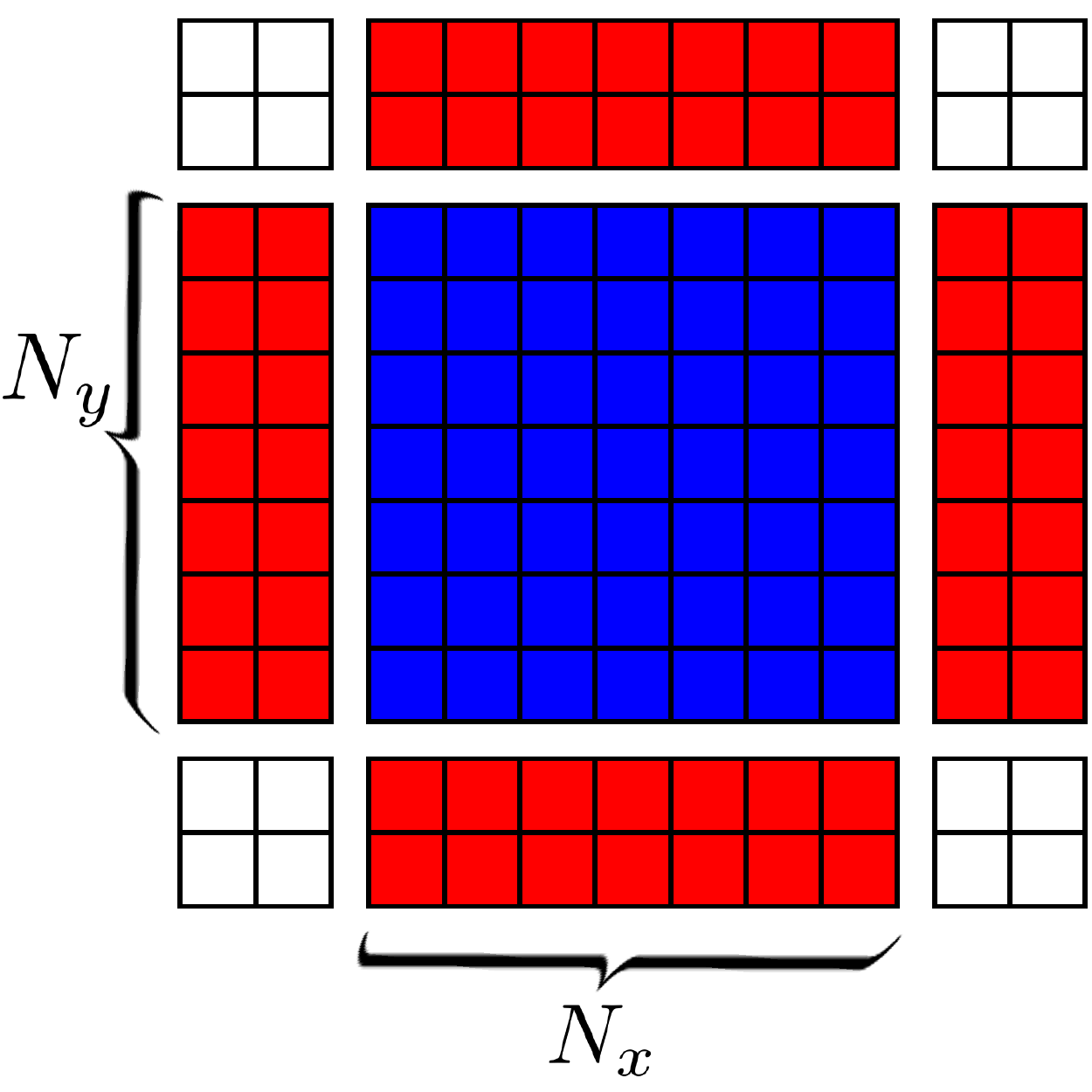} 
  \end{tabular}
\end{center}
\vspace{-7mm}
\caption{Depiction of our computational lattice used in GPU-VH. The blue region indicats the physical volume of the simulation. The red cells are ghost cells that are added to allow the information in the cells on the edge of the blue region to be evolved in time with the same set of instructions (using the fluxes at the boundary) as for interior cells. The left plot shows the three-dimensional lattice (with the ghost cells in the front plane removed for visualization purposes) while the right plot shows a two-dimensional slice of the 3D data. White cells are never accessed by the algorithm.}
\label{fig:compGrid}
\end{figure}
In our program, the numbers of physical grid points in our 3D lattice are $N_x$, $N_y$, and $N_\eta$. The grid point triad $(i,j,k)$ runs over $i=2,\dots,N_x+1$; $j=2,\dots,N_y+1$; and $k=2,\dots,N_\eta+1$. The fluxes calculated using the KT algorithm couple each grid point to its nearest and next-to-nearest neighbors. In particular, the flux in the $x$-direction at a given grid point $(i,j,k)$ requires knowledge of the conserved variables at $(i\pm 2,j,k)$ and $(i\pm 1,j,k)$. To close the discrete system at $i=2$ and $i=N_x+1$, we impose boundary conditions
\begin{align}
\q_{0,j,k}&\equiv\q_{1,j,k}\equiv\q_{2,j,k}\;,\label{ghostCells_1}\\
\q_{N_x+3,j,k}&\equiv\q_{N_x+2,j,k}\equiv\q_{N_x+1,j,k}\;,
\label{ghostCells_2}
\end{align}
These boundary conditions could easily be implemented by branching using \texttt{if}-\texttt{else} statements; this leads, however, to branch divergences which are known to be slow on GPUs. This can be avoided by introducing the red ghost cells shown in Fig.~\ref{fig:compGrid} and initializing these according to Eqs.~(\ref{ghostCells_1}) and (\ref{ghostCells_2}); these equations implement content in the red cells in Fig.~\ref{fig:compGrid} that is identical to the content of the outermost cell layers of the blue region. The algorithm for advancing the cell content within the physical blue region is then identical for all blue cells. The algorithm evolves all blue cells in a cube of size $N_x\times N_y\times N_\eta$, using information on the larger cube of size $(N_x+2)\times (N_y+2)\times (N_\eta+2)$ whose outer four (red) layers are not evolved but initialized at each time step according to Eqs.~(\ref{ghostCells_1}) and (\ref{ghostCells_2}). 
The white cells in Fig.~\ref{fig:compGrid} indicate grid points in the three-dimensional cube that are never accessed by the algorithm. When calculating the spatial derivatives of the conserved/primary variables in the source terms, all grid points are treated formally the same. That is, instead of dealing with the left and right boundaries of the blue region with forward and backward difference equations, $\partial_{x}A_{i}=(A_{i+1}-A_{i})/\Delta x$ and $\partial_{x}A_{i}=(A_{i-1}-A_{i-2})/\Delta x$, we eliminate branching by using the second-order center differences for all grid points. We point out that this is the most efficient implementation for parallelism in finite volume schemes; namely, each thread (or process) computes the same set of instructions on a different grid point element and only writes to that element.

\subsection{Memory arrangement}
\label{sec:memory}
On GPUs the memory layout is a crucial aspect to consider for optimal performance. First off, we need storage for the conserved and primary variables at the current and updated time step, as well as for intermediate results. Minimally, this requires the conserved variables at the current and updated time steps $n$ and $n+1$ to be stored in vectors $\q^{n}$ and $\q^{n+1}$, denoted in our algorithm as \texttt{q} and \texttt{Q}. The intermediate variables for the time integration Eqs.~(\ref{step1})-(\ref{finalStep}) are stored in $\q^{*}$, denoted as \texttt{qS} in the code. For the primary variables energy density and pressure, we only need \texttt{e} and \texttt{p}, but for the fluid velocity we need $(u^{\mu})^{n}$, $(u^{\mu})^{n+1}$, and $(u^{\mu})^{*}$, denoted as \texttt{u}, \texttt{up}, \texttt{uS}, due to the time derivatives of $u^\mu$ appearing in the source terms in Eq.~(\ref{relEqs}). All of these variables are written to global memory. \texttt{q} and \texttt{u}, etc., are implemented as a structure of arrays where their fields are the components of $\q$ in Eq.~(\ref{qComponents}) and of $u^\mu$ in Eq.~(\ref{fluidVelocity}), respectively. This seems counterintuitive, since on a CPU one normally uses an array of structures, where each grid point would be associated with a scalar field corresponding to the components of $\q$ and $u^\mu$. On GPUs the former is actually the more optimal data structure, while the latter is preferred on CPUs~\cite{Cook}. Intermediate variables from the KT algorithm such as the fluxes are stored locally for each thread, to avoid the extra cost of read/write accesses from global memory caused by the slower memory bandwidth.

In GPU-VH, three-dimensional data are stored linearly in one-dimensional arrays packed in column-major order, where $z$ is the slowest varying direction. This allows us to access array elements that are contiguous in memory, since in CUDA multidimensional threads are implemented with the third dimension (i.e. $z$-direction) being the slowest varying. The maximum number of threads that can be launched in the $x$, $y$, and $z$ directions are 1024, 1024, and 64, respectively. Obviously, these are just labels; one could also have our data in the $x$ direction indexed by the $z$ direction in CUDA threads. However, taking $z\equiv\eta_s$ as the slowest varying dimension is most natural for heavy ion collisions because at high energies the evolution is approximately boost-invariant near midrapidity ($\eta_s=0$), and this choice allows for an almost trivial reduction of the code to $2{+}1$-dimensional evolution with longitudinal boost-invariant if desired. We also expect to use a smaller number of grid points in the rapidity direction than the transverse directions and thus would want to launch more threads in $x$ and $y$.

Let \texttt{N}$\equiv$\texttt{(Nx+4)*(Ny+4)*(Nz+4)} be the total number of grid cells and \texttt{bytes} denoting the number of bytes in computer memory of the floating-point format that the data type uses (4 or 8 for single and double-precision floating point formats, respectively). The total number of bytes of memory needed to store a variable at all grid points is the total number of grid cells times the number of bytes of the precision format, i.e.~\texttt{varBytes=N*bytes}. For the conserved variables, we need storage arrays for all 15 conserved currents and four arrays to store the components of the fluid velocity, denoted by \texttt{qBytes} and \texttt{uBytes}, respectively. The total amount of global memory usage by our program (through the storage of \texttt{q, qS, Q, up, uS, u, e,} and \texttt{p}) is then \texttt{3*qBytes+3*uBytes+2*varBytes}. This amounts to $\sim{4}$ GB of global memory for the largest number of grid points \texttt{Nx=Ny=Nz=256} considered for the simulations we conducted in this paper. For lower end graphics cards, designed for mobile/laptop architecture's, this becomes problematic because their memory sizes are usually less than 4 GB. However, we only used this grid spacing for performance testing purposes, and in practice we do not expect to have to resolve structures at such fine resolutions; all other grid sizes used herein took up less than 3 GB of global memory usage. We point out that if the amount of global memory used by GPU-VH becomes too large to fit on an individual card, it can be reduced by simply propagating only the independent components of $\pi^{\mu\nu}$ (five) and $u^\mu$ (three), with the remaining components determined algebraically.

\subsection{Program flow}
\label{sec:ProgramFlow}
\tikzstyle{decision} = [diamond, draw, fill=yellow!20, 
    text width=7em, text badly centered, node distance=3cm, inner sep=0pt]
\tikzstyle{block} = [rectangle, draw, fill=blue!20, 
    text width=25em, text centered, rounded corners, minimum height=4em]
\tikzstyle{gpu_block} = [rectangle, draw, fill=red!20, 
    text width=25em, text centered, rounded corners, minimum height=4em]
\tikzstyle{cloud} = [draw, ellipse,fill=green!20, node distance=3cm,
    minimum height=2em]
\tikzstyle{line} = [draw, very thick, color=black, -latex']

\begin{figure}
\begin{center}
\scalebox{0.8}{
\begin{tikzpicture}[node distance = 2.5cm, scale = 0.5]
    \node [cloud] (start) {Start};
    \node [block, below of=start] (config) {System Configuration (C/CUDA) \\ (Allocate host and device memory)};
    \node [block, below of=config] (init) {Fluid dynamics initialization \\ (Generate or read in initial conditions)};
    \node [block, below of=init] (cons) {Calculate conserved quantities and copy to GPU. \\ (Set up boundary conditions/ghost cells)};
    \node [gpu_block, below of=cons] (evolve) {Evolve the system one full time step: \\ $\tau{=}\tau_0{+}n{\cdot}\Delta\tau$};
    \node [decision, below of=evolve, yshift=-0.5cm] (freezeout) {$T<T_f$ or $\tau>\tau_f$};
    \node [block, below of=freezeout, yshift=-0.5cm] (copy) {Copy variables back to CPU host memory and output to disk};
    \coordinate[right=2.5cm of copy]  (c1)  {};
    \node [decision, left of=freezeout, xshift=-3cm] (out) {$n\,\mathrm{\texttt{mod}}\,m=0$};
    \node [cloud, right of=freezeout, xshift=2.5cm] (stop) {End program};
    \path [line] (start) -- (config);
    \path [line] (config) -- (init);
    \path [line] (init) -- (cons);
    \path [line] (cons) -- (evolve);
    \path [line] (evolve) -- (freezeout);
    \path [line] (freezeout) -- node[anchor=south] {no} (out);
    \path [line] (freezeout) -- node[anchor=south] {yes}(stop);
    \path [line] (out) |- node[anchor=east] {yes} (copy);
    \path [line] (out) |- node[anchor=east] {no} (evolve);
    \path [line](copy) -- (c1) |- (evolve);
\end{tikzpicture}
}
\end{center}
\caption{The program flow chart for GPU-VH.\label{fig:HydroPluginFlowChart}} 
\end{figure}
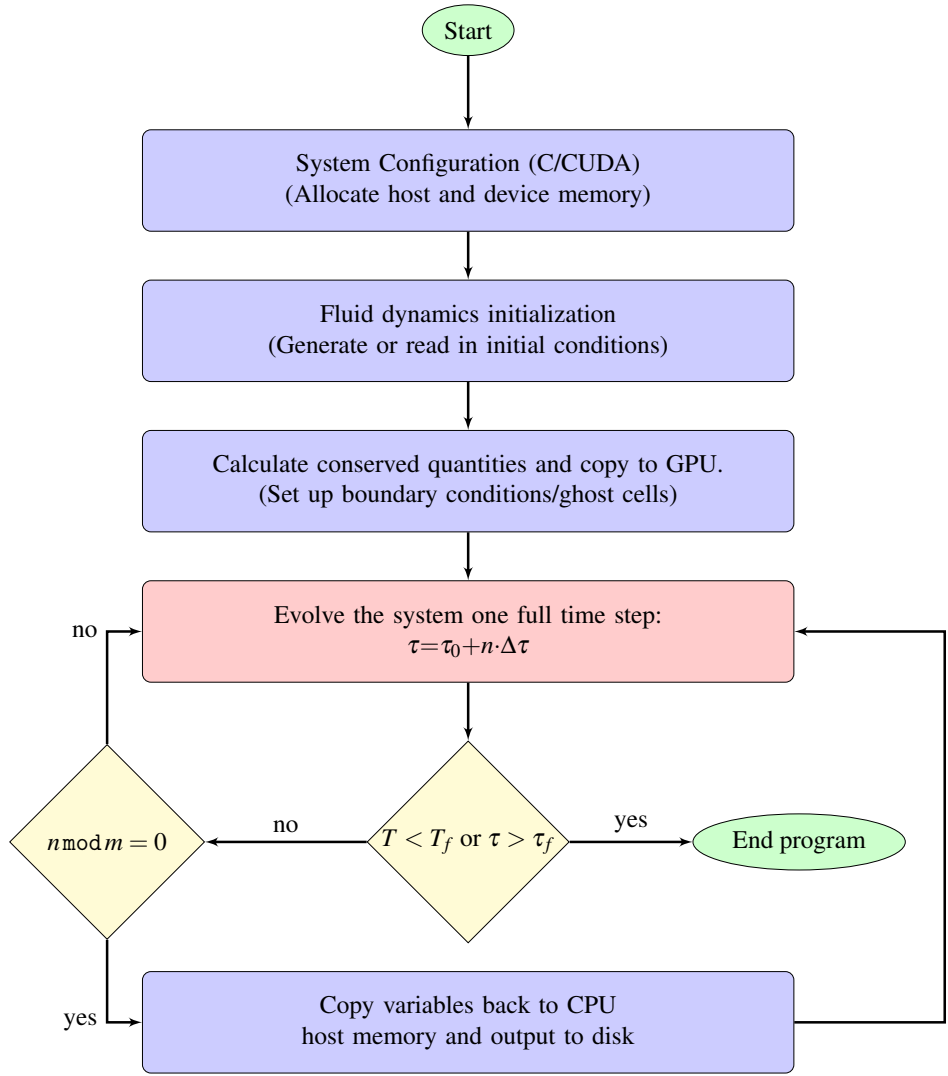
%
The program flow of GPU-VH is sketched in Fig.~\ref{fig:HydroPluginFlowChart}. Below we enumerate the steps:
\begin{enumerate}
\setlength{\itemsep}{0pt}
\item The system is initialized, and various configuration files are read in for future use.
\item Allocate the necessary memory on the host and device (CUDA global memory).
\item Setup the initial conditions for the inferred/primary variables ($\ed$, $u^\mu$) and dissipative currents (if any) of the specific fluid dynamic test/problem. (We note that we explicitly propagate all four components of $u^\mu$, namely $u_\tau$; $u_x$; $u_y$; and $u_\eta$, rather than forcing the normalization condition to determine $u_\tau$. Making sure that $|u\cdot u-1|<\epsilon$ is used as another runtime check of the code.)   
\item The conserved currents are then calculated from the inferred variables: $\q^{0}$ at all physical grid points. The boundary conditions are enforced by adding ghost cells. 
\item The initial conditions of the conserved and primary variables are then copied to device memory.
\item On the GPU, the fluid dynamic system of ordinary differential equations is evolved one full time step. 
\item $\tau=\tau_{0}+n\Delta\tau$
\item \texttt{If} the temperature $T$ is less than the freeze-out temperature $T_f$ everywhere in the simulation volume or $\tau$ reaches the final time $\tau_f$ (i.e.~$n$ has reached the externally specified maximal number of proper time points), exit the program \texttt{else} check if the results should be outputted (see next step).
\item \texttt{If} $n$ \texttt{mod} $m$ (where $m$ is an integer that sets after how many time steps $\Delta\tau$ we output the variables), then copy back the inferred/conserved variables back to CPU host memory and write to file \texttt{else} go to step (6) and repeat the procedure. 
\end{enumerate}
In Fig.~\ref{fig:HydroPluginFlowChart} we denote as usual all processing steps with a rectangular box and decision branches with a diamond. Additionally, we color coded the rectangles to distinguish between processing being done on the CPU (blue) and GPU (red). The decision trees (yellow diamonds) are always performed on the CPU. Step (6) (the red box in Fig.~\ref{fig:HydroPluginFlowChart}) is the main workhorse for GPU-VH, and we will now describe it in more detail. 

The current time is given by $\tau^{n}=\tau_{0}+n\Delta\tau$. To evolve the system one full time step to time $\tau^{n+1}$, the needed Euler steps Eqs.~(\ref{step1})-(\ref{finalStep}) can be written as $\q^{n+1}\equiv(\q^n+\q^{**})/2$, where $\q^{**}\equiv{\bf E}[\q^{*}]$ and $\q^{*}\equiv{\bf E}[\q^{n}]$. Rather than allocate additional memory to store the values of $\q^{**}$, we set first $\q^{n+1}\equiv{\bf E}[\q^{*}]$, and then overwrite it by  $\q^{n+1}\equiv(\q^n+\q^{n+1})/2$. In device memory we have stored the current values of $\q^{n}$, $\ed^{n}$, $\peq^{n}$, and $(u^{\mu})^{n}$ in \texttt{q, e, p} and \texttt{u}. In addition we also will need the fluid velocity at the previous time $(u^{\mu})^{n-1}$ (stored in \texttt{up}) in order to compute time derivatives in the source terms using the backwards Euler step.\footnote{%
To start the process, we initialize the system by setting $(u^{\mu})^{-1}=(u^{\mu})^{0}$.
} Intermediate results $\q^{*}$, $\q^{**}$, $(u^\mu)^{*}$, and $(u^\mu)^{**}$ are stored in \texttt{qS, Q, uS} and \texttt{u}, respectively. The updated values $\q^{n+1}$ and $(u^{\mu})^{n+1}$ are then stored in \texttt{Q} and \texttt{u}. All current, intermediate, and updated results for the energy density and pressure, e.g. $\ed^{n}$, $\ed^{*}$, $\ed^{n+1}$, are all stored in \texttt{e} and \texttt{p}. This reduces the total global memory usage of the program. We remark that all needed memory is already allocated in step (2) of Fig.~\ref{fig:HydroPluginFlowChart}. Our time integration contains the following steps:
\begin{enumerate}
\item Compute the forward Euler step to estimate the slope of the variables at the current time step $\tau^{n}$:
\begin{equation*}
\q^{*}\equiv
{\bf E}[\q^{n};\ed^{n},\peq^{n},(u^\mu)^{n},(u^\mu)^{n-1}]\;.
\end{equation*}
Here, the variables at the current time step $n$ are \texttt{q, e, p, u} and the fluid velocity at the previous time step $n-1$ is \texttt{up}. The Euler step gives the predicted value \texttt{qS}.
\item Set the intermediate values of $\ed^{*}$, $\peq^{*}$, and $(u^\mu)^{*}$. $\ed^{*}$ is obtained by solving Eq.~(\ref{eInferred}) from $\q^{*}$, and then the equation of state gives $\peq^{*}$. Eq.~(\ref{utInferred}) and (\ref{uiInferred}) determines the fluid velocity at the current time step $(u^\mu)^{*}$. In our code, this process gives \texttt{e, p} and \texttt{uS}.
\item The dissipative currents in $\q^{*}$ are regulated according to  Eq.~(\ref{regPimunu}). The regularization parameter $\rho$ defined in Eq.~(\ref{rho}) is computed from $\q^{*}$, $\ed^{*}$, $\peq^{*}$, and $(u^{\mu})^{*}$.
\item The boundary conditions are then imposed by setting the ghost cells to the faces of the 3D cube of physical data. This procedure is done for $\q^{*}$, $\ed^{*}$, $\peq^{*}$, and $(u^\mu)^{*}$. 
\item Compute the forward Euler step to estimate the corrected slope at time $\tau^{n}+\Delta\tau$:
\begin{equation*}
\q^{**}\equiv
{\bf E}[\q^{*};\ed^{*},\peq^{*},(u^\mu)^{*},(u^\mu)^{n}]\;.
\end{equation*}
For our numerical purposes, \texttt{Q} is the corrected value obtained from the Euler step of \texttt{qS, e, p, uS}, and \texttt{u}.
\item The two stage RK scheme is concluded by the {\it convex} combination of the predicted and corrected forward Euler steps:
\begin{equation*}
\q^{n+1}\equiv\frac{1}{2}(\q^{n}+\q^{**})\;.
\end{equation*}
That is, \texttt{Q} is averaged over itself and \texttt{q} which contains the conserved variables at time step $n$.
\item The fluid velocities \texttt{up} and \texttt{u} are swapped. Now \texttt{up} contains the values of $u^\mu$ at time step $n$. This is needed for this whole process to continue to be iterated. We need to always know the value of the fluid velocity at the previous time step. 
\item The inferred variables $\ed^{n+1}$, $\peq^{n+1}$, and $(u^\mu)^{n+1}$ are determined from $\q^{n+1}$ by Eqs.~(\ref{eInferred})-(\ref{uiInferred}). The system has now been evolved one full time step and the updated values $\q^{n+1}$, $\ed^{n+1}$, $\peq^{n+1}$ and $(u^\mu)^{n+1}$ are all stored in \texttt{Q, e, p}, and \texttt{u}. Then the dissipative current in \texttt{Q} are regulated, followed by setting the ghost cells of \texttt{Q, e, p}, and \texttt{u}. 
\item Swap the values of \texttt{q} and \texttt{Q} to prepare for the next iteration.
\end{enumerate}
This one step time integration sub-program flow of GPU-VH is shown in Fig.~\ref{fig:twoStepRungeKutta} where each separate process is implemented with its own CUDA kernel indicated by red.~(The repeated processes such as the predicted and corrected Euler steps are the same implementation.) The Euler step in the flow chart is implemented as four separate kernels; one calculates all of the source terms (other than the gradients of the dissipative currents) as described in Algorithm~\ref{Algorithm:eulerStepKernelSource} below, followed by three kernels that calculate the Euler steps in the $x$, $y$, and $\eta_s$ directions, respectively, shown in Algorithm~\ref{Algorithm:eulerStepKernelX}. These kernels calculate the corresponding derivatives of the dissipative currents as well the fluxes of the primary variables, implemented in Algorithm~\ref{Algorithm:HFlux}. Algorithm~\ref{Algorithm:eulerStepKernelX} shows the Euler step kernel in the $x$-direction; the implementation for $y$ and $\eta_s$ fluxes is formally the same, and we do not show them here. Although this involves more read/writes from global memory it is actually more optimal than using fused kernels because it significantly reduces the register usage which limits the number of resources that can be launched for a given kernel. 
\tikzstyle{decision} = [diamond, draw, fill=yellow!20, 
    text width=7em, text badly centered, node distance=3cm, inner sep=0pt]
\tikzstyle{block} = [rectangle, draw, fill=blue!20, 
    text width=15cm, text centered, rounded corners, minimum height=4em]
\tikzstyle{gpu_block} = [rectangle, draw, fill=red!20, 
    text width=5cm, text centered, rounded corners, minimum height=4em]
\tikzstyle{green_block} = [rectangle, draw, fill=green!20, 
    text width=8.5cm, text centered, rounded corners, minimum height=4em]
\tikzstyle{cloud} = [draw, ellipse,fill=green!20, node distance=3cm,
    minimum height=2em]
\tikzstyle{line} = [draw, very thick, color=black, -latex']
\begin{figure}
\label{fig:twoStepRungeKutta2}
\begin{center}
\scalebox{0.75}{
\begin{tikzpicture}[node distance = 2cm, auto]
    \node [block] (setup) {\textbf{Predicted step:} Estimate the slope of the variables at the current time step $\tau^{n}$};
    \node [green_block, below of=setup, xshift=-3.25cm] (input) {$\q^{n}$, $\ed^{n}$, $\peq^{n}$, $(u^\mu)^{n}$, $(u^\mu)^{n-1}\to$ \texttt{q, e, p, u, up}};
    \node [gpu_block, below of=setup, xshift=5cm] (evolve) {Euler step};
    \node [green_block, below of=input] (ginf) {Get \texttt{e, p}, and \texttt{uS} are from \texttt{qS} computed as\\
    $\q^{*}{\equiv}{\bf E}[\q^{n};\ed^{n},\peq^{n},(u^\mu)^{n},(u^\mu)^{n-1}]$};
    \node [gpu_block, below of=evolve] (inf) {Set inferred variables};
    \node [green_block, below of=ginf] (greg) {$\pi^{\mu\nu}\to\frac{\tanh\rho}{\rho}\pi^{\mu\nu}$};
    \node [gpu_block, below of=inf] (reg) {Regulate dissipative currents};
    \node [green_block, below of=greg] (gghost) {Apply Eqs.~(\ref{ghostCells_1}) and (\ref{ghostCells_2}) to \texttt{qS, e, p, uS}};
    \node [gpu_block, below of=reg] (ghost) {Set ghost cells};
    \node [block, below of=setup, yshift=-8cm] (setup2) {\textbf{Corrected step:} Estimate the corrected slope at time $\tau^{n}+\Delta\tau$};
    \node [green_block, below of=setup2, xshift=-3.25cm] (gevolve) {store 
    $\q^{**}{\equiv}{\bf E}[\q^{*};\ed^{*},\peq^{*},(u^\mu)^{*},(u^\mu)^{n}]$
    in \texttt{Q}, \\
    with $\ed^{*},\,\peq^{*},\,(u^\mu)^{*},\,(u^\mu)^{n}$ given by \texttt{e, p, uS, u}};
    \node [gpu_block, below of=setup2, xshift=5cm] (evolve2) {Euler step};
    \node [green_block, below of=gevolve] (gcomb) {Average predicted and corrected slopes for the conserved variables in Eq.~(\ref{finalStep}) implemented via \texttt{Q = (q+Q)/2}};
    \node [gpu_block, below of=evolve2] (comb) {Convex combination of forward Euler steps};
    \node [green_block, below of=gcomb] (ginf2) {Swap fluid velocities \texttt{up} and \texttt{u}, then set \texttt{e, p, u} from \texttt{Q}};
    \node [gpu_block, below of=comb] (inf2) {Set inferred variables};
    \node [green_block, below of=ginf2] (greg2) {$\pi^{\mu\nu}\to\frac{\tanh\rho}{\rho}\pi^{\mu\nu}$};
    \node [gpu_block, below of=inf2] (reg2) {Regulate dissipative currents};
    \node [green_block, below of=greg2] (gghost2) {Apply Eqs.~(\ref{ghostCells_1}) and (\ref{ghostCells_2}) to \texttt{Q, e, p, u}};
    \node [gpu_block, below of=reg2] (ghost2) {Set ghost cells};
    \node [block, below of=setup2, yshift=-10cm] (setup3) {\textbf{Result:} Swap the values of the current and updated conserved variables \texttt{q} and \texttt{Q}};
	\coordinate[right=0.5cm of setup]  (c)  {};
	\coordinate[right=0.5cm of setup3]  (c3)  {};
	\coordinate[left=0cm of setup]  (e)  {};
	\coordinate[left=0cm of setup2]  (e2)  {};
	\coordinate[left=0cm of setup3]  (e3)  {};
	\path [line] (c3) -- (c);
	\path[->,very thick, color=black, -latex']
	(e) edge[bend right=10] node [left] {} (e2);
	\path[->,very thick, color=black, -latex']
	(e2) edge[bend right=10] node [left] {} (e3);
\end{tikzpicture}
}
\end{center}
\caption{Program flow chart for the two-step Runge-Kutta algorithm.}
\label{fig:twoStepRungeKutta}
\end{figure}
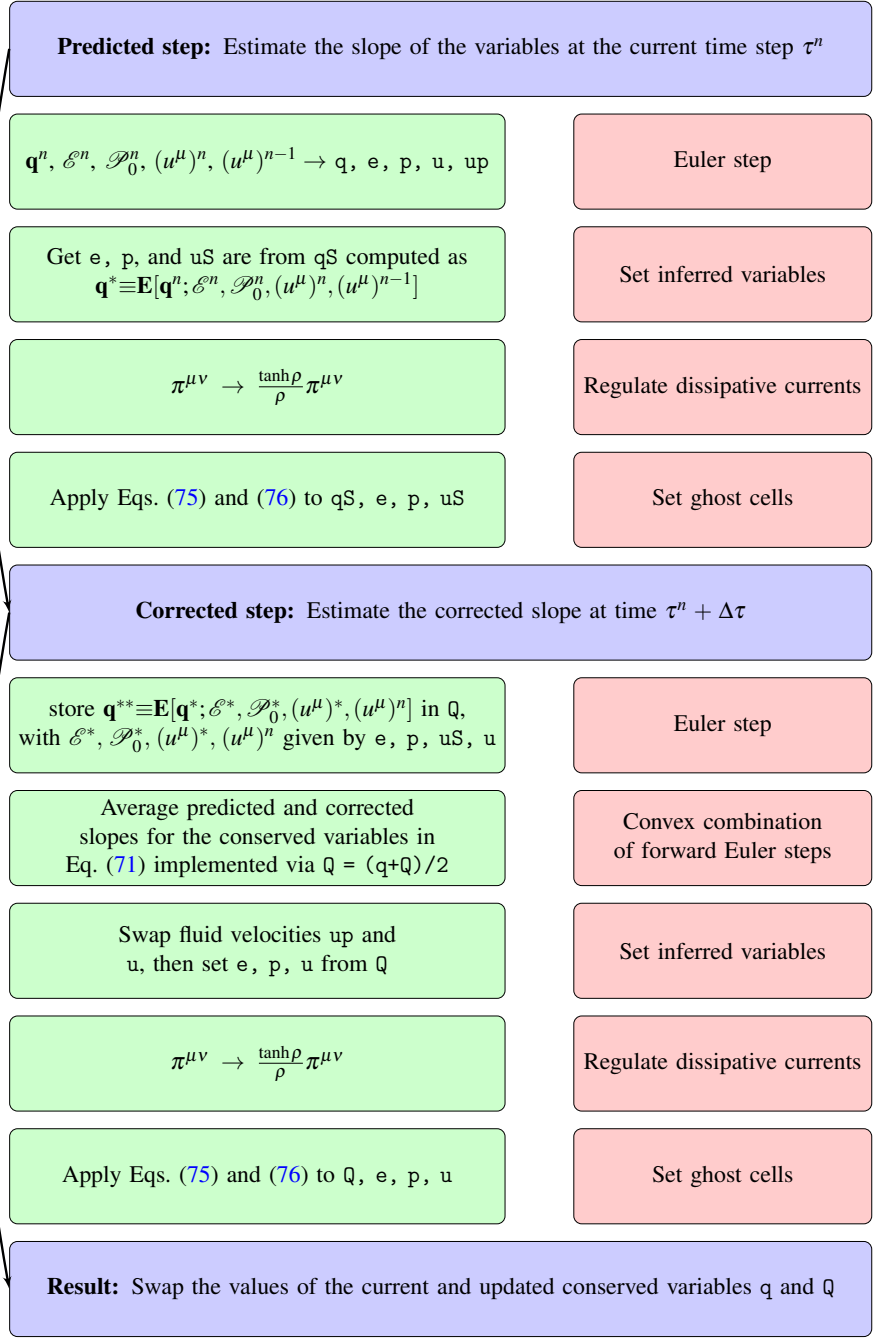
%

\alglanguage{pseudocode}
 \renewcommand{\algorithmicrequire}{\textbf{Input:}}
\renewcommand{\algorithmicensure}{\textbf{Output:}}
\begin{algorithm}[t!]
\small
\caption{Euler step for the source terms}
\label{Algorithm:eulerStepKernelSource}
\begin{algorithmic}[1]
\Require \texttt{currentVars} - struct of conserved variables at the current time step 
 \renewcommand{\algorithmicrequire}{\textbf{      }}
\Require \hspace{.75cm}\texttt{e} - energy density values at the current time step 
\Require \hspace{.75cm}\texttt{p} - pressure at the current time step 
\Require \hspace{.75cm}\texttt{u} - struct of fluid velocity components at the current time step 
\Require \hspace{.75cm}\texttt{up} - struct of fluid velocity components at the previous time step 
\Ensure \texttt{updatedVars} - struct of updated conserved variables at the time step 
\Function{$\mathbf{eulerStepKernelSource}$}{\texttt{currentVars, updatedVars, e, p, u, up}}
\State \texttt{i,j,k}$\leftarrow$ unique 3D indices from the CUDA thread adjusted for the number of ghost cells in the negative direction (here, \verb|N_GHOST_CELLS_M=2|)
\State \texttt{s=i+Nx*(j+Ny*k)}$\leftarrow$ linear index to access column packed data
\State static local memory allocation for the storage of  intermediate results:
\texttt{Q,\,S} have \texttt{N} array elements
	\For {each conserved quantity \texttt{n} in \texttt{currentVars} }
    		\State \texttt{Q}$\leftarrow$ store the data in \texttt{currentVars} at \texttt{s}
    \EndFor
\State \texttt{loadSourceTerms2(Q, S, u, up, e, p, s)}$\leftarrow$ load source terms into \texttt{S}
\For {\texttt{n} in \texttt{N}}
	\State \texttt{Q(n)+=dt*S(n)}
	\EndFor
    \For {each conserved quantity \texttt{n} in \texttt{currentVars}}
    		\State \verb|updatedVars->conserved_quantity_n=Q(n)| $\leftarrow$ set the updated variables to the source terms
    \EndFor    
\EndFunction
\Statex
\end{algorithmic}
  \vspace{-0.4cm}%
\end{algorithm}
%
\alglanguage{pseudocode}
 \renewcommand{\algorithmicrequire}{\textbf{Input:}}
\renewcommand{\algorithmicensure}{\textbf{Output:}}
\begin{algorithm}[th!]
\small
\caption{Euler step for spatial gradients in the $x$ direction}
\label{Algorithm:eulerStepKernelX}
\begin{algorithmic}[1]
\Require \texttt{currentVars} - struct of conserved variables at the current time step 
 \renewcommand{\algorithmicrequire}{\textbf{      }}
\Require \hspace{.75cm}\texttt{e} - energy density values at the current time step 
\Require \hspace{.75cm}\texttt{u} - struct of fluid velocity components at the current time step 
\Ensure \texttt{updatedVars} - struct of updated conserved variables at the time step 
\Function{$\mathbf{eulerStepKernelX}$}{\texttt{currentVars, updatedVars, e, u}}
\State \texttt{i,j,k}$\leftarrow$ unique 3D indices from the CUDA thread adjusted for the number of ghost cells in the negative direction (here, \verb|N_GHOST_CELLS_M=2|)
\State \texttt{s=i+Nx*(j+Ny*k)}$\leftarrow$ linear index to access column packed data
\State static local memory allocation for the storage of  intermediate results:
\texttt{I} has \texttt{5*N} array elements and \texttt{Q,\,H} have \texttt{N} array elements
	\For {each conserved quantity \texttt{n} in \texttt{currentVars} }
    		\State \texttt{I}$\leftarrow$ store the 5-point stencil data at \texttt{s} and its 
    		neighbors \texttt{i$\pm$1,i$\pm$2} for all current conserved variables in 
    		\verb|currentVars->conserved_quantity_n|
    \EndFor
\State \texttt{flux(I,H)} $\leftarrow$ calculates flux in the $x$-direction at the cell interface \texttt{i+1/2} 
\For {\texttt{n} in \texttt{N}}
	\State \texttt{Q(n)=-H(n)}
	\EndFor
\State \texttt{flux(I,H)} $\leftarrow$ calculates flux in the $x$-direction at the cell interface \texttt{i-1/2} 
\For {\texttt{n} in \texttt{N}}
	\State \texttt{Q(n)+=H(n)}
		\State \texttt{Q(n)/=dx}
	\EndFor
\State \texttt{loadSourceTermsX(I, H, u, s)}$\leftarrow$ load source terms with \texttt{x} derivatives in to \texttt{H}
\For {\texttt{n} in $\mathrm{\texttt{N}}_\mathrm{consv}$}
	\State \texttt{Q(n)+=H(n)}
		\State \texttt{Q(n)*=dt}
	\EndFor
    	\For {\texttt{n = 0:N}}
    		\State \texttt{Q(n)*=dt}
    \EndFor
    \For {each conserved quantity \texttt{n} in \texttt{currentVars}}
    		\State \verb|updatedVars->conserved_quantity_n+=Q(n)| $\leftarrow$ add intermediate result to the updated variables
    \EndFor    
\EndFunction
\Statex
\end{algorithmic}
  \vspace{-0.4cm}%
\end{algorithm}
%
\alglanguage{pseudocode}
 \renewcommand{\algorithmicrequire}{\textbf{Input:}}
\renewcommand{\algorithmicensure}{\textbf{Output:}}
\begin{algorithm}[th!]
\small
\caption{Calculate numerical flux function in Eq.~(\ref{Hflux})}
\label{Algorithm:HFlux}
\begin{algorithmic}[1]
\Require \texttt{data} - linear storage of the 5-point stencil (in the $r=x$, $y$ or $\eta$-direction) for each conserved variable 
\renewcommand{\algorithmicrequire}{\textbf{      }}
\Require \hspace{.75cm}\texttt{*rightHalfCellExtrapolation} - function pointer to Eq.~(\ref{rightHalfCellExtrapolationForward} or (\ref{rightHalfCellExtrapolationBackwards})
\Require \hspace{.75cm}\texttt{*leftHalfCellExtrapolation} - function pointer to Eq.~(\ref{leftHalfCellExtrapolationForward} or (\ref{leftHalfCellExtrapolationBackwards})
\Require \hspace{.75cm} \texttt{*fluxFunction} - function pointer to the flux function ${\bf F}_{r}$
\Require \hspace{.75cm} \texttt{*spectralRadius} - function pointer to the spectral radius $\rho_{r}\equiv |v_r|$
\Ensure \texttt{result} - stores the result of the numerical flux function
\Function{$\mathbf{flux}$}{\texttt{data, result}}
	\For {\texttt{n = 0:N}}
		\State Extract the values of the grid point and neighbor cells of the \texttt{n}-th conserved variable:
    		\State \texttt{qmm = data(5$\cdot$n)} 
    		\State \texttt{qm = data(5$\cdot$n+1)} 
		\State \texttt{q = data(5$\cdot$n+2)}
		\State \texttt{qp = data(5$\cdot$n+3)}
    		\State \texttt{qpp = data(5$\cdot$n+4)} 	
    \State{Right and left half cell extrapolation}
    \State \texttt{qR(n)	= rightHalfCellExtrapolation(qmm, qm, q, qp, qpp)}		    \State \texttt{qL(n)	= leftHalfCellExtrapolation(qmm, qm, q, qp, qpp)}    	
    \EndFor
\State \texttt{utR,uxR,uyR,unR} $\leftarrow$ the right extrapolated values of the primary variables obtained from \texttt{qR}
\State \texttt{utL,uxL,uyL,unL} $\leftarrow$ the left extrapolated values of the primary variables obtained from \texttt{qL}
\State \texttt{a} $\leftarrow$ local propagation speed  
    	\For {\texttt{n = 0:N}}
    		\State \texttt{FqR} $\leftarrow$ flux function from right half extrapolated values
    		\State \texttt{FqL} $\leftarrow$ flux function from left half extrapolated values   
    		\State \texttt{result(n) = (FqR+FqL-a*(qR(n)-qL(n)))/2} 		
    \EndFor
\EndFunction
\Statex
\end{algorithmic}
  \vspace{-0.4cm}%
\end{algorithm}
%
\section{Numerical tests}
\label{sec:num_tests}
In order to validate our code, we will perform various numerical tests where (for most cases) analytic and semi-analytic solutions exist. In order to test the fluid dynamic part of the code we will compare to Riemann problem for the Euler equations. Then we proceed to test the effects of the expansion geometry, dissipation, and the microscopic QCD parametrization of the EoS and specific bulk viscosity. All tests described in this section can be repeated by the user by setting corresponding parameters in the input/configuration files and preprocessor macros. The corresponding input files are included in the code package.
\subsection{Relativistic Sod shock tube}
\label{sec:shockTube1d}
\begin{figure}[t!]
\begin{center}
\includegraphics[width=\linewidth]{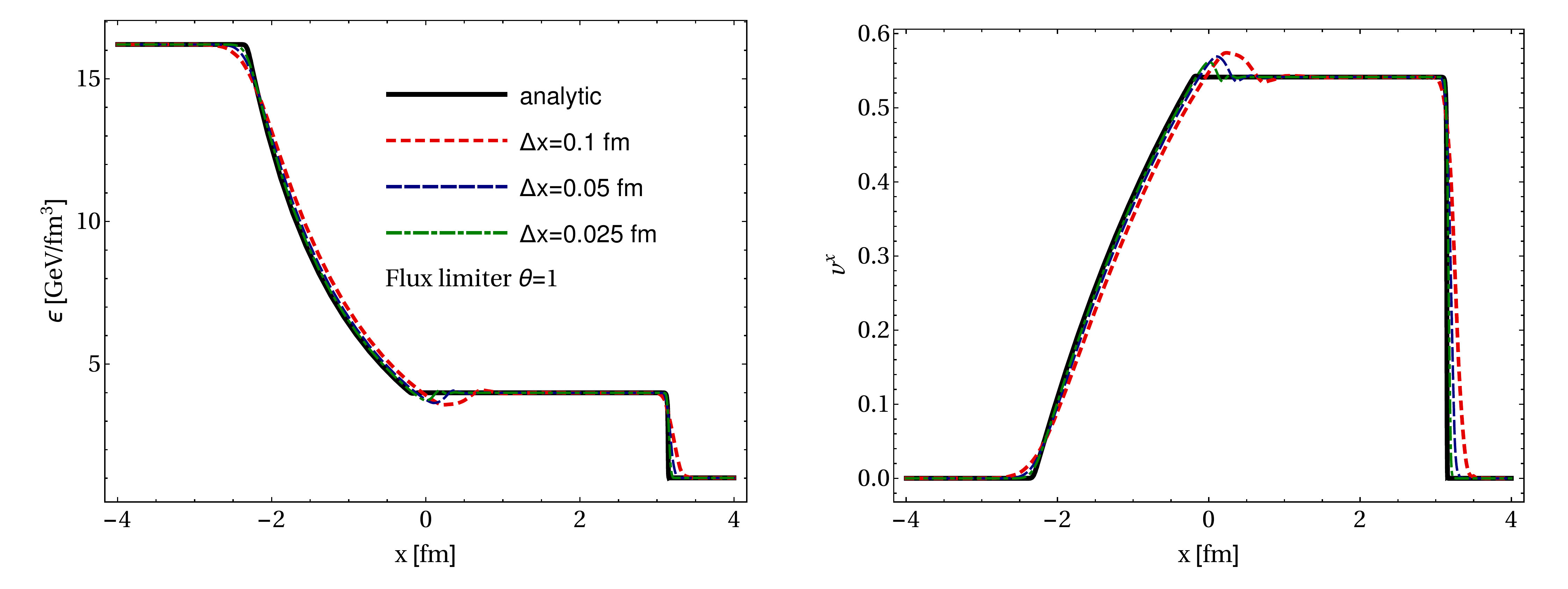}
\end{center}
\vspace{-7mm}
\caption{Relativistic Sod shock tube test where the flux limiter parameter $\theta=1$ is fixed and the spatial resolution $\Delta x=\{0.1,0.05,0.025\}$ fm is varied. The left panel shows the energy density and the right panel shows the $x$ component of the fluid velocity.}
\label{fig:sodTestDeltaXFig}
\end{figure}
\begin{figure}[t!]
\begin{center}
\includegraphics[width=\linewidth]{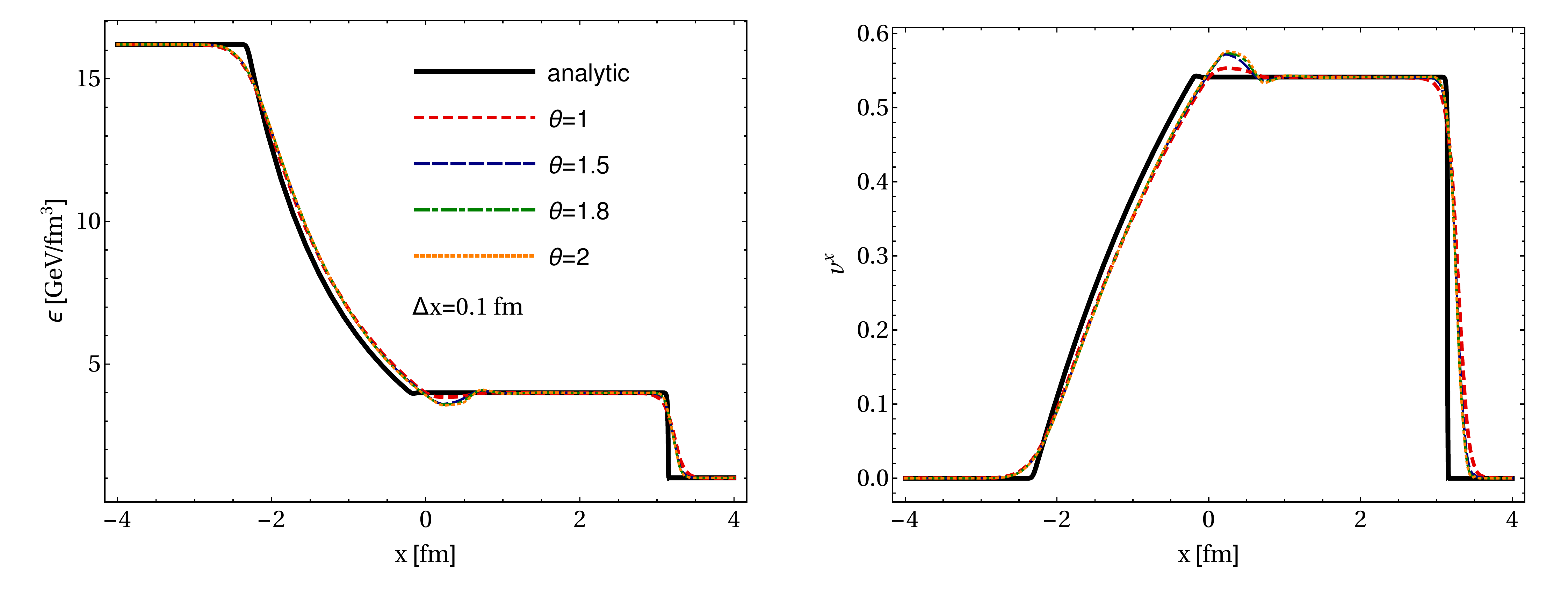}
\end{center}
\vspace{-7mm}
\caption{Same as Fig.~\ref{fig:sodTestDeltaXFig}, but now $\Delta x=0.1$ fm is fixed and the flux limiter parameter $\theta=\{1,1.5,1.8,2\}$ is varied.}
\label{fig:sodTestThetaFig}
\end{figure}
The first test we perform is the standard comparison to relativistic Riemann problems -- the one-dimensional shock tube test which admits an analytical solution. (The semi-analytic solution to the relativistic Riemann problem is given in Ref.~\cite{SCHNEIDER199392}.) Here, two ideal fluids in thermal equilibrium at constant pressures $\p_\mathrm{L}>\p_\mathrm{R}$ are placed in a box separated by a physical membrane at $x=0$. At time $t=0$ the membrane is suddenly removed creating a surface of discontinuity. A shock wave emerges propagating into the region of lower pressure (to the right), and rarefaction wave moves to the left (into the higher pressure region). The shock plateau is the region between the two and moves at a constant velocity. It should be pointed out that the fluid on both sides is initially at rest and the resulting motion of the fluid is entirely due to the discontinuities in the initial conditions. The initial state on the left ($x\le 0$) and right ($x>0$) sides of the membrane are taken to be $\ed_0=0.0246\,\mathrm{GeV/fm}^{3}$ and $\ed_0=0.0015\,\mathrm{GeV/fm}^{3}$, respectively. (In this test we use the conformal equation of state $\ed=3\peq$.) This problem provides a good test of a numerical schemes ability to capture shocks and contact discontinuities (e.g. see Ref.~\cite{Molnar:2009tx}). The comparison between the numerical results obtained from our GPU-VH code run in its ideal fluid dynamic mode using Cartesian coordinates (i.e. setting the geometrical source terms to zero) and the analytic solution is shown in Fig.~\ref{fig:sodTestDeltaXFig}. We use this test to make sure that the code converges to the analytical result in the infinite spatial resolution limit ($\Delta x\to 0$). 
Fig.~\ref{fig:sodTestThetaFig} shows the dependence of varying the value for the flux limiter $\theta$ parameter for a fixed $\Delta x=0.1\,\mathrm{fm}$. Based on this, we choose $\theta=1$ as the default. We have checked that placing the initial discontinuity along the other two spatial axes does not change the results presented here for the $x$-direction. 

\subsection{(2+1)-dimensional shock tube}
\label{sec:2dShockTube}
\begin{figure*}[h!]
  \centering
  \begin{tabular}{cc}
  \includegraphics[width=0.45\linewidth]{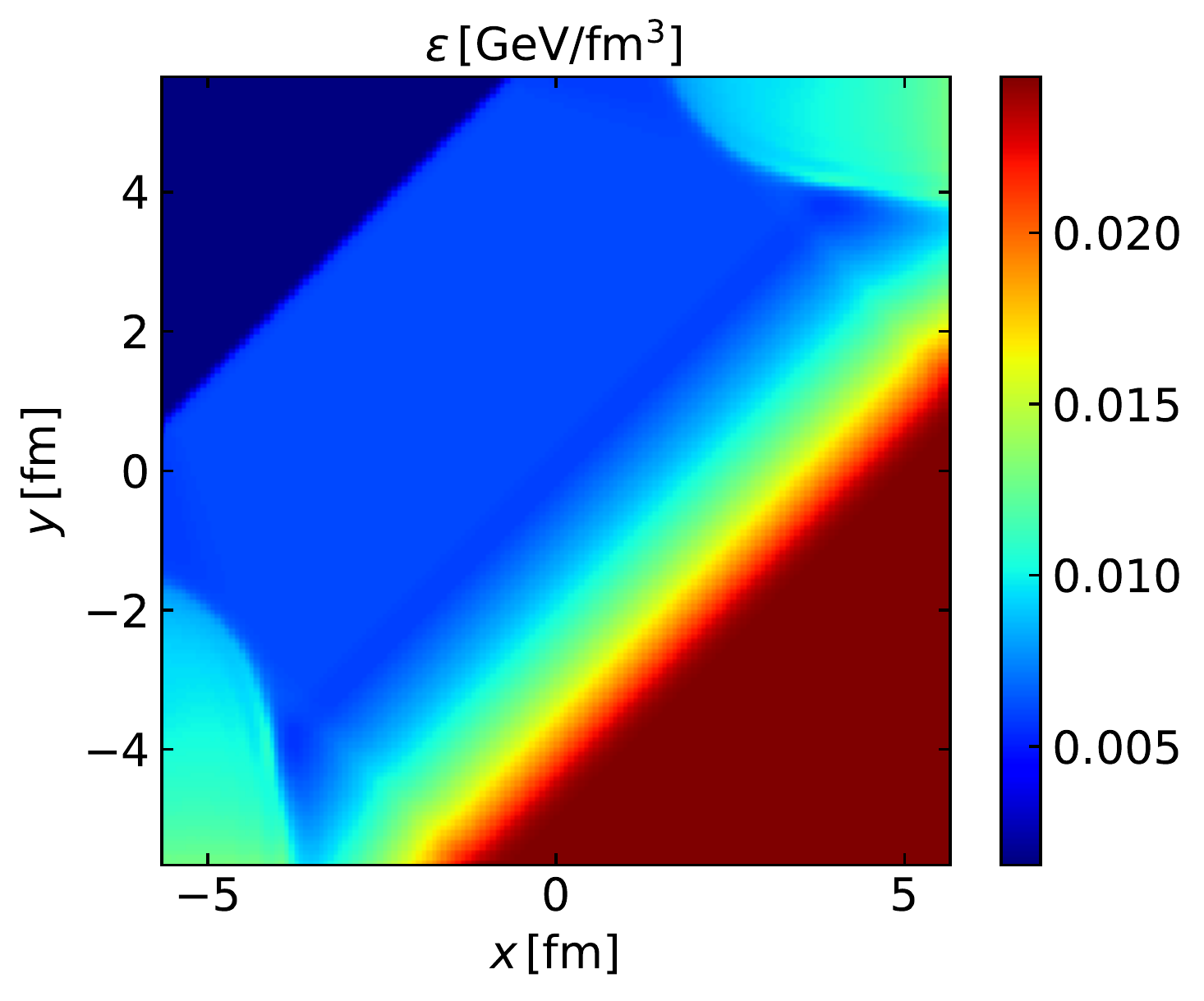} &
  \includegraphics[width=0.45\linewidth]{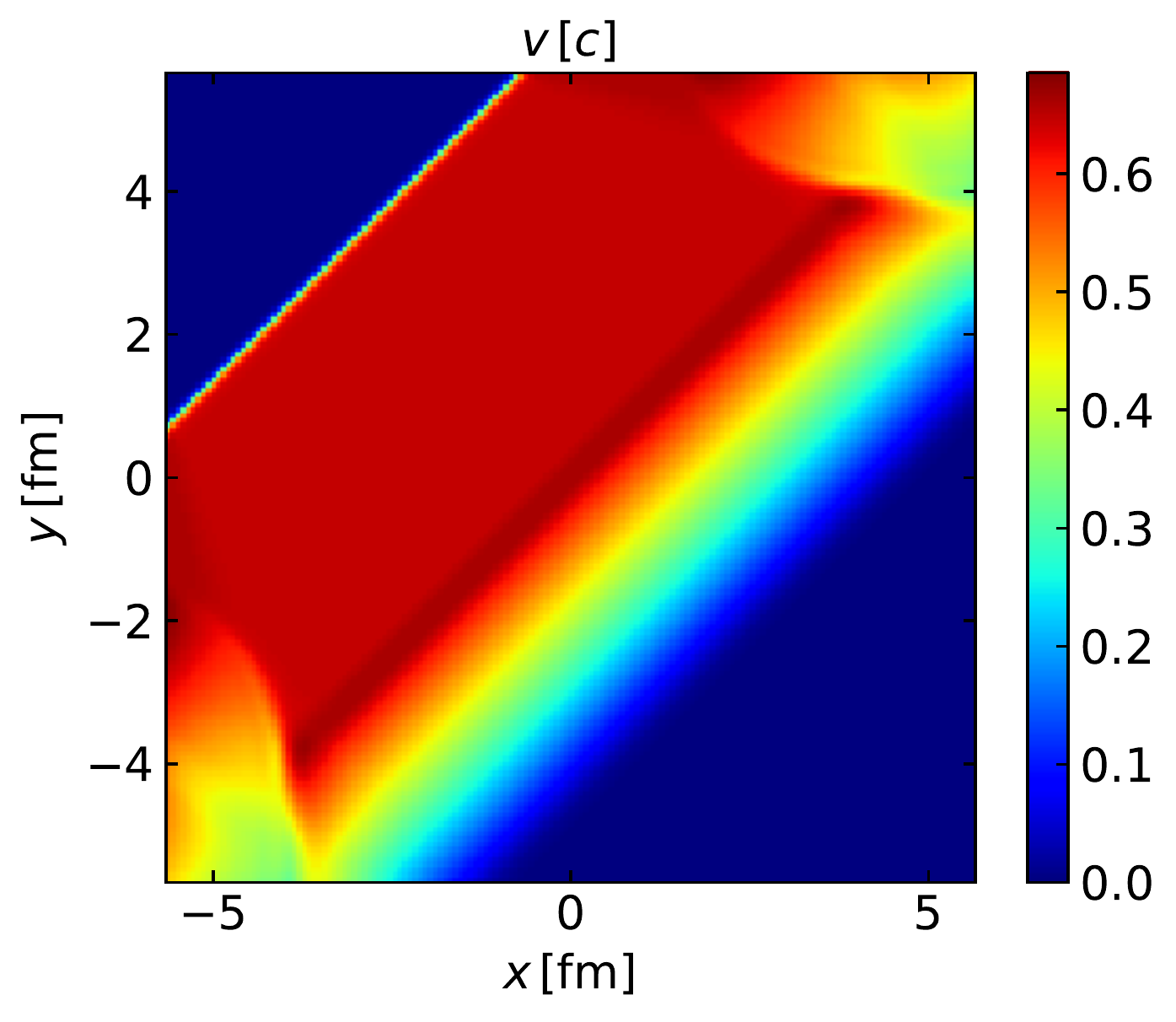} \\
  \includegraphics[width=0.45\linewidth]{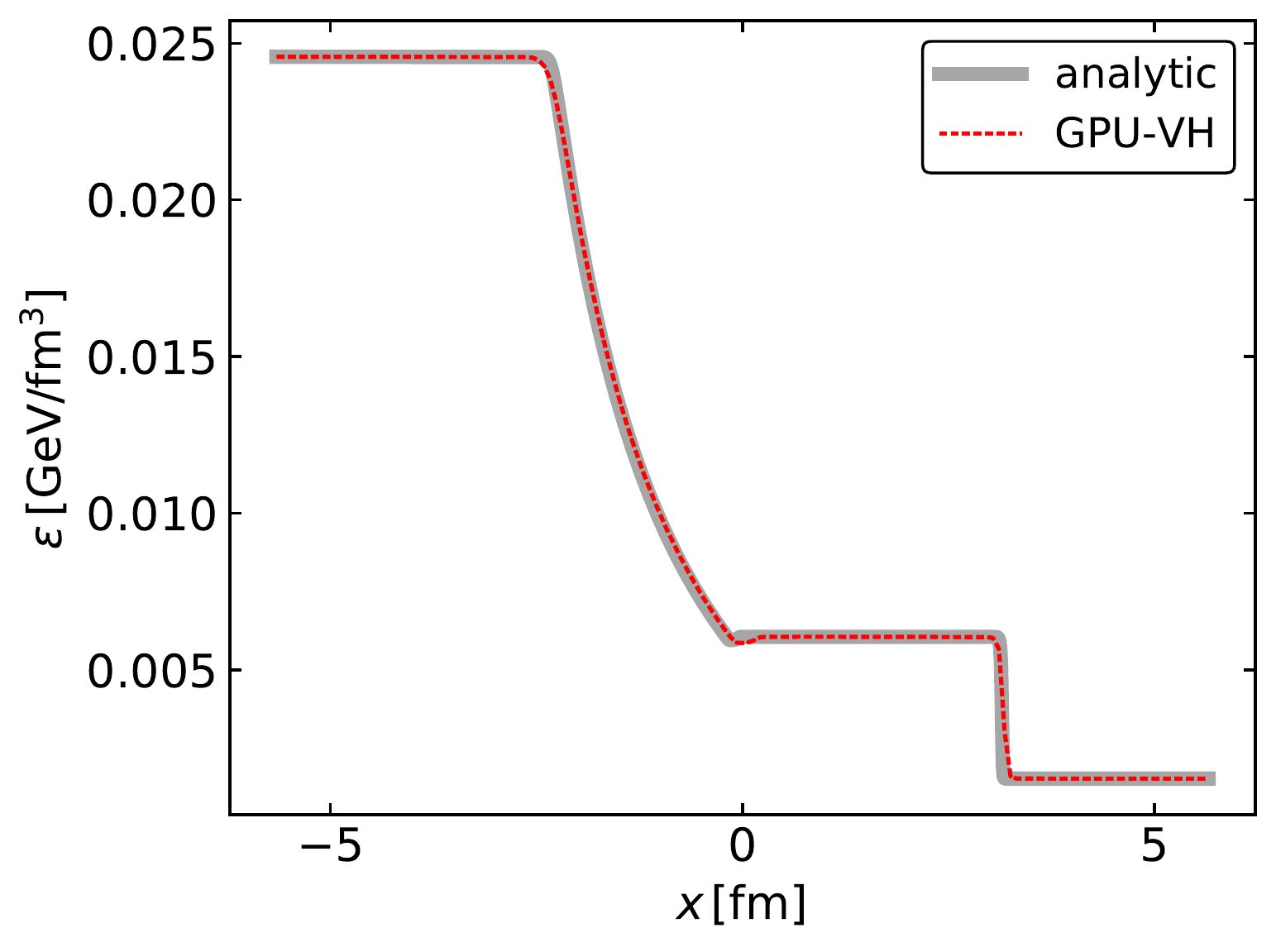} &
  \includegraphics[width=0.45\linewidth]{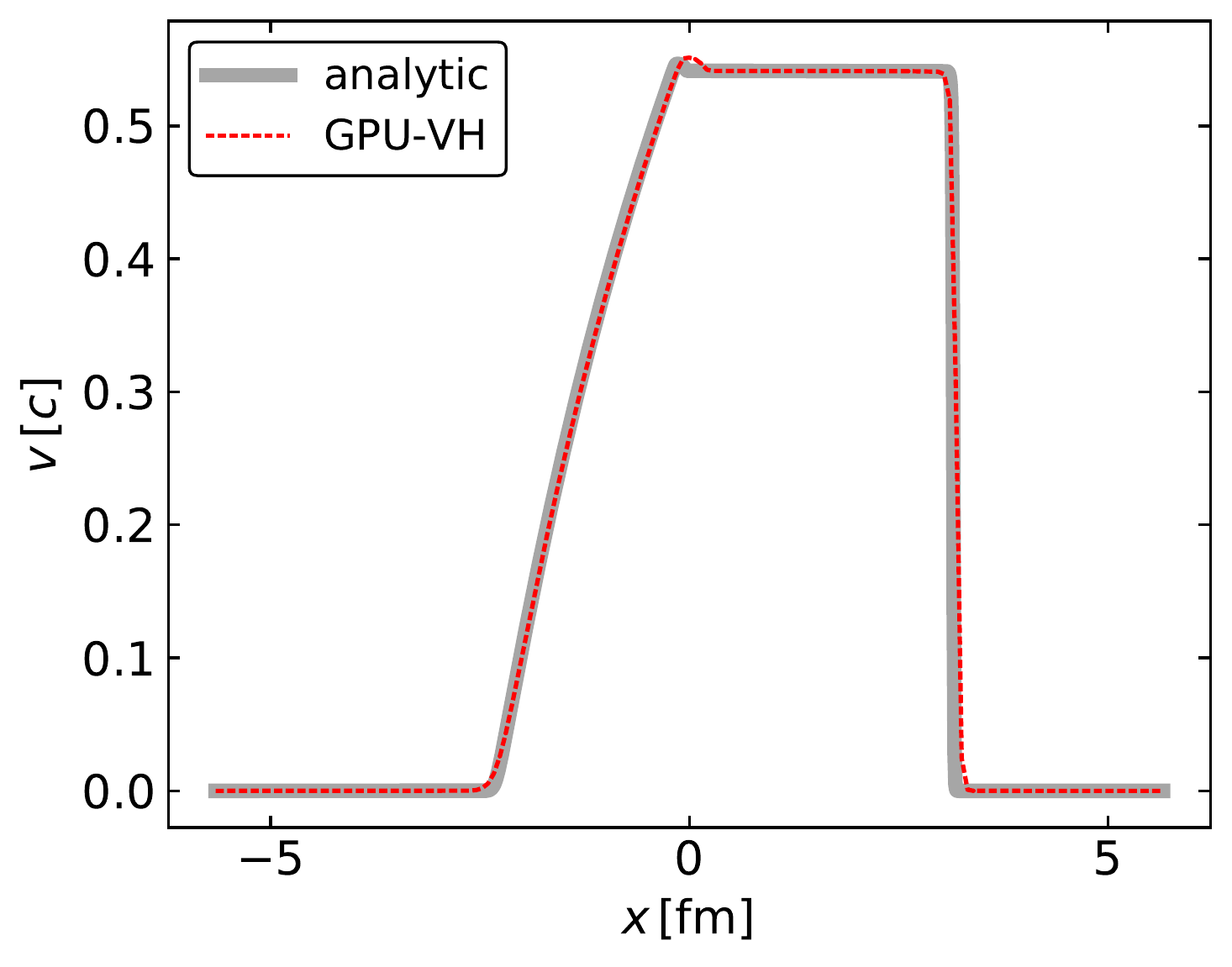}
  \end{tabular}
  \caption{Relativistic Sod shock tube test with the discontinuity in the initial condition placed along $y=x$. The top panels shows the energy density (left) and velocity $v\equiv\sqrt{v^{2}_x+v^{2}_y}$ in the transverse plane. The bottom panels show the same results, with the axis rotated by $\pi/4$, in order to make comparisons to the analytical one-dimensional result (i.e. the initial discontinuity placed along $x=0$).}
  \label{SodShockTube2dFig}
\end{figure*}
By placing the initial discontinuity along the main diagonal of the transverse computational grid (i.e. along $y=x$ plane), we can test the accuracy of the spatial gradients and fluid velocity components in the $x$ and $y$ directions simultaneously. The result of this test is shown in Fig.~\ref{SodShockTube2dFig}. The initial energy densities on both sides of the membrane as well as the EoS were chosen the same as in Sec.~\ref{sec:shockTube1d} before. The top row shows colormaps of the two-dimensional energy density and fluid velocity $v\equiv\sqrt{v^{2}_x+v^{2}_y}$ in the transverse plane. In order to make comparison to the one-dimensional analytic solution discussed in Sec.~\ref{sec:shockTube1d} we rotate the solution by an angle $\pi/4$ by applying the rotation matrix $R(\theta)$ on the grid points $(x,y)$, i.e. $(x^\prime,y^\prime)^\mathrm{T}=R(\pi/4)(x,y)^\mathrm{T}$. We then plot the result of this transformation in the bottom row of Fig.~\ref{SodShockTube2dFig} at $y^\prime=0$ as a function of $x^\prime$, dropping the prime.

\subsection{Cylindrical explosion}
\label{sec:2dSphericalExplosion}
\begin{figure*}[h!]
  \centering
  \begin{tabular}{cc}
    \includegraphics[width=0.45\linewidth]{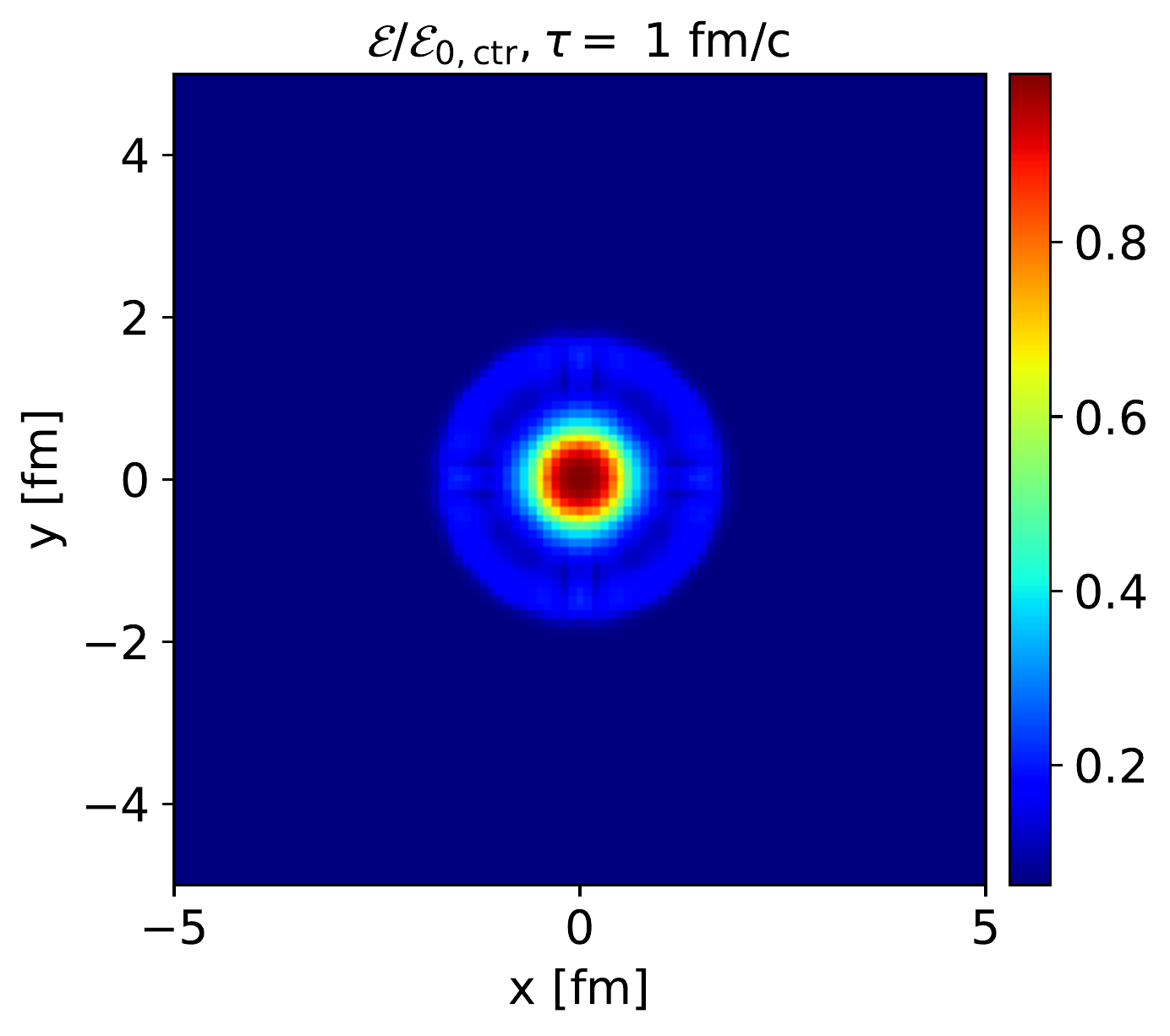} &
  	\includegraphics[width=0.45\linewidth]{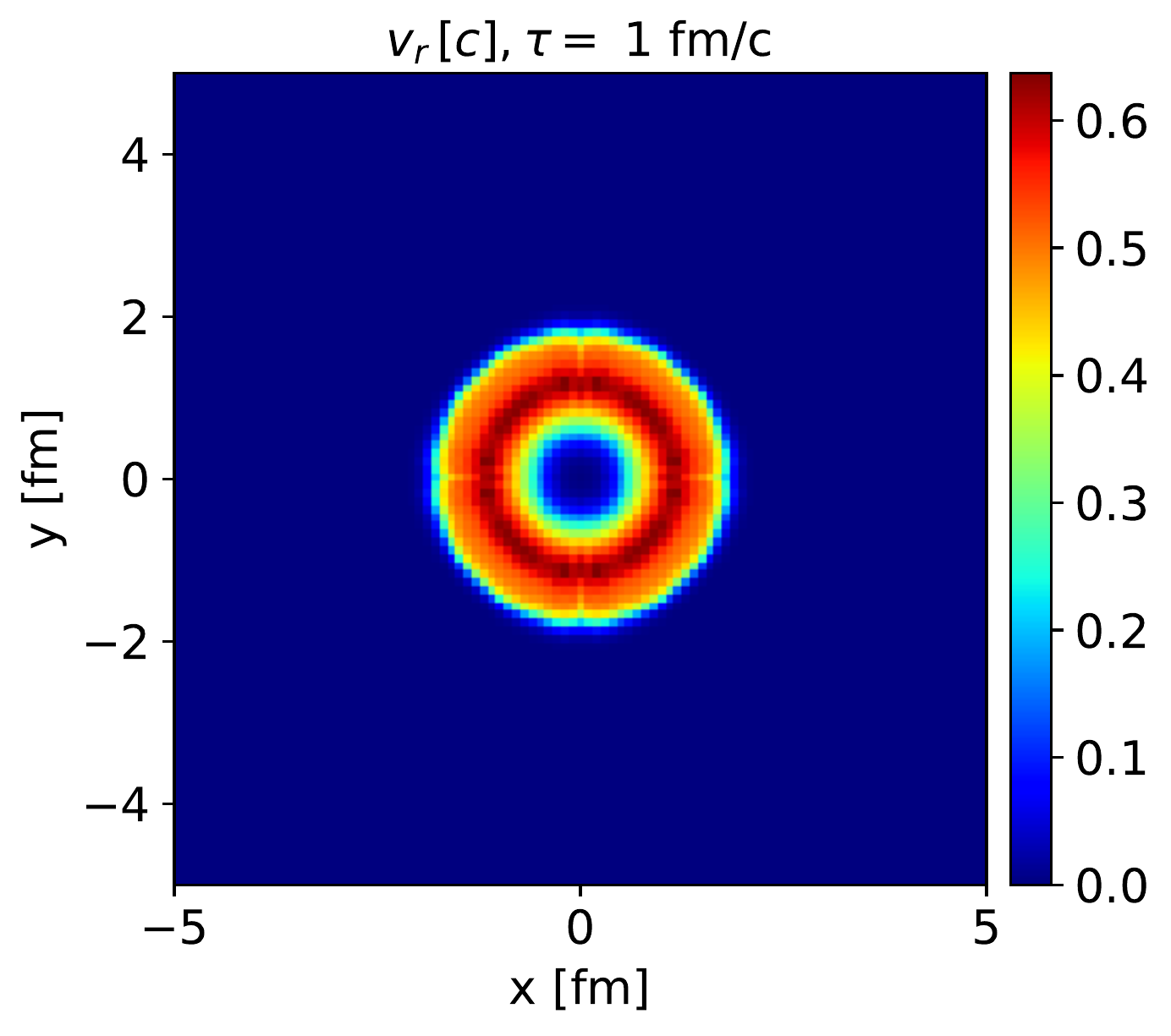} \\
  	\includegraphics[width=0.45\linewidth]{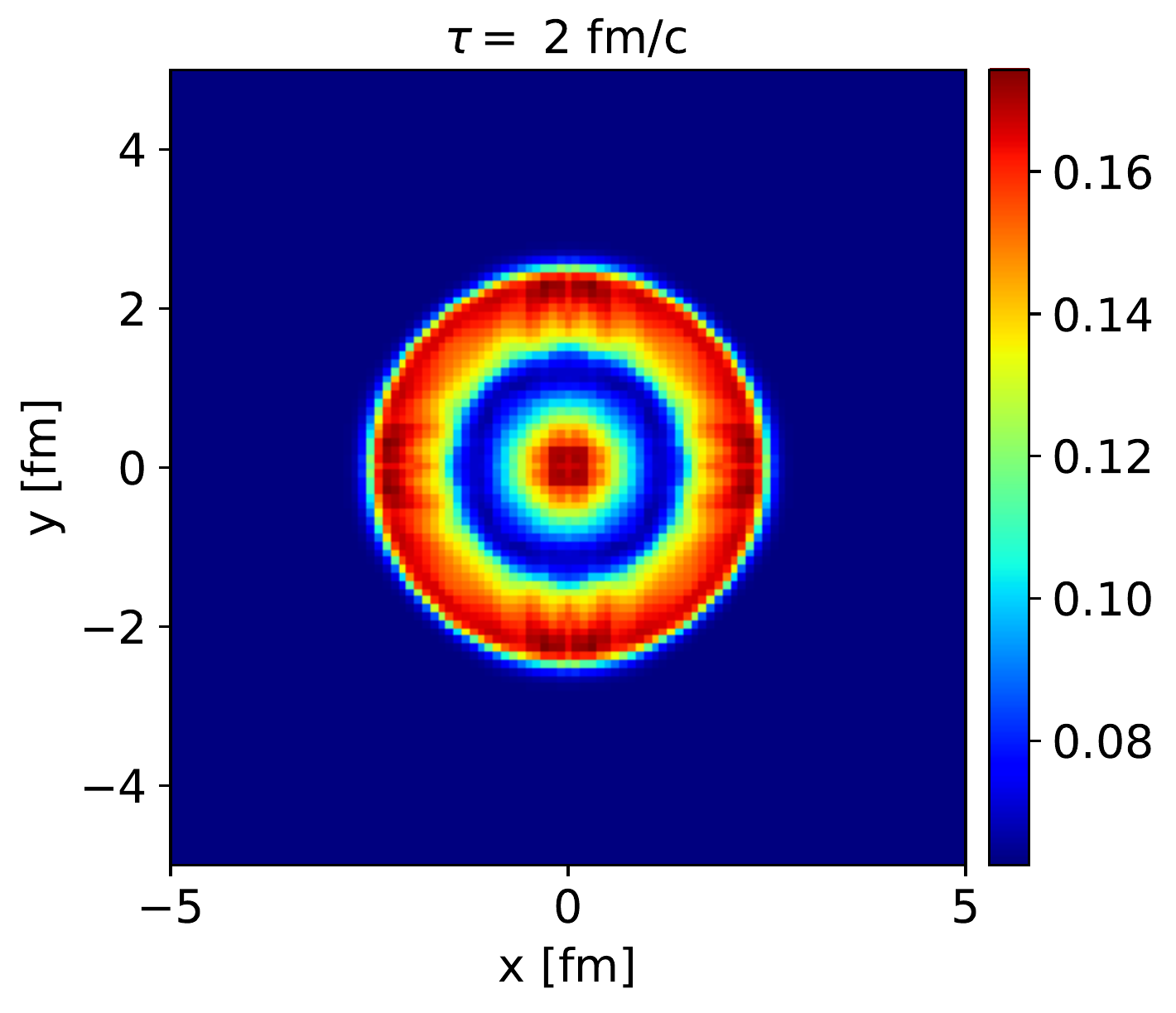} &
  	\includegraphics[width=0.45\linewidth]{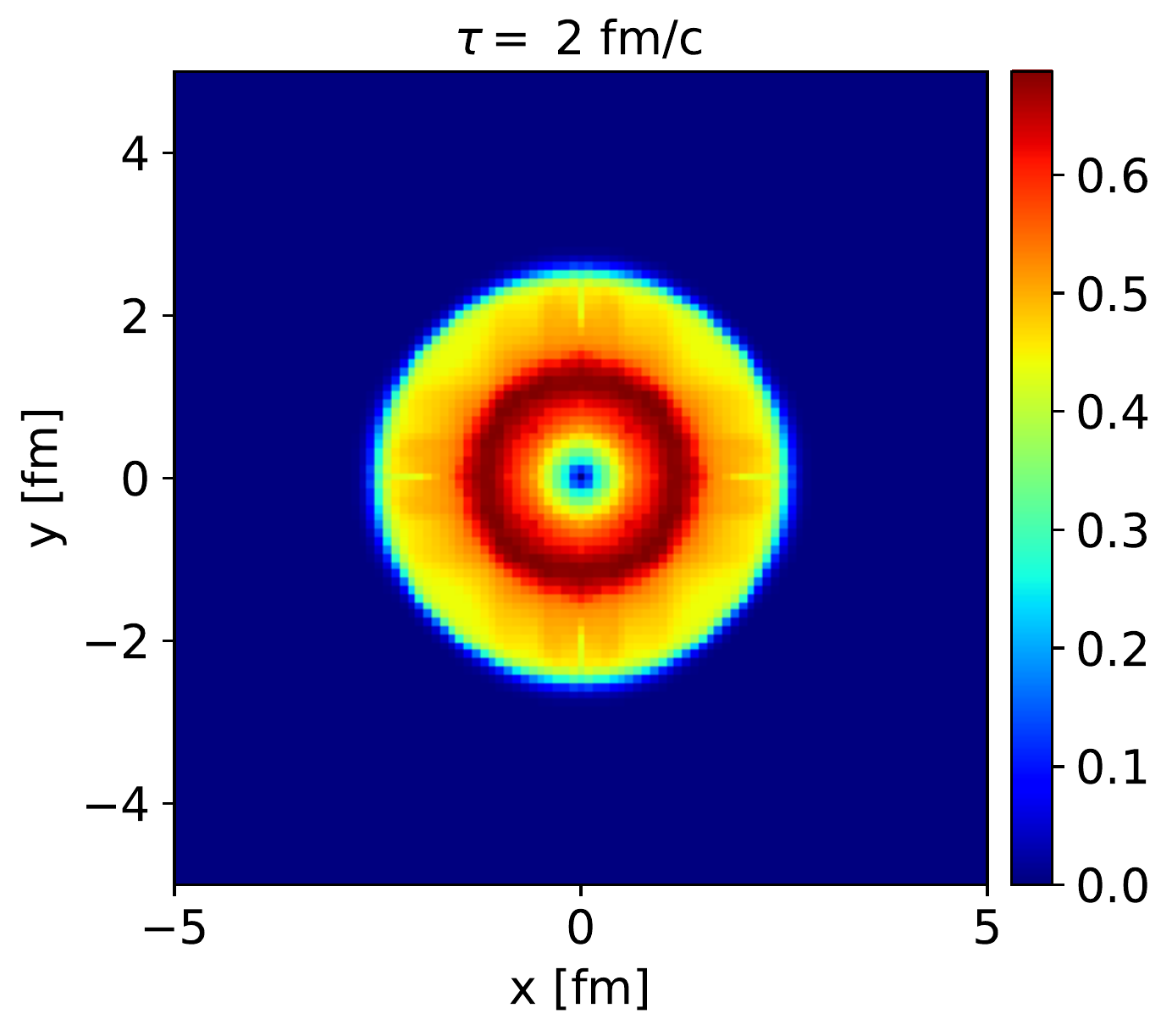} \\
    \includegraphics[width=0.45\linewidth]{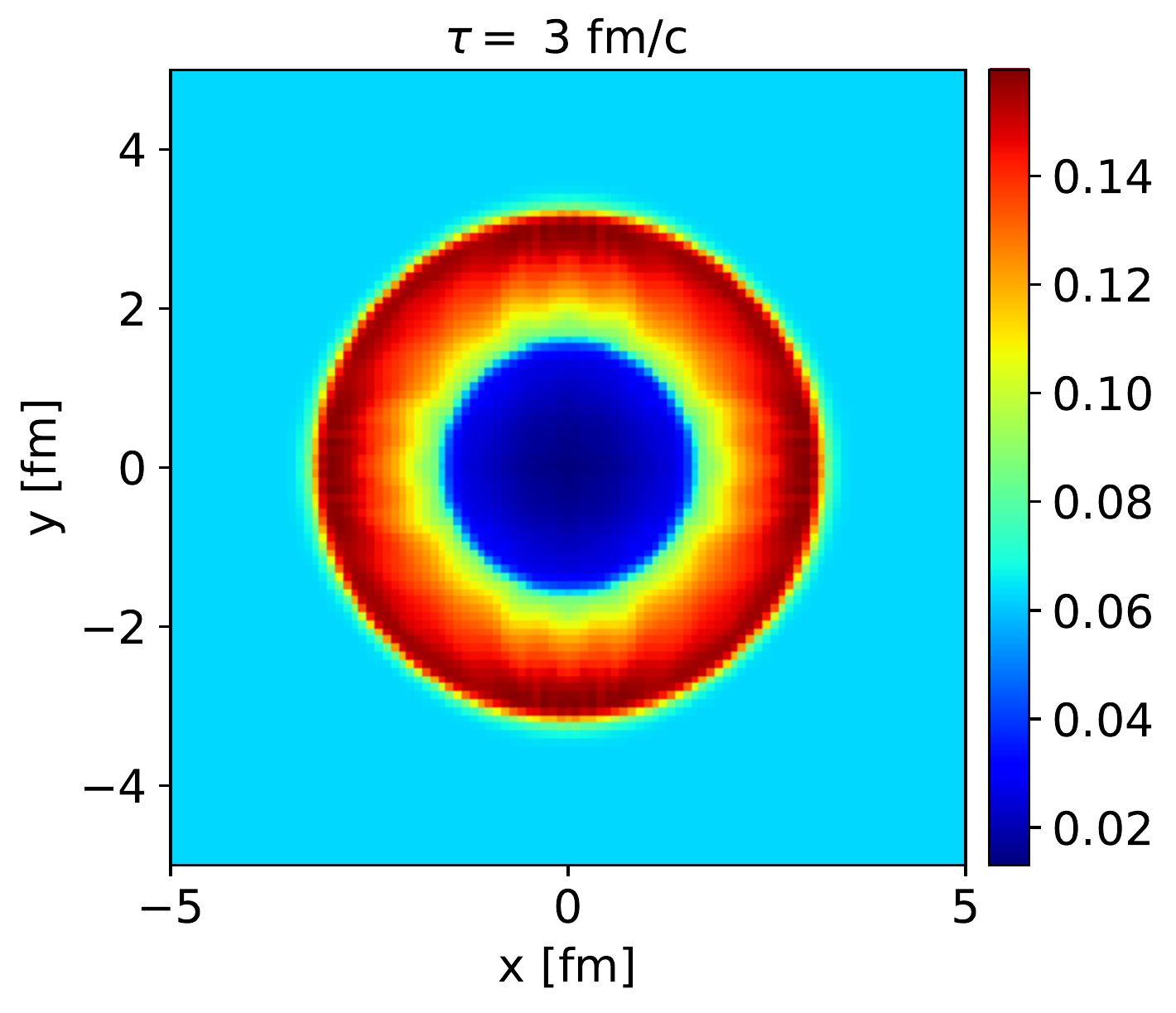} &
  	\includegraphics[width=0.45\linewidth]{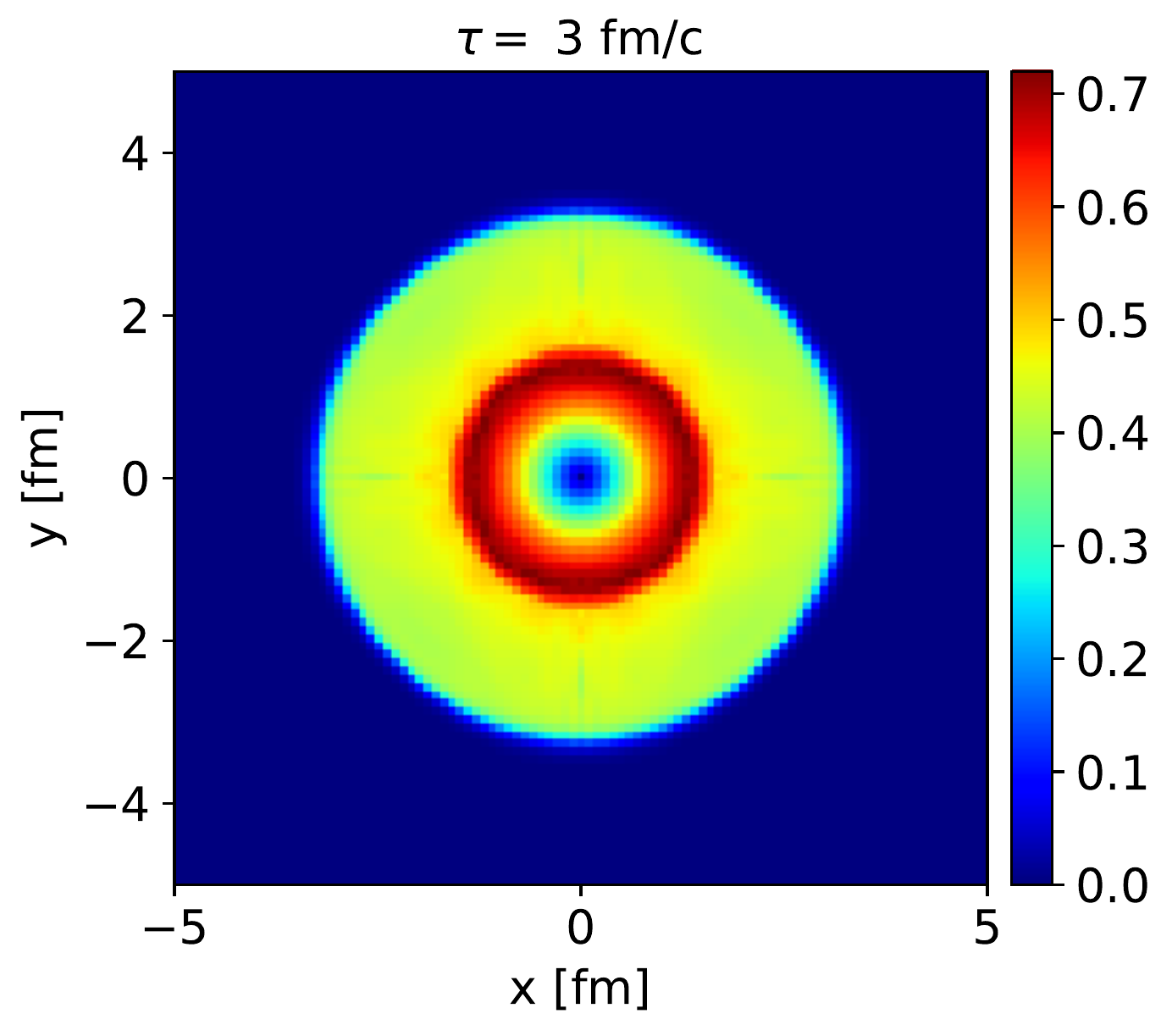} 
  \end{tabular}
\caption{Energy density (normalized by the initial energy density inside the cylindrical membrane ${\cal E}_{0,ctr}$, left panels) and the radial velocity $v_r$ (right panels) in the transverse plane at times 1, 2, and 3 fm/$c$ (top, middle, and bottom rows, respectively).}
\label{fig:cylindricalExplosionTestFig}
\end{figure*}
The next test we perform is a cylindrical blast wave from a small initial pressure (or energy density) perturbation inside a homogenous medium.
For this test problem the initial energy density in a $10^{2}$ fm$^2$ box is set to $\ed_0=0.0015$ GeV/fm$^{3}$. A fluid of much higher pressure ($\ed_0=0.0246$ GeV/fm$^{3}$) is placed inside a cylindrical membrane of radius $R=1$ fm defined by $r\equiv\sqrt{x^2+y^2}\le R$. We again use the equation of state $\ed=3\peq$. When the membrane is suddenly removed at $t=0$, a shock wave should propagate radially outwards and a rarefaction wave inwards. In Fig.~\ref{fig:cylindricalExplosionTestFig} we plot the energy density (normalized by the initial energy density inside the cylindrical membrane ${\cal E}_{0,ctr}$) and the radial velocity $v_r\equiv \frac{x}{r}v_x+\frac{y}{r}v_y$ in the transverse plane at the times $t=1$, 2, and 3 fm$/c$ (top, middle, and bottom rows, respectively). 
The first row is at a snapshot in time ($t=1$ fm/$c$) where the rarefaction wave has reached a radius $0<r<R$ resulting in a homogenous cylinder of radius $r$ with the original energy density of the cylinder (corresponding to 1 in Fig.~\ref{fig:cylindricalExplosionTestFig}), surrounded by by lower-density matter expanding with a velocity $v_r$. In the second row and third rows the rarefaction wave has reached the center and reflected off of the origin resulting in an additional shock front propagating radially outwards.
This test demonstrates the ability of the code to preserve cylindrical symmetry while using a Cartesian coordinate system. 

\subsection{Conformal Bjorken flow}
\label{sec:conformalBjorken}
\begin{figure}[t!]
\begin{center}
\includegraphics[width=\linewidth]{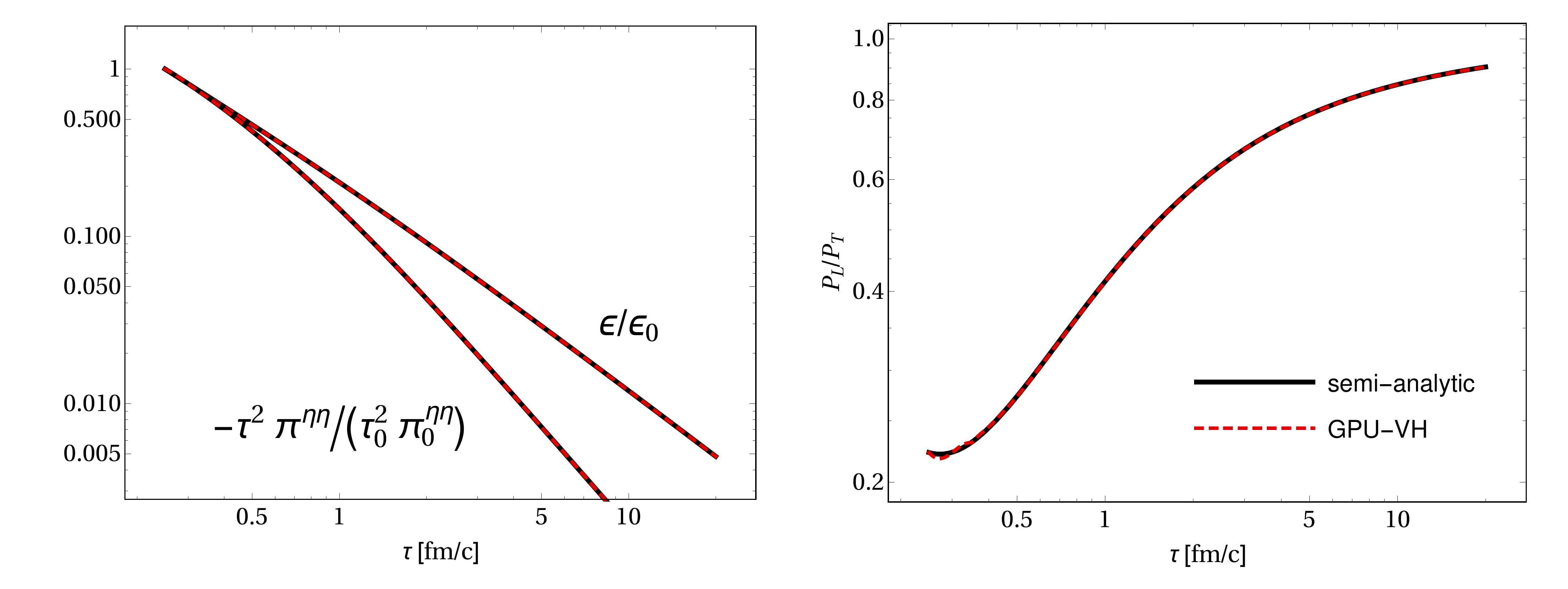}
\end{center}
\vspace{-7mm}
\caption{Comparison of the energy density (scaled by its initial value $\ed_{0}$) between the semi-analytical solution (black line) and the numerical result from GPU-VH (red dashed line) for Bjorken flow. The initial conditions in this figure are $T_0=0.6\,\mathrm{GeV}$, $\tau_{0}=0.25$\, fm, $\pi^{\mu\nu}_{0}=\pi^{\mu\nu}_\mathrm{NS}$, and the specific shear viscosity is $\eta/s=0.2$.}
\label{fig:bjorkenTestFig}
\end{figure}
Having checked the fluid dynamic part of our code against various Riemann problems in Cartesian coordinates, we will now test the code in Milne coordinates. The first test we consider is a conformal system ($\peq=\ed/3$) undergoing Bjorken flow -- longitudinally boost-invariant and transversely homogenous expansion. The resulting equations of motion are a well known set of coupled ordinary differential equations for $\ed_0$ and $\pi\equiv -\tau^{2}\pi^{\eta\eta}$ that can easily be solved by high accuracy ODE solvers:
\begin{eqnarray}
\dot{\ed} &=&-\frac{\ed+\peq-\pi }{\tau }\,,
\label{evolutionEd0p1} \\
\tau _{\pi }\dot{\pi}+\pi &=&\frac{4}{3}\frac{\eta}{\tau} -\left( \frac{1}{3}\tau
_{\pi \pi }+\delta _{\pi \pi }\right) \frac{\pi }{\tau }  \,,  \label{evolutionpi1}
\end{eqnarray}
where $\dot{f}\equiv\frac{df}{d\tau}$.
Fig.~\ref{fig:bjorkenTestFig} compares the semi-analytical solution (black line) to the numerical solution obtained with the GPU-VH code (red dashed line) as a function of proper time. The system was initialized at $\tau_0=0.25\,\mathrm{fm}$ with a constant temperature $T_{0}=0.6\,\mathrm{GeV}$. At this time the system is assumed to be highly anisotropic in momentum-space; we implement this by setting the initial shear stress tensor to its Navier-Stokes value, $\pi^{\mu\nu}_{0}=\pi^{\mu\nu}_\mathrm{NS}$, considering a value of $\eta/s=0.2$ for the specific shear viscosity. The left panel in Fig.~\ref{fig:bjorkenTestFig} shows the normalized energy density and shear stress tensor and the right panel shows the pressure anisotropy $P_\mathrm{L}/P_\perp=\frac{\peq-\pi}{\peq+\pi/2}$. The overall agreement between the semi-analytical solution and the numerical simulation is excellent.  

\subsection{Nonconformal Bjorken flow}
\label{sec:nonconformalBjorken}

\begin{figure*}[t!]
  \centering
  \begin{tabular}{cc}
  \includegraphics[width=0.45\linewidth]{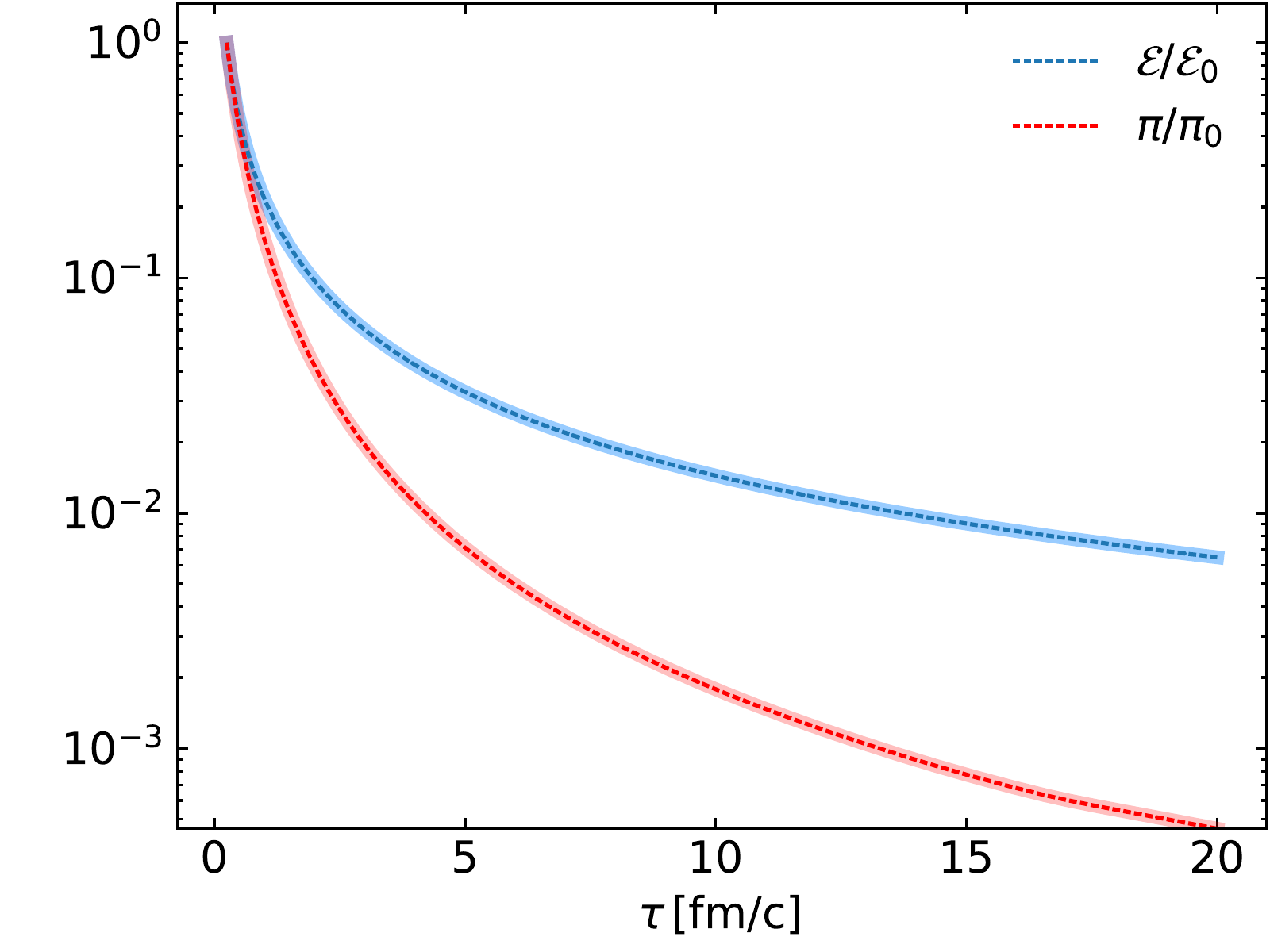} &
  \includegraphics[width=0.45\linewidth]{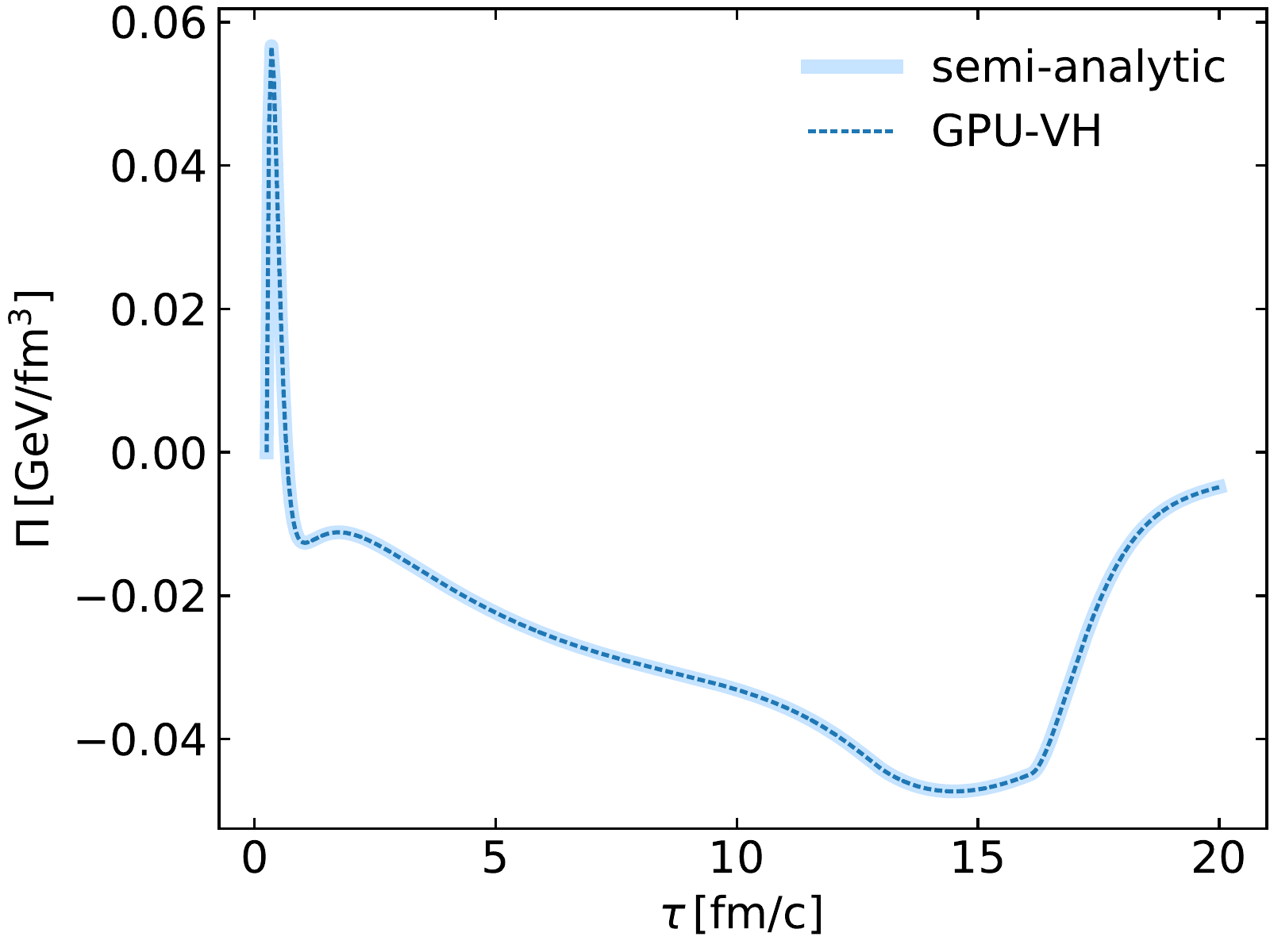} 
  \end{tabular}
  \caption{Same initial conditions as Fig.~\ref{fig:bjorkenTestFig}, but with an equation of state defined via Eq.~(\ref{traceAnomaly} and a nonvanishing bulk viscous pressure (initialized to zero, $\Pi_0=0$).}
  \label{nonconformalBjorkenTest}
\end{figure*}

\begin{figure}[t!]
\begin{center}
\includegraphics[width=\linewidth]{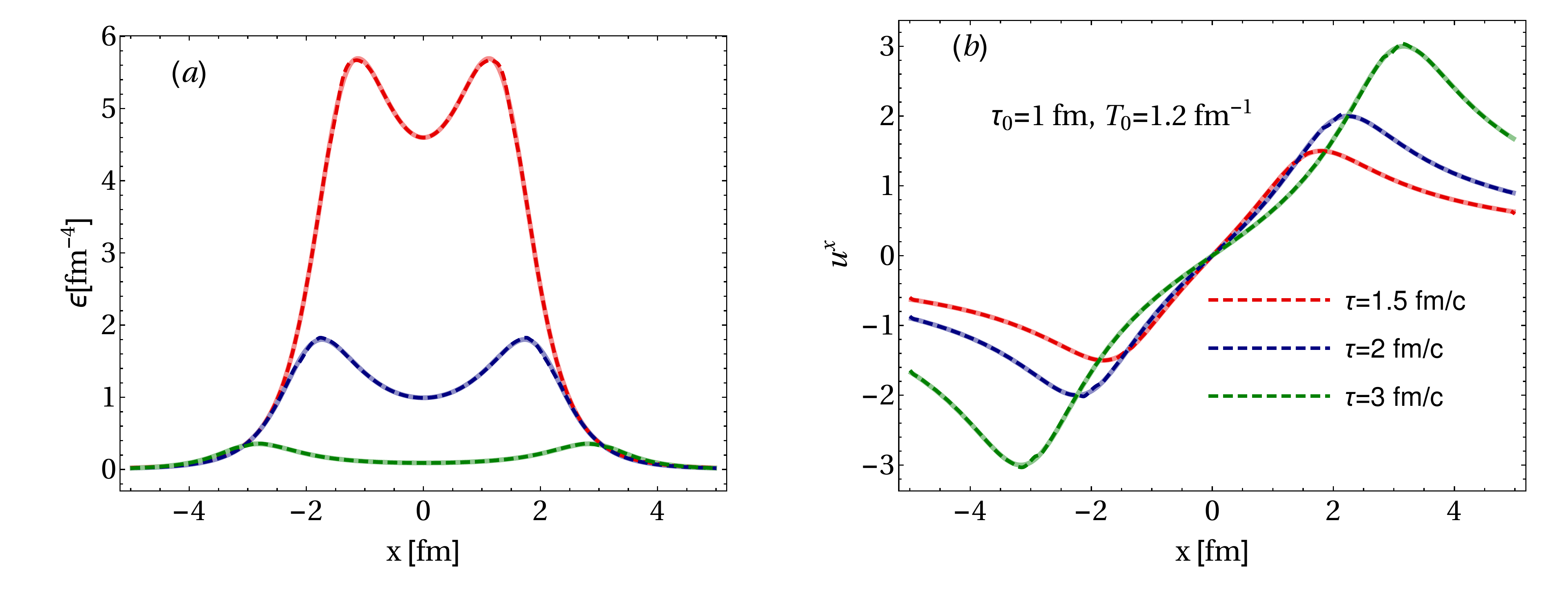}
\end{center}
\vspace{-7mm}
\caption{Comparison between Gubser's ideal analytical solution (black line) and the numerical result from GPU-VH run in its ideal fluid dynamics mode (red dashed line) plotted at three different point in proper time $\tau=\{1.5,\,2,\,3\}$\, fm. Panel (a) shows the energy density, and panel (b) the $x$ component of the fluid velocity in the transverse plane at $y=0$. The initial conditions in this figure are $\tau_{0}=1$\, fm, $T_0=1.2\,\mathrm{fm}^{-1}$.}
\label{fig:idealGubserTestFig}
\end{figure}

We will now test the evolution of the bulk viscous pressure $\Pi$ in Bjorken flow. This is done by replacing the conformal EoS $\peq=\ed/3$ by the nonconformal equation of state for QCD matter (see Eqs.~(\ref{traceAnomaly}) and (\ref{peqDividedByT})) and self-consistently allowing $\Pi$ to propagate as an additional degree of freedom. In this case the 0+1d fluid dynamic equations are
\begin{eqnarray}
\dot{\ed} &=&-\frac{\ed+\peq+\Pi -\pi }{\tau }\,,
\label{evolutionEd} \\
\tau _{\Pi }\dot{\Pi}+\Pi &=&-\frac{\zeta }{\tau }-\delta _{\Pi \Pi }\frac{%
\Pi }{\tau }+\lambda _{\Pi \pi }\frac{\pi }{\tau }\,,  \label{evolutionPi} \\
\tau _{\pi }\dot{\pi}+\pi &=&\frac{4}{3}\frac{\eta}{\tau} -\left( \frac{1}{3}\tau
_{\pi \pi }+\delta _{\pi \pi }\right) \frac{\pi }{\tau }+\frac{2}{3}\lambda _{\pi \Pi }%
 \frac{\Pi}{\tau}  \,,  \label{evolutionpi}
\end{eqnarray}
where the transport coefficients are defined in Eqs.~(\ref{betaPi})-(\ref{lambdapiPi}) and $\pi\equiv -\tau^{2}\pi^{\eta\eta}$.%
\footnote{The conformal fluid dynamic equations for Bjorken flow used above can be obtained by setting the trace anomaly term to zero in the equation of state. The approximation made in the transport coefficients does not, however, reduce Eq.~(\ref{evolutionPi}) to the trivial equation $0=0$ as it should. Instead we must explicitly define a vanishing bulk viscous pressure, $\Pi\equiv 0$.} 
The result of this test is shown in Fig.~\ref{nonconformalBjorkenTest}. The same initial conditions as in Sec.~\ref{sec:conformalBjorken} are used along with $\Pi_0=0$. Fig.~\ref{nonconformalBjorkenTest} compares the numerical simulation with the semi-analytic solution obtained by solving the set (\ref{evolutionEd})-(\ref{evolutionpi}). The left panel shows the normalized energy density (blue solid line) and shear stress tensor (red solid line) as a function of proper time compared to their semi-analytic solution (light banded lines). The right panel shows the numerical evolution of $\Pi$; its qualitative behavior is largely determined by the parametrization used for the bulk viscosity $\zeta/s$ (given by Eq.~\ref{eq:zetas}). $\Pi$ initially starts at zero and then becomes positive even though bulk viscous pressure in expanding systems is normally negative. This is due to the bulk-shear coupling term~\cite{Denicol:2014mca}, $\lambda _{\Pi \pi }\pi/\tau$, in Eq.~(\ref{evolutionPi}) which dominates the dynamical evolution of $\Pi$ at all times when the system is not near $T_c$ where $\zeta/s$ peaks.

\subsection{Analytic solution to ideal fluid dynamics under Gubser symmetry}
\label{sec:GubserIdeal}

In order to test the transverse dynamics of our code, we will compare against ($1+1$)-dimensional solutions of relativistic fluid dynamics subject to Gubser flow which includes both longitudinal and transverse radial expansion~\cite{Gubser2010,Gubser2011469}. For ideal fluid dynamics, the analytical solution is:
\begin{align}
T(\tau,r)&=\frac{\hat{T}_0(2q\tau)^{2/3}}{\tau\left[1+2q^2(\tau^2+r^2)+q^{4}(\tau^2-r^2)^{2}\right]^{1/3}}\;,
\\
u_{x}&=\frac{x}{r}\sinh\kappa(\tau,r)\;,
\\
u_{y}&=\frac{y}{r}\sinh\kappa(\tau,r)\;,
\\
\kappa(\tau,r)&=\tanh^{-1}\frac{2q^2\tau r}{1+q^2\tau^2+q^2r^2}\;,
\end{align}
where $q$ is an arbitrary inverse length scale which we set to $q=1\,\mathrm{fm}^{-1}$, and $u_\eta\equiv 0$. We start the hydrodynamical evolution at $\tau_0=1\,\mathrm{fm}$ with an initial temperature $\hat{T}_{0}=1.2\,\mathrm{fm}^{-1}$.
Fig.~\ref{fig:idealGubserTestFig} shows excellent agreement of GPU-VH (in its ideal fluid dynamic mode) with the analytic solution at various values of proper time.
\subsection{Semi-analytic solution to dissipative fluid dynamics under Gubser symmetry}
\label{sec:GubserIS}
\begin{figure}[t!]
\begin{center}
\includegraphics[width=\linewidth]{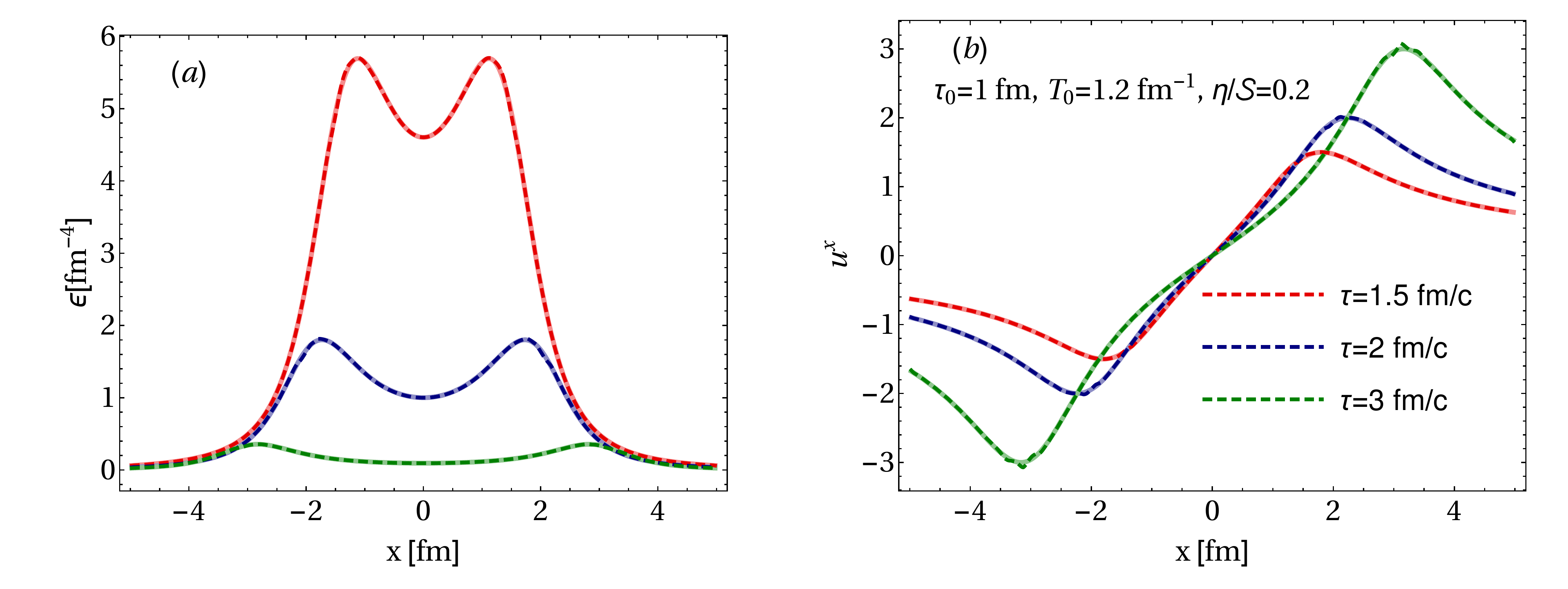}
\includegraphics[width=0.45\linewidth]{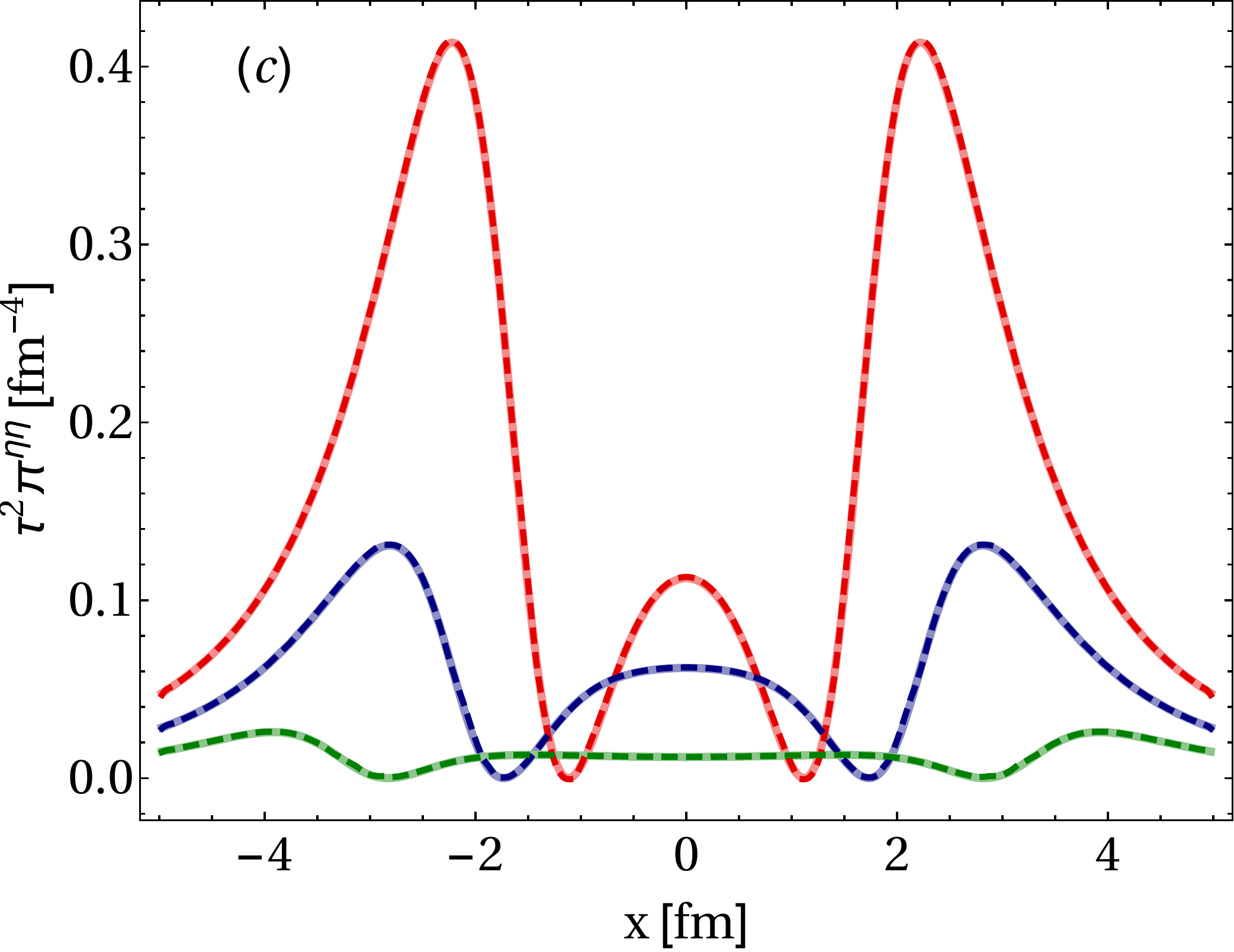}
\end{center}
\vspace{-7mm}
\caption{Comparision between the semi-analytical solution (thin solid line) and the numerical result from GPU-VH (thicker dashed line) plotted at three different points in proper time $\tau=\{1.5,\,2,\,3\}$ fm$/c$. Panel (a) shows the energy density, panel (b) the $x$ component of fluid velocity, and panel (c) the $\eta\eta$ component of the shear stress tensor in the transverse plane at $y=0$. The initial conditions in this figure are $\tau_{0}=1$\, fm, $T_0=1.2~\mathrm{fm}^{-1}$, and $\eta/s=0.2$. }
\label{fig:viscousGubserTestFig}
\end{figure}

Marrochio et al.~\cite{Marrochio:2013wla} derived solutions to the relativistic Israel-Stewart equations subject to Gubser flow. We refer the reader to Eqs. (11) and (12) in  Ref.~\cite{Marrochio:2013wla} for this semi-analytical solution. In order to make comparisons we must set the transport coefficient $\tau_{\pi\pi}\equiv 0$. We take the same initial conditions as Ref.~\cite{Marrochio:2013wla}: $\tau_{0}=1$\, fm, $T_0=1.2~\mathrm{fm}^{-1}$, $\pi^{\mu\nu}_0=0$ and $\eta/s=0.2$.
In Fig.~\ref{fig:viscousGubserTestFig} we show the energy density, $x$-component of the fluid velocity, and the shear stress tensor for three different times $\tau=1.5,\,2$, and 3 fm$/c$. Once again, the agreement between the semi-analytical solution and the numerical simulation is excellent. The results shown here are obtained with a much less dissipative flux limiter parameter $\theta=1.8$.  (This value of the flux limiter gives the ``best" agreement with the semi-analytic solution; the same value was also used in Ref.~\cite{Marrochio:2013wla}.)

\begin{figure*}[h!]
  \centering
  \scalebox{0.95}{
  \hspace{-5mm}
  \begin{tabular}{ccc}
  	\includegraphics[width=0.33\linewidth]{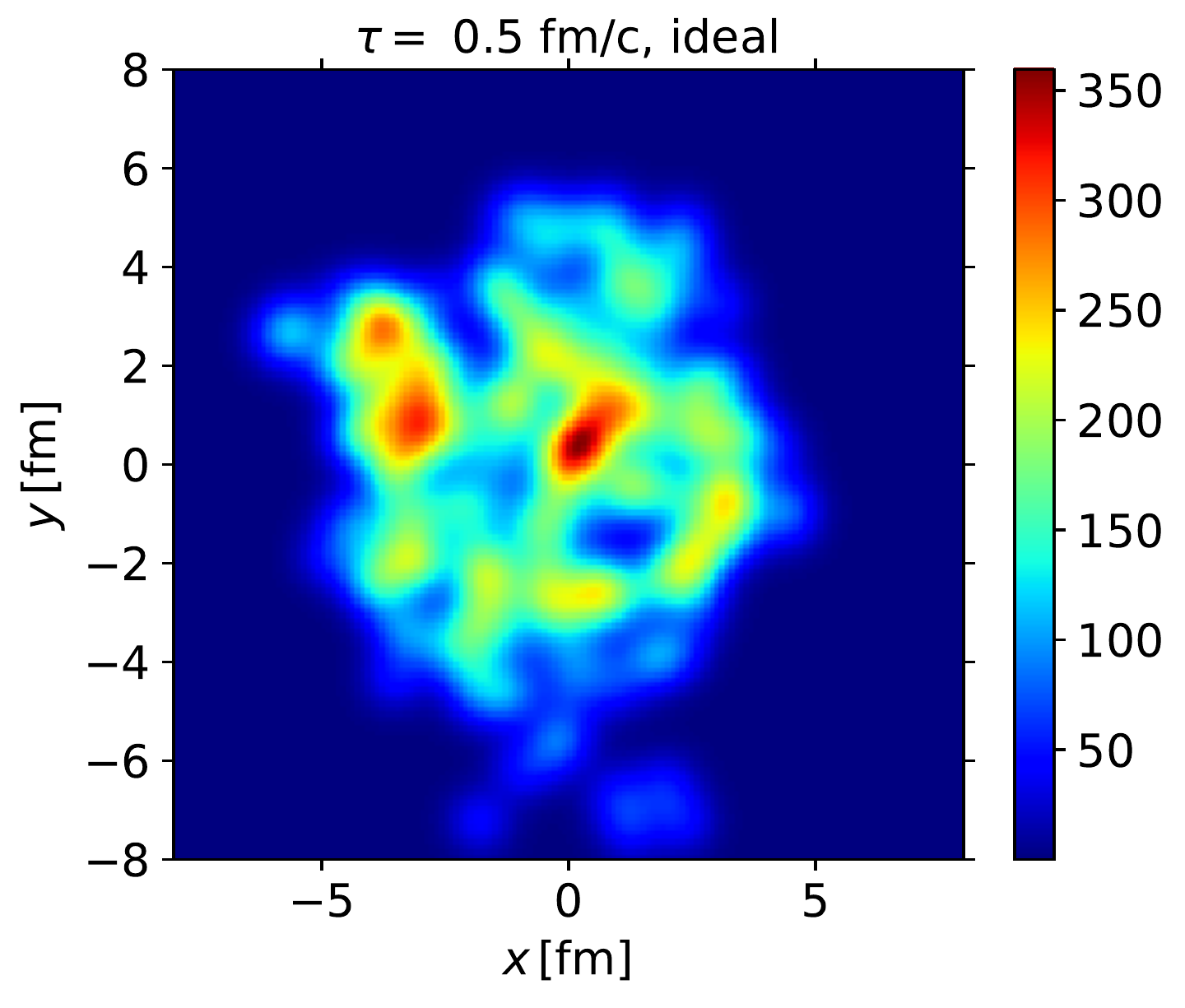} &
  	\includegraphics[width=0.33\linewidth]{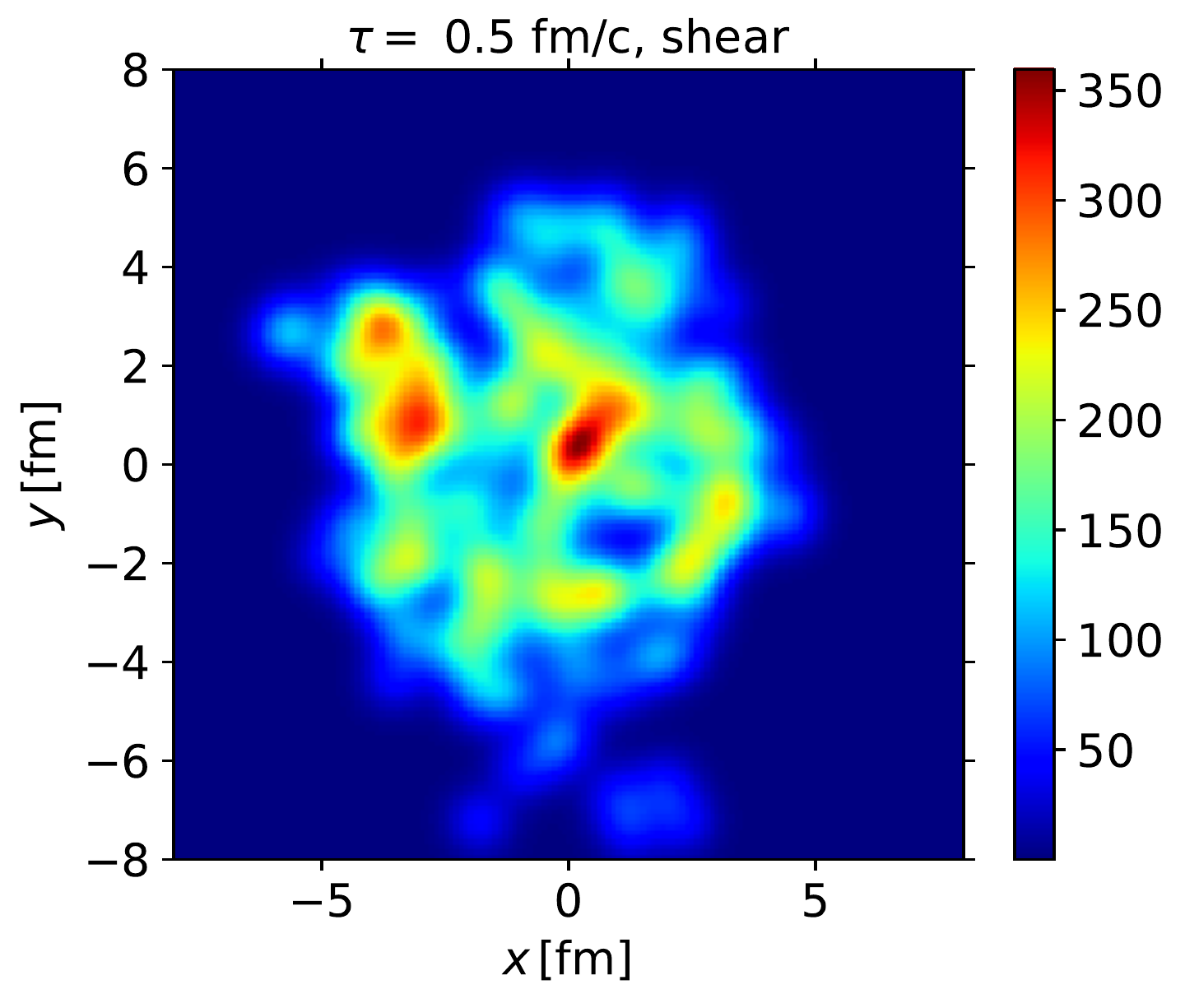} &
  	\includegraphics[width=0.33\linewidth]{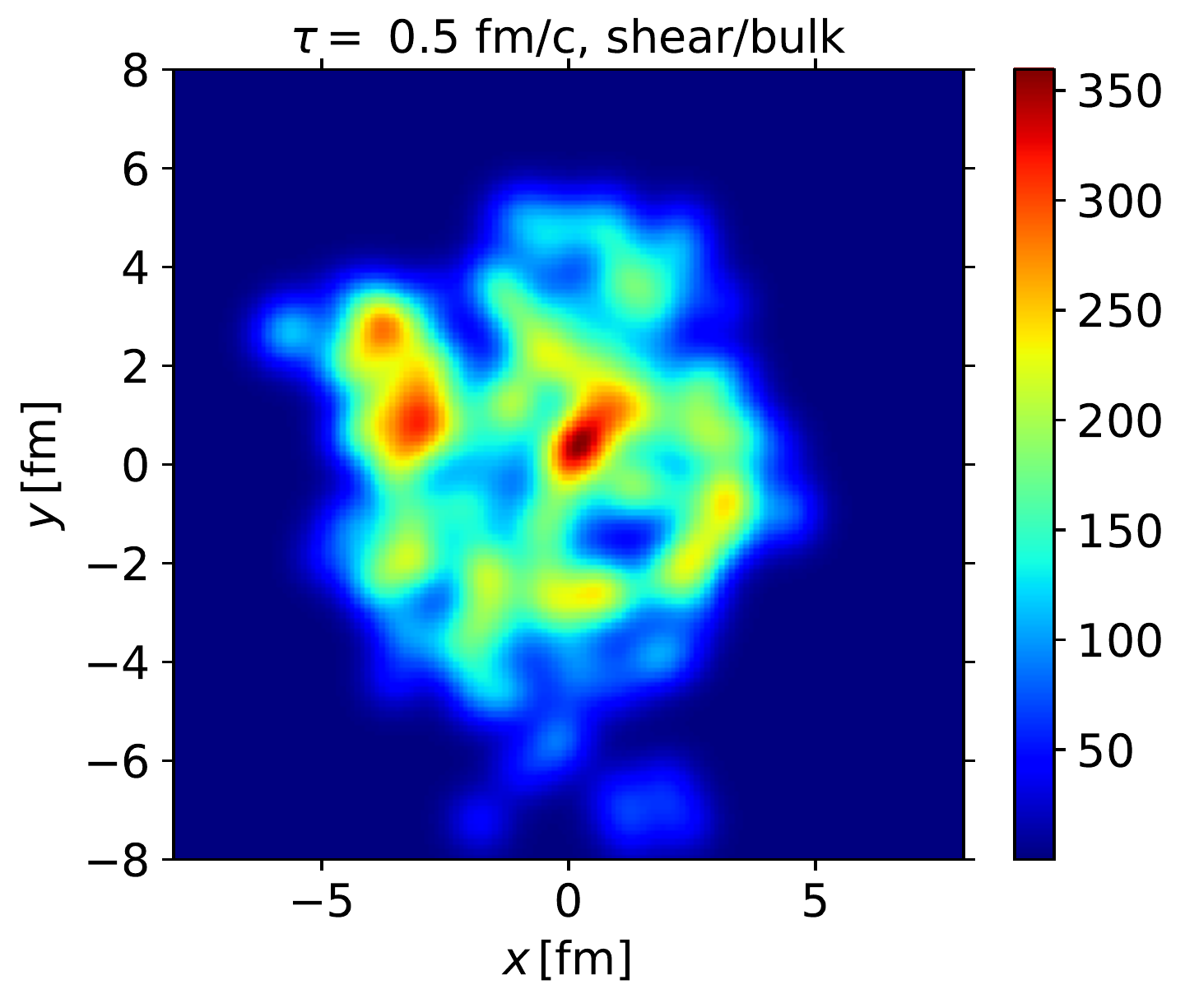} \\
  	\includegraphics[width=0.33\linewidth]{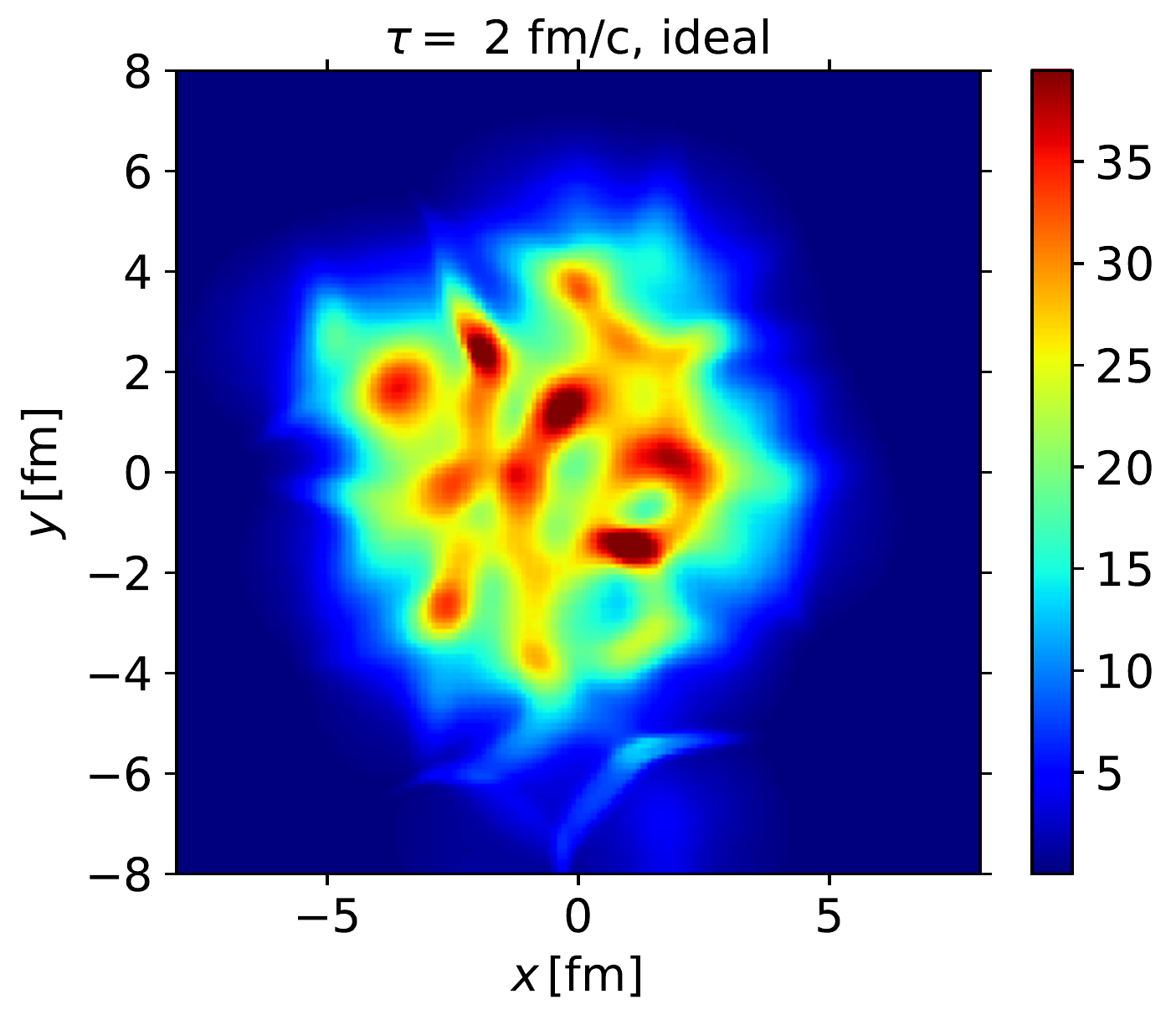} &
  	\includegraphics[width=0.33\linewidth]{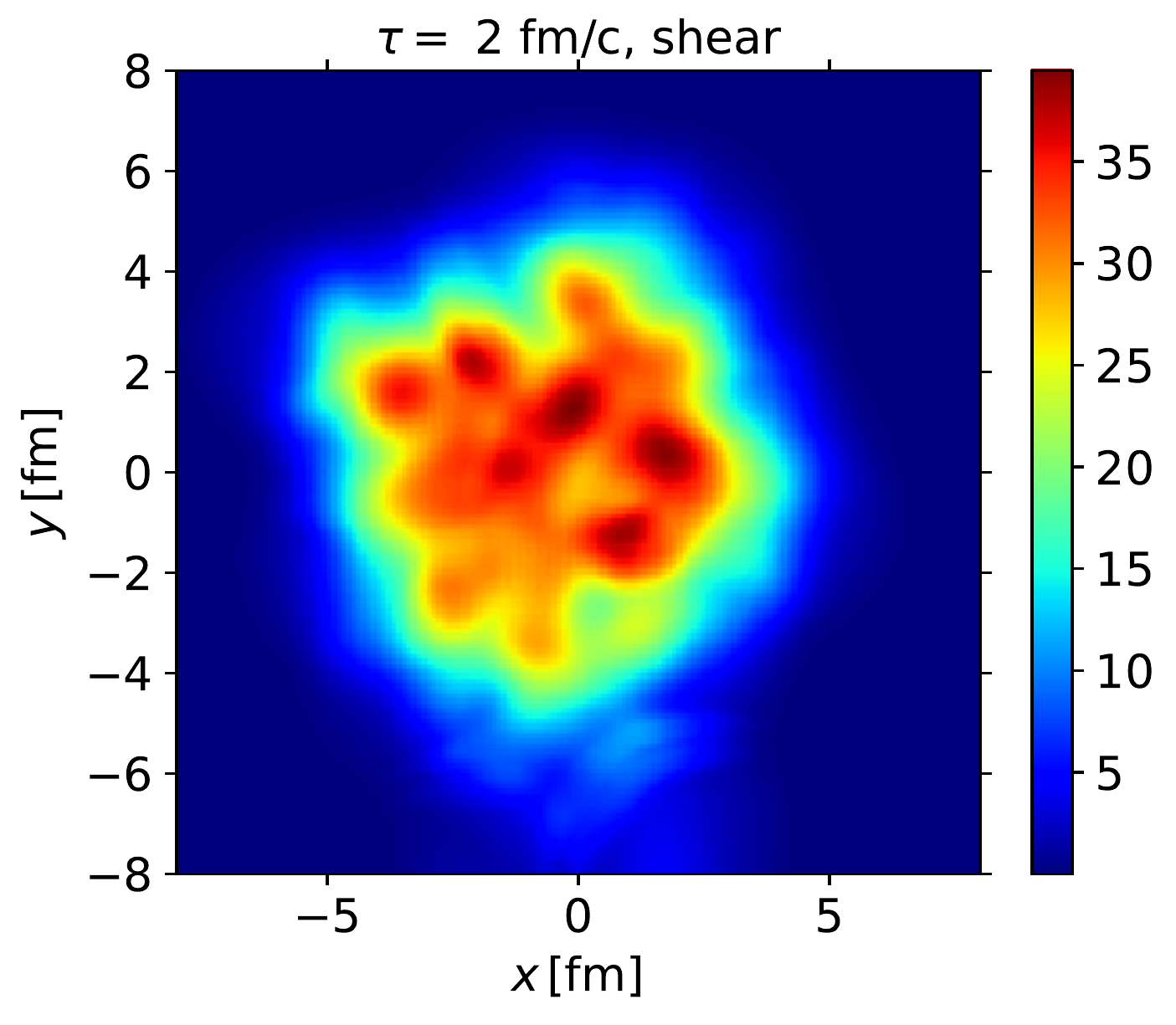} &
  	\includegraphics[width=0.33\linewidth]{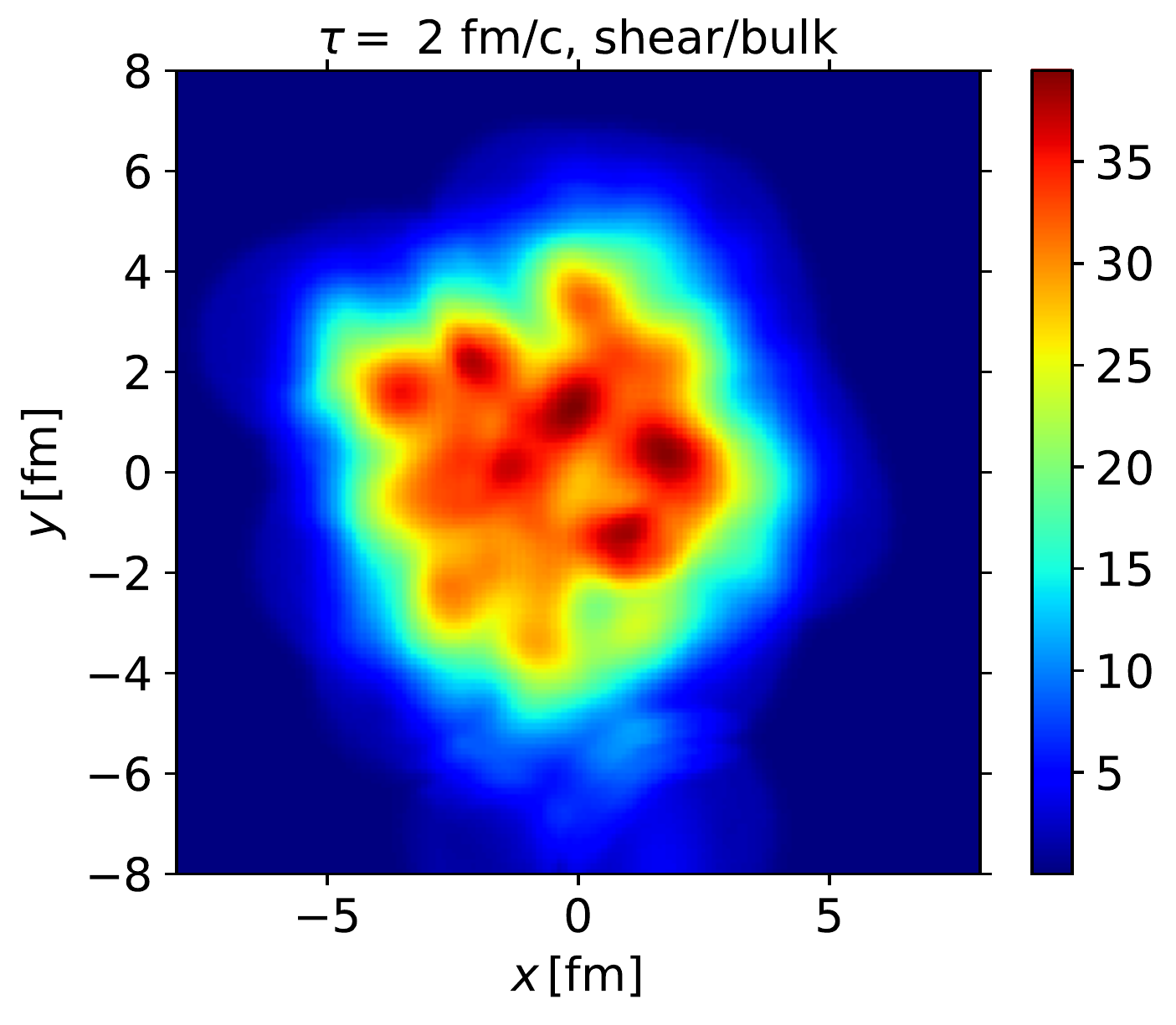} \\
  	\includegraphics[width=0.33\linewidth]{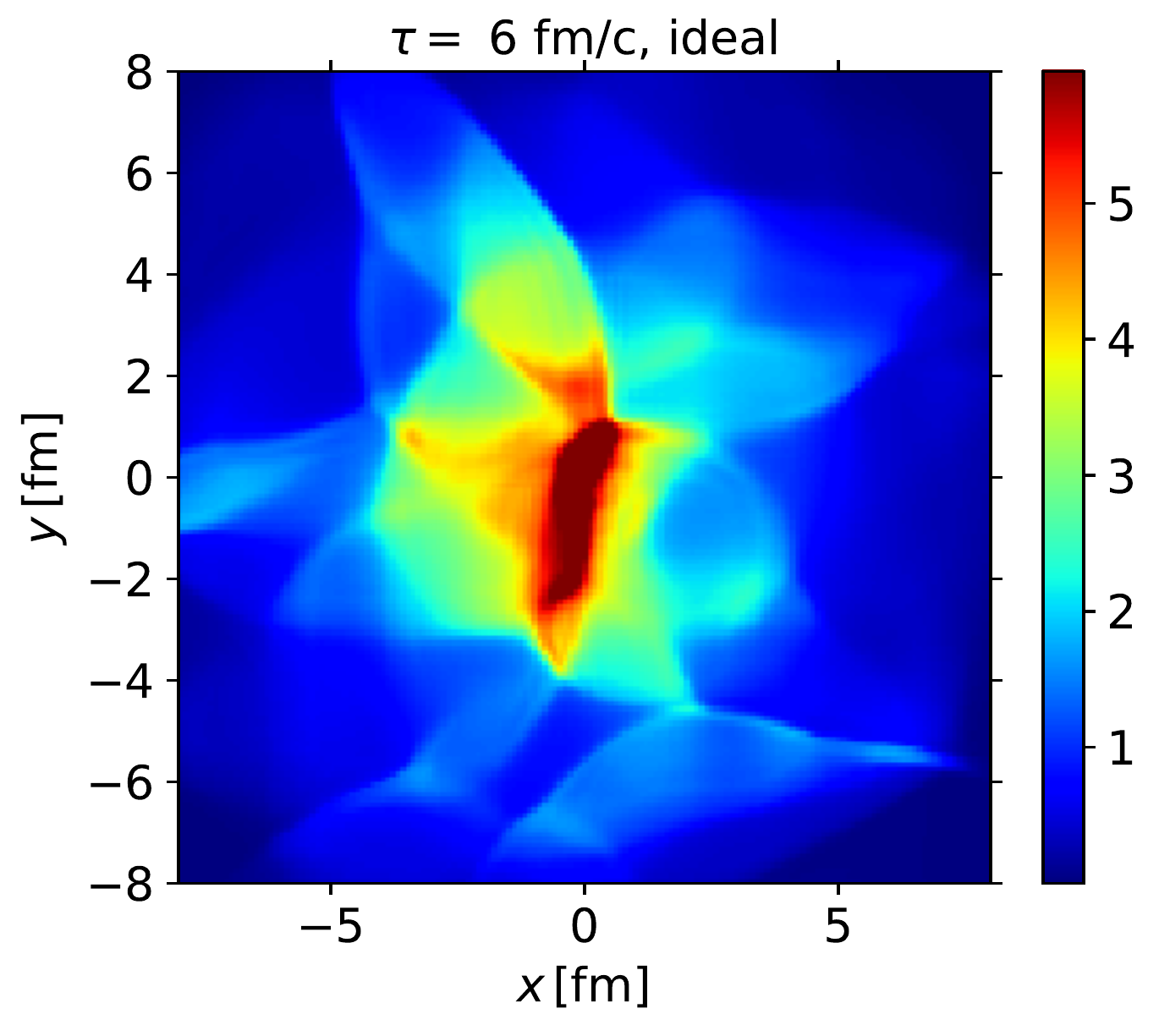} &
  	\includegraphics[width=0.33\linewidth]{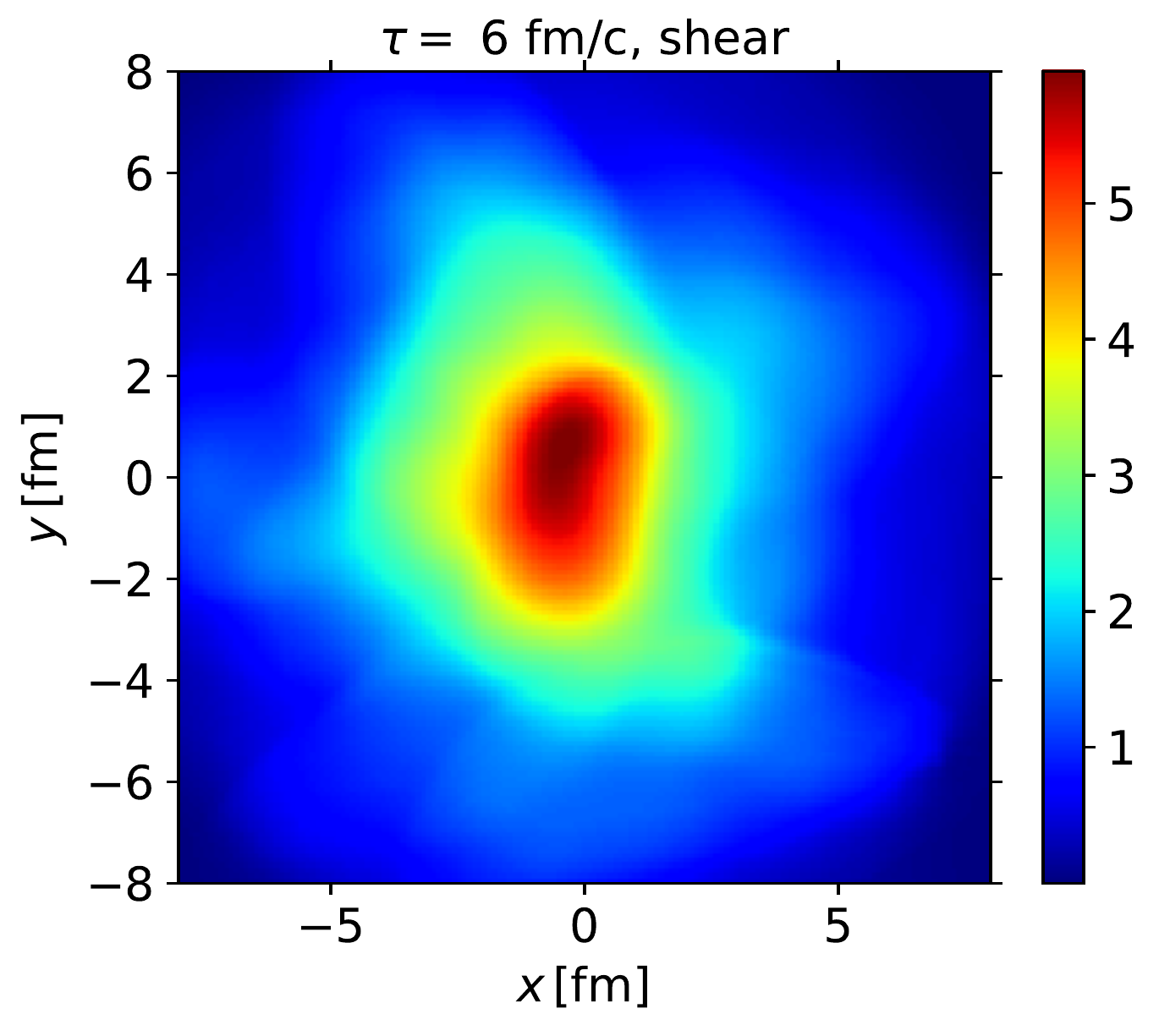} &
  	\includegraphics[width=0.33\linewidth]{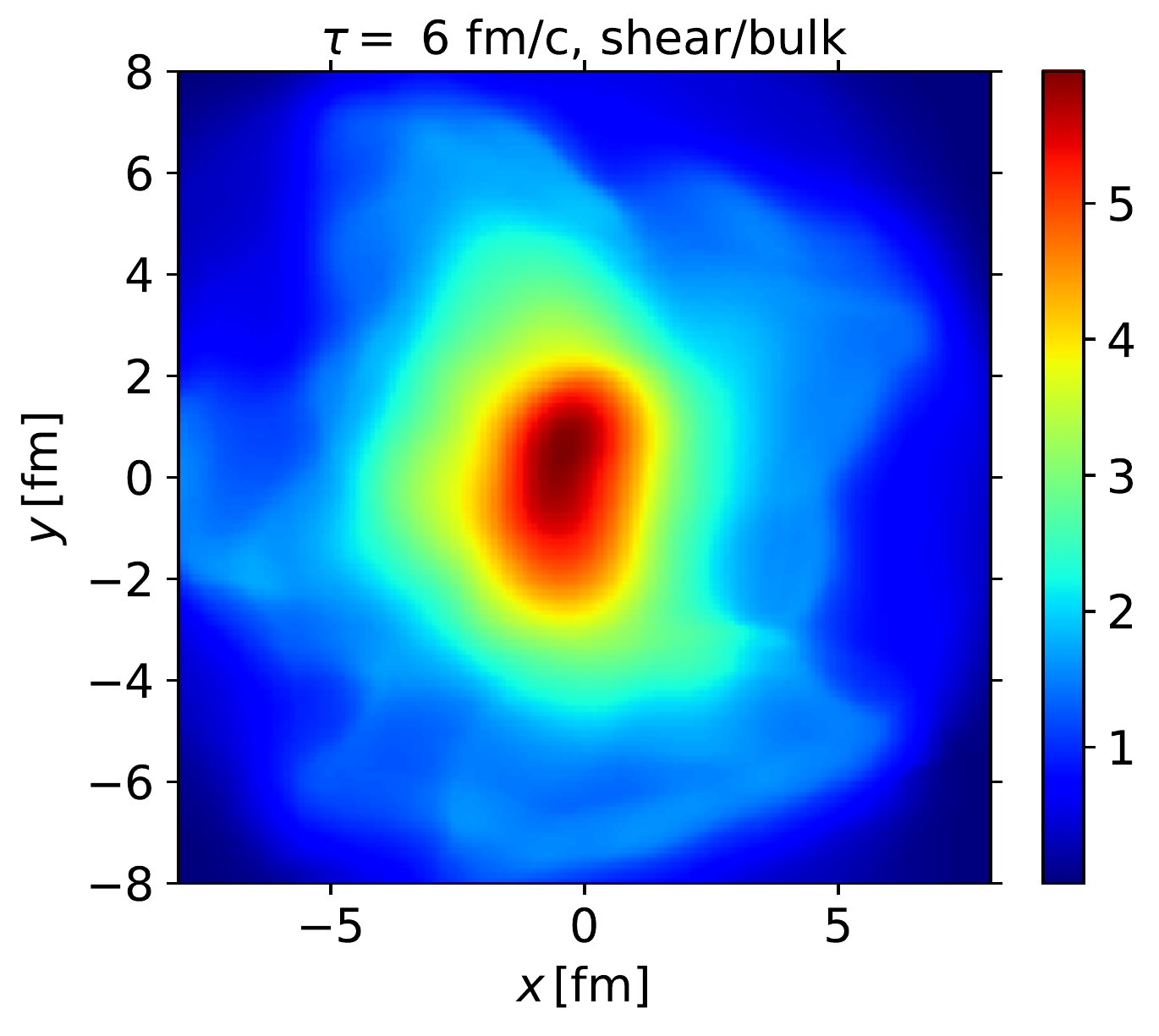} \\
  	\includegraphics[width=0.33\linewidth]{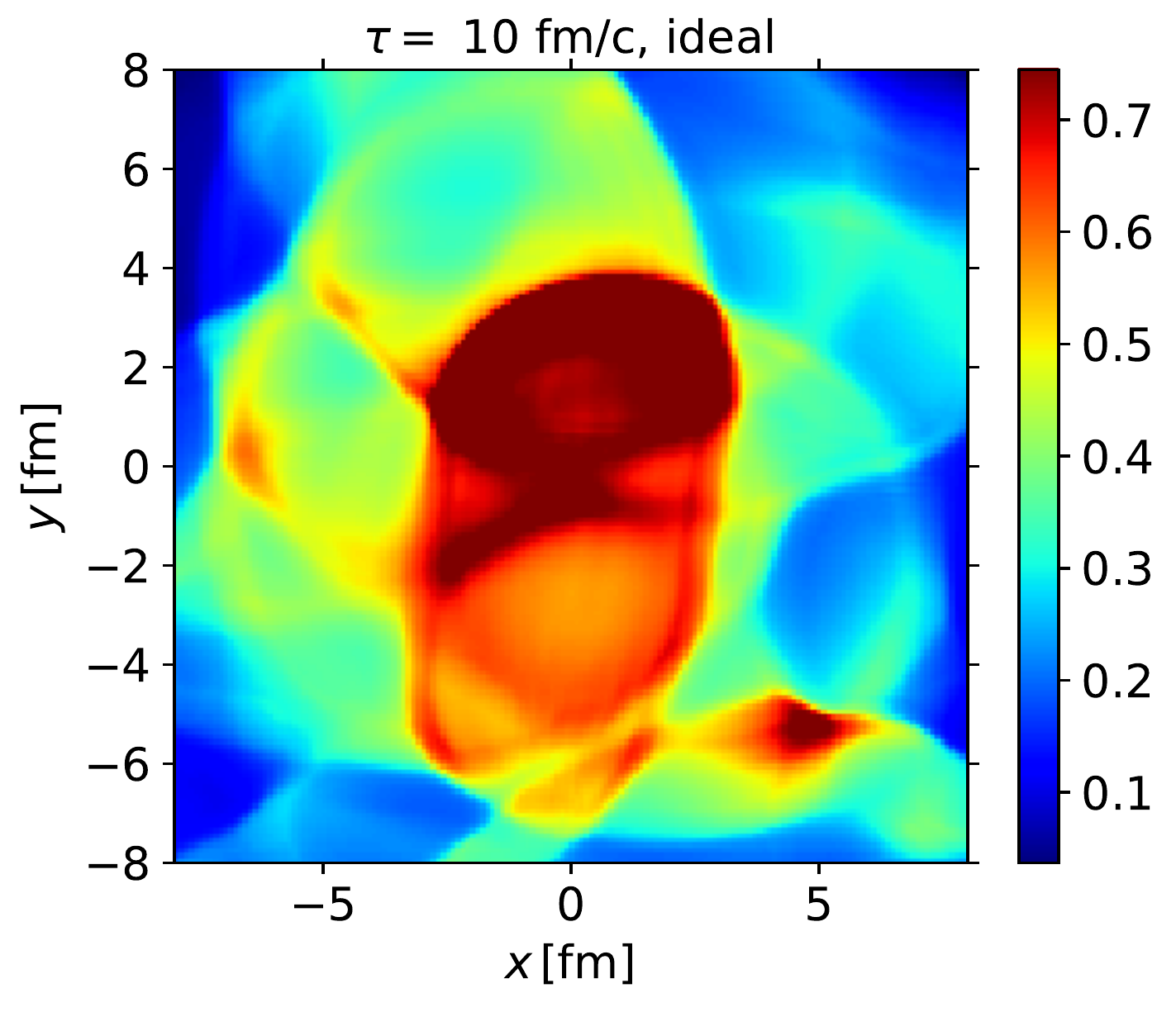} &
  	\includegraphics[width=0.33\linewidth]{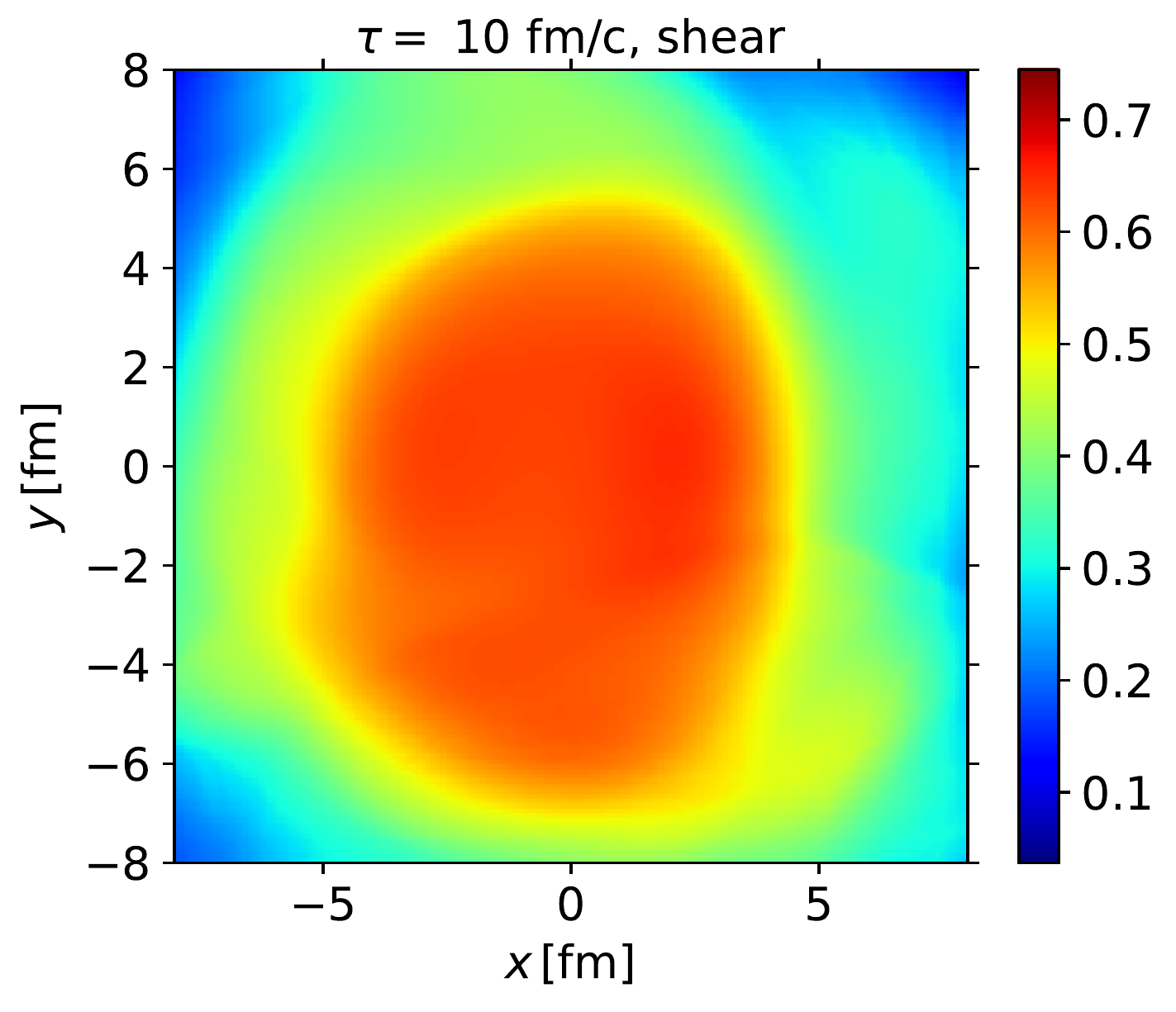} &
  	\includegraphics[width=0.33\linewidth]{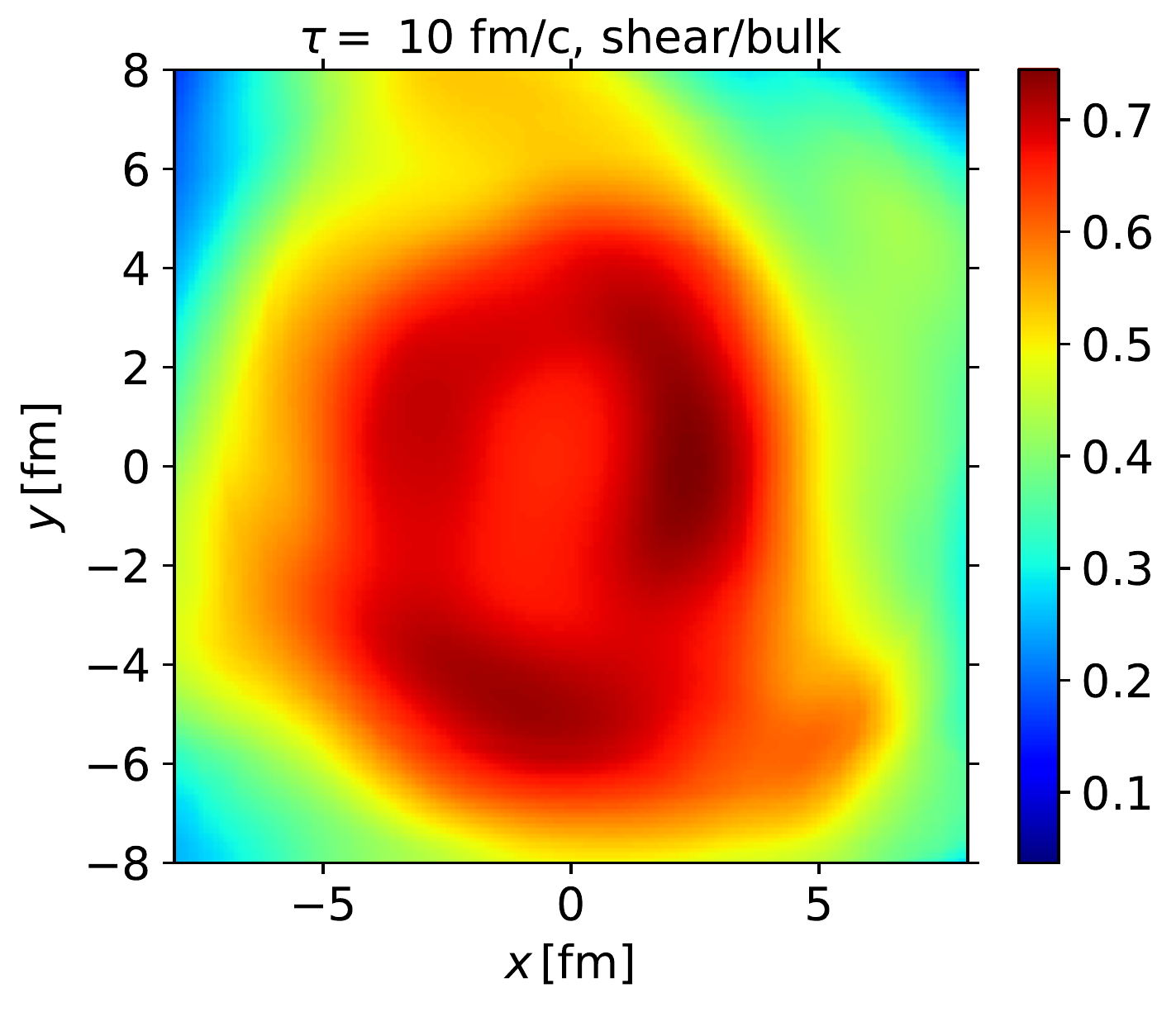}
  \end{tabular}
  }
  \caption{Visualization of the energy density (in units of $\mathrm{GeV/fm}^{3}$) at different proper times for ideal fluid dynamics (left panels), with a finite shear viscosity $\eta/s=0.2$ and zero bulk viscosity (middle panels), and  a finite bulk (see Eq.~(\ref{eq:zetas})) and shear viscosity (right panel). The system was initialized at a proper time $\tau_0=0.5$ fm$/c$ using a Monte-Carlo Glauber wounded nucleon profile for the energy density. $\Pi$ and $\pi^{\mu\nu}$ are initialized at their Navier-Stokes values. }
  \label{visCompFig}
\end{figure*}
\section{Visualization of a (3+1)-dimensionally expanding quark-gluon plasma}
\label{sec:vis}

\subsection{Initial conditions}
\label{sec:ic}
For the results presented below we use Monte-Carlo Glauber (MC-Glauber) initial conditions for the initial energy density profile in the transverse plane~\cite{Miller:2007ri}. This is factored with a  longitudinal initial profile following Ref.~\cite{Schenke:2010nt}:
\begin{equation}
\ed_\mathrm{L}(\eta_s)\equiv e^{-\frac{(|\eta_s|-\eta_\mathrm{flat})^2}{2\sigma^2_\eta}
\theta(|\eta_s|-\eta_\mathrm{flat})}\;.
\end{equation}
The full energy density profile is
\begin{equation}
\ed(\tau_0,x,y,\eta_s)=\ed_{0}\ed_\perp(x,y)\ed_\mathrm{L}(\eta_s)\;,
\end{equation}
where $\ed_\perp(x,y)$ is the MC-Glauber initial condition and $\ed_0\equiv\ed(T_0)$. We take parameter values quoted in Ref.~\cite{Schenke:2010nt}: $\eta_\mathrm{flat}\equiv 5.9$ and $\sigma^2_\eta\equiv 0.16$. The parameter values in Ref.~\cite{Schenke:2010nt} were tuned to experimental data using a continuous optical Glauber initial transverse energy density profile and ideal fluid dynamics.  For dissipative simulations, these parameters should be re-tuned to data; however, since we are not interested here in making connection to experiment, these values should suffice.

\subsection{Results}
\label{sec:results}

\begin{figure*}[t!]
  \centering
    \begin{tabular}{ccc}
  \includegraphics[width=0.45\linewidth]{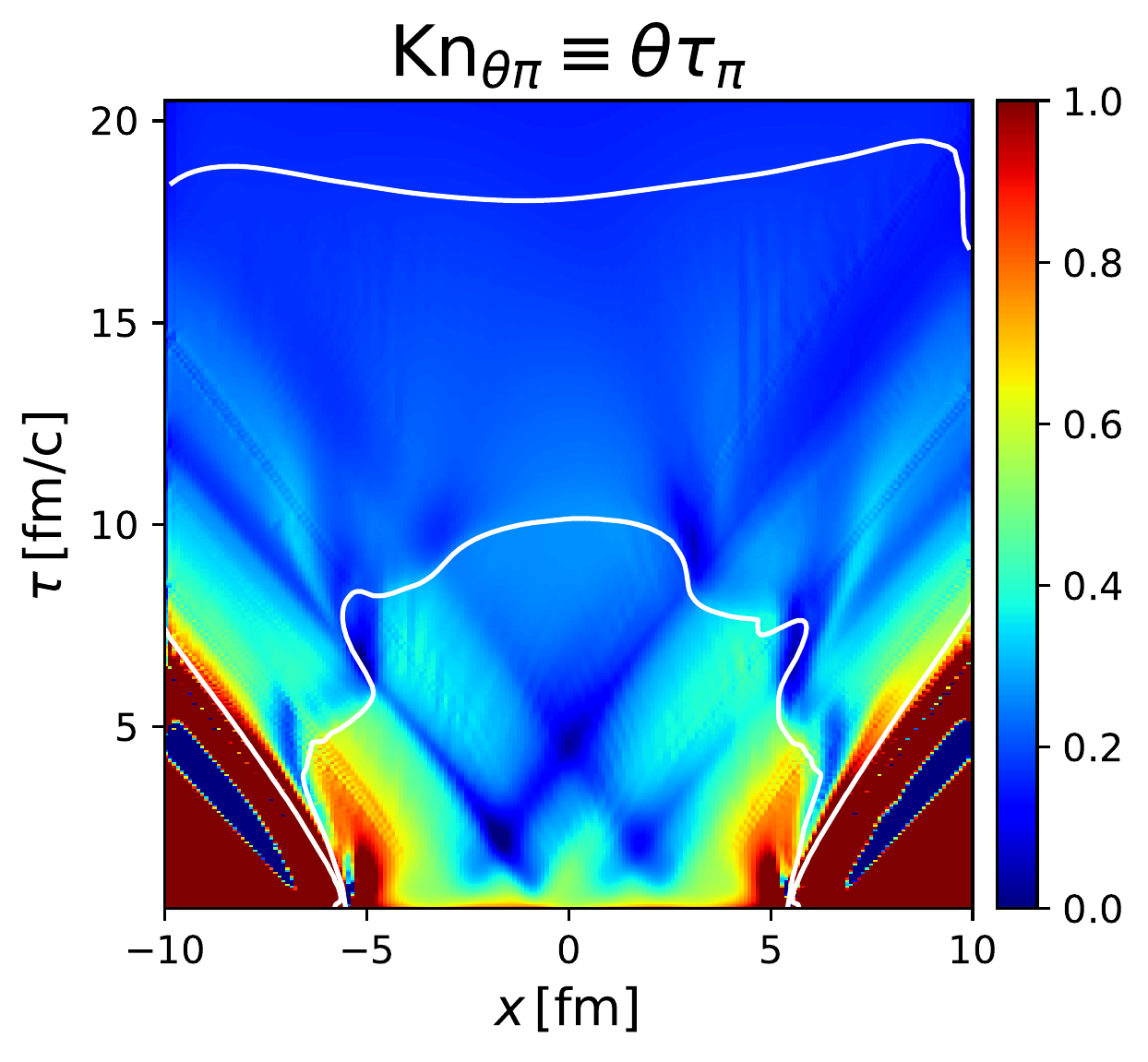} &
  \includegraphics[width=0.45\linewidth]{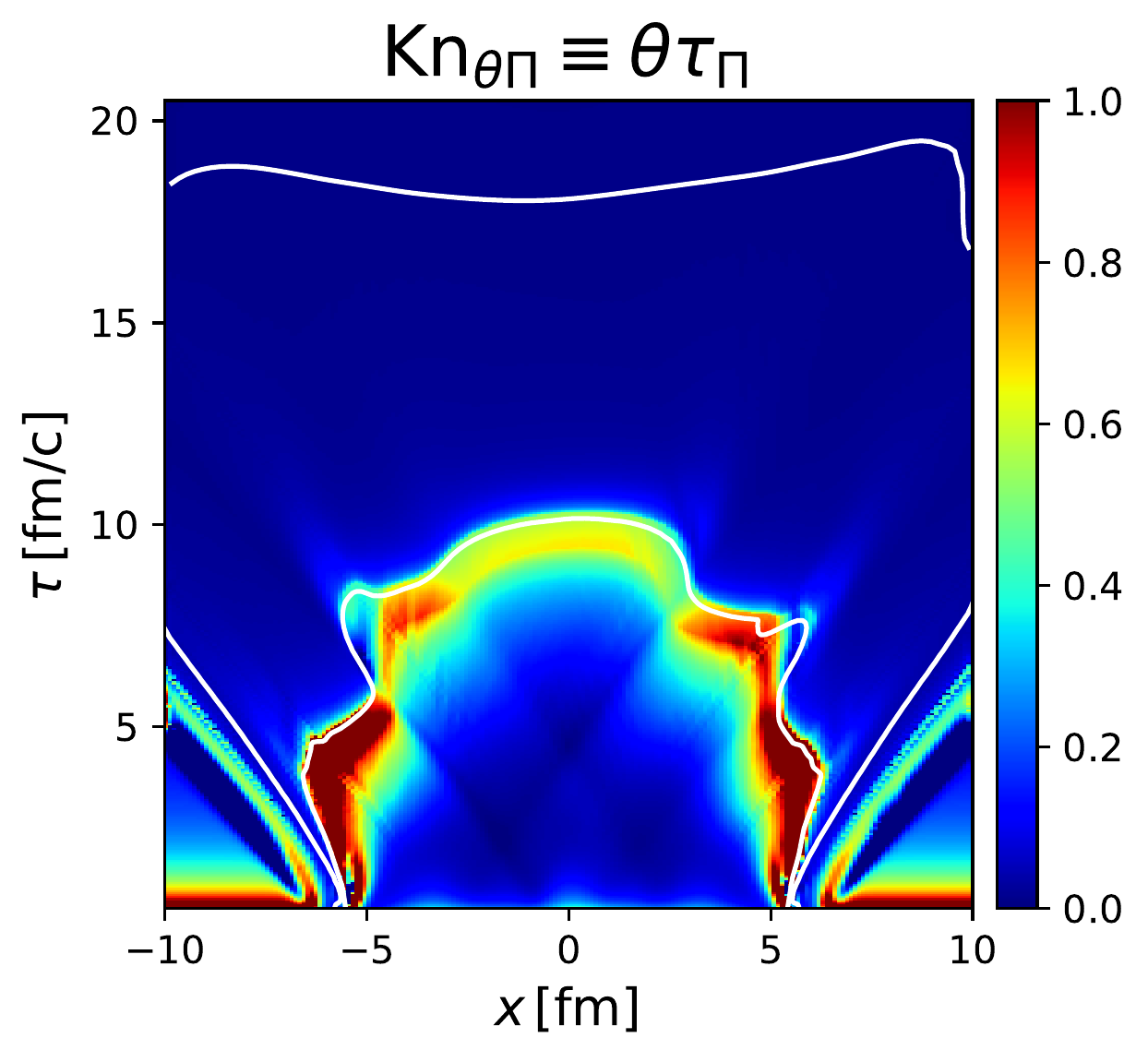} 
    \end{tabular}
  \vspace{-3mm}
  \caption{Proper time evolution along the $x$ axis of the Knudsen numbers $\mathrm{Kn}\equiv\tau_\pi\theta$ and $\mathrm{Kn}\equiv\tau_\Pi\theta$. The white isothermal contours represents $T_c=200$ MeV (inner contour) and the freezeout surface at $T_f=155$ MeV (outer contour). }
  \label{KnudsenNumberFig}
\end{figure*}

\begin{figure*}[t!]
  \centering
  \begin{tabular}{ccc}
  \includegraphics[width=0.45\linewidth]{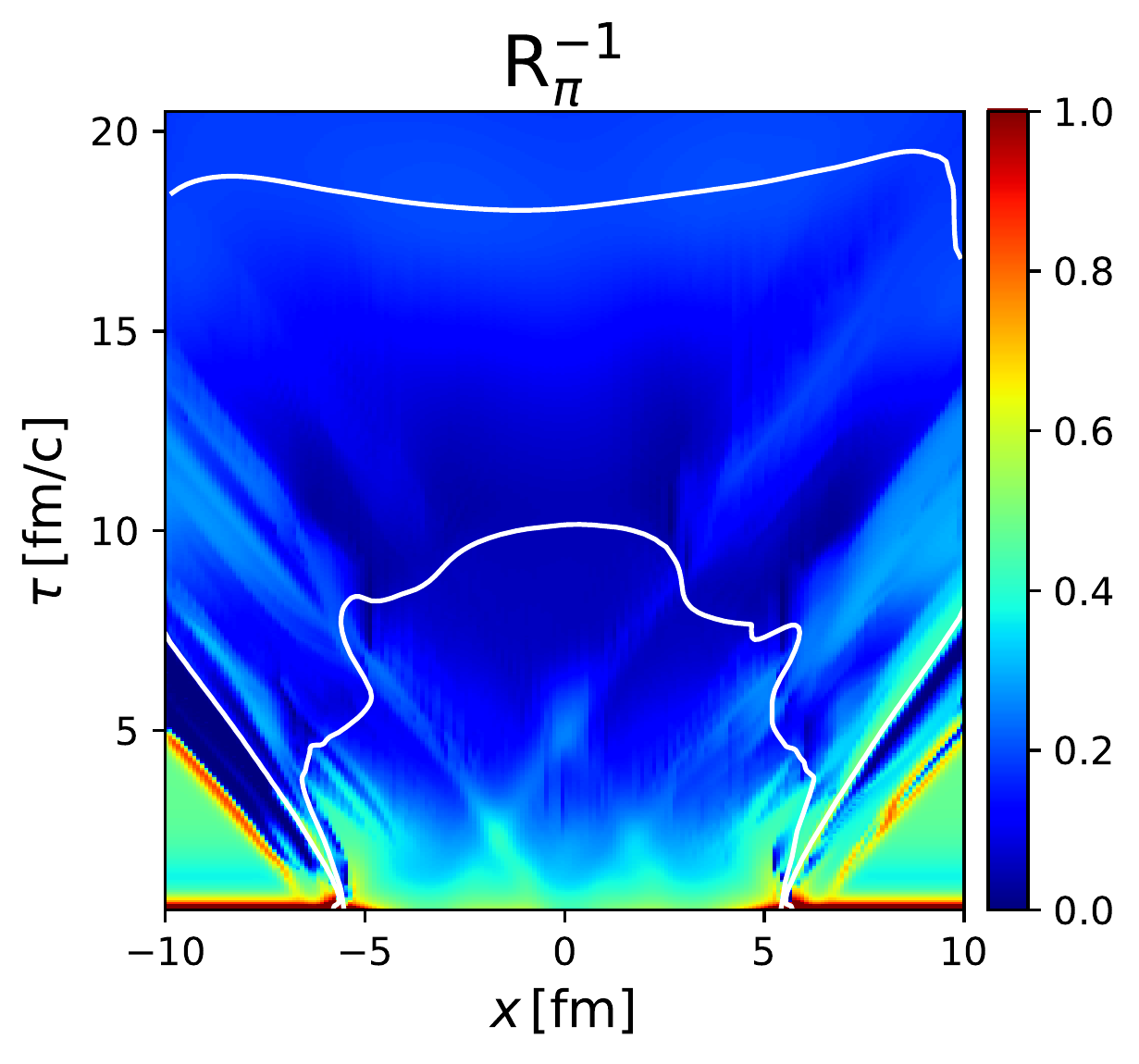} &
  \includegraphics[width=0.45\linewidth]{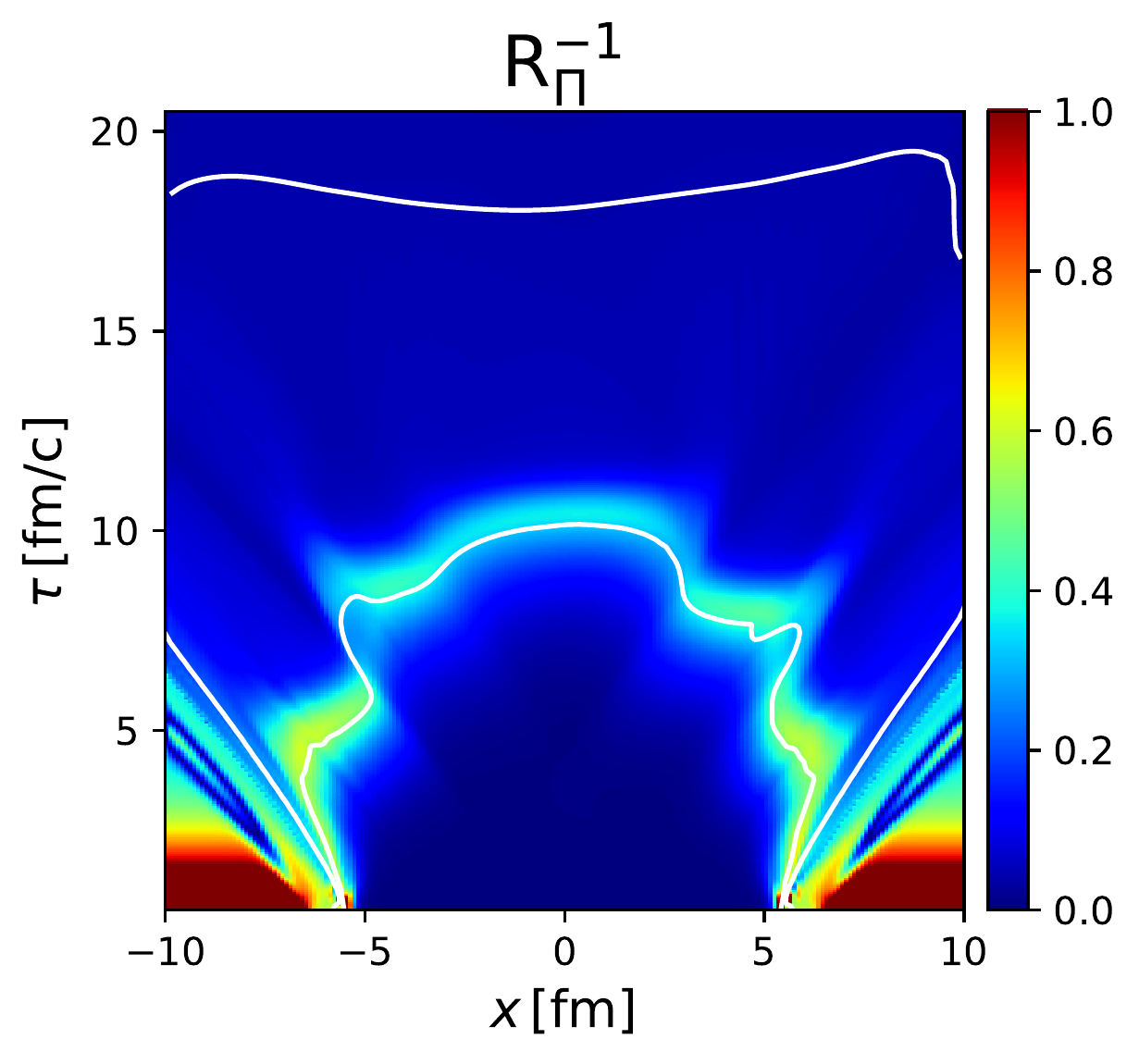} \\
    \includegraphics[width=0.45\linewidth]{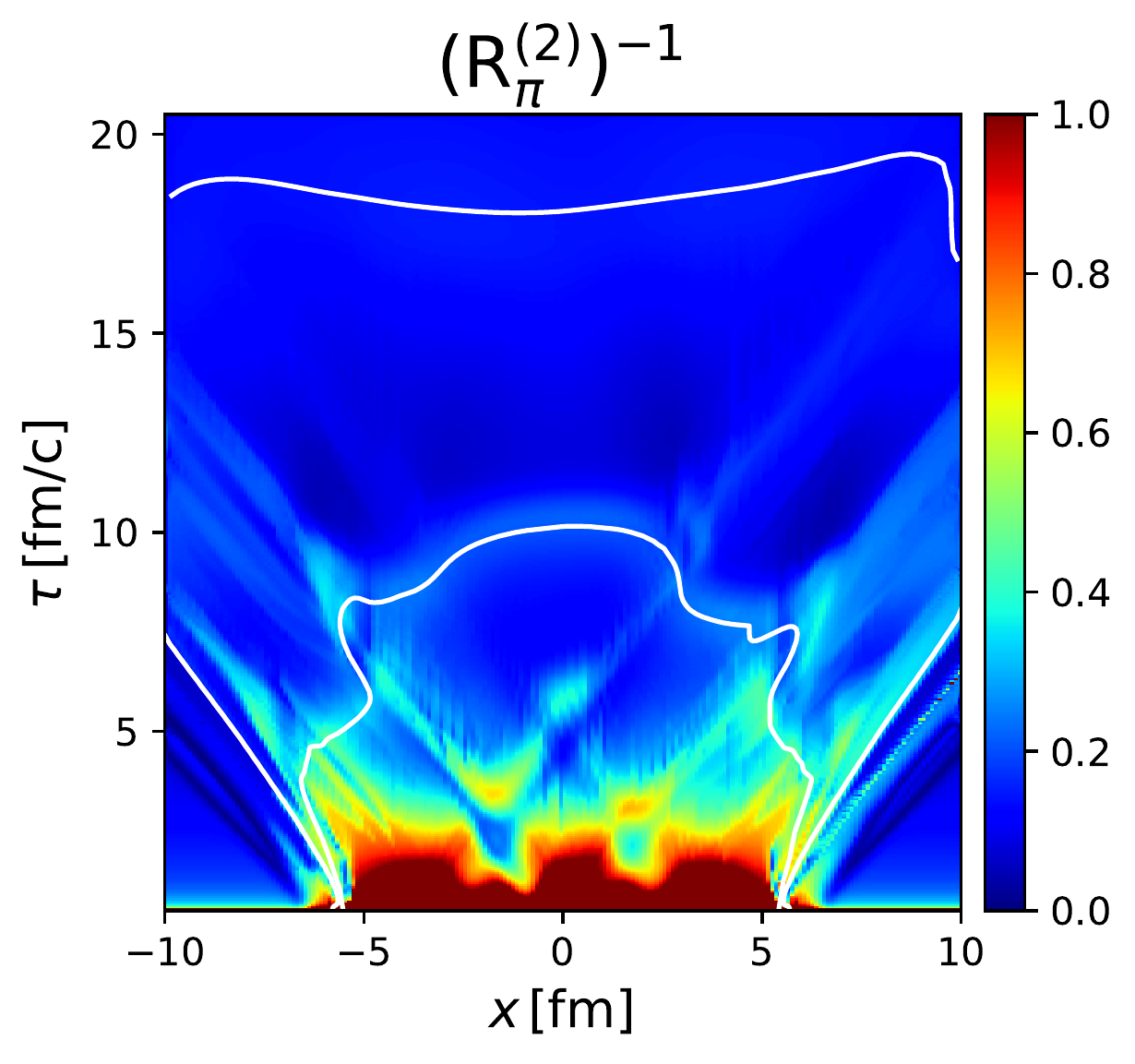} &
  \includegraphics[width=0.45\linewidth]{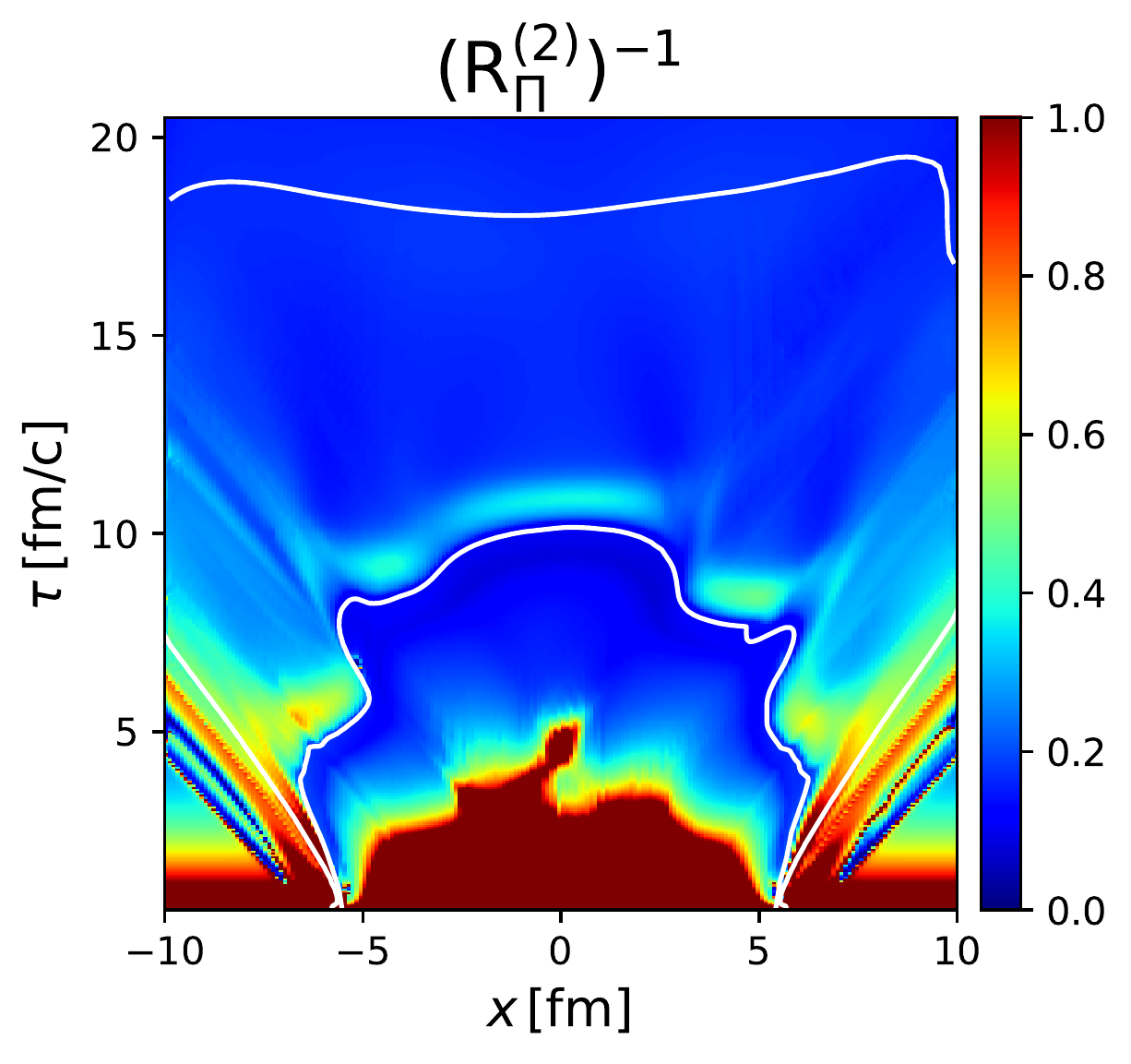}
  \end{tabular}
  \vspace{-3mm}
  \caption{Proper time evolution along the $x$ axis of Eqs.~(\ref{inverseReynoldsNumbers}) and (\ref{inverseReynoldsNumbers2}).}
  \label{inverseReynoldsNumberFig}
\end{figure*}

\begin{figure*}[t!]
  \centering
  \includegraphics[width=0.7\linewidth]{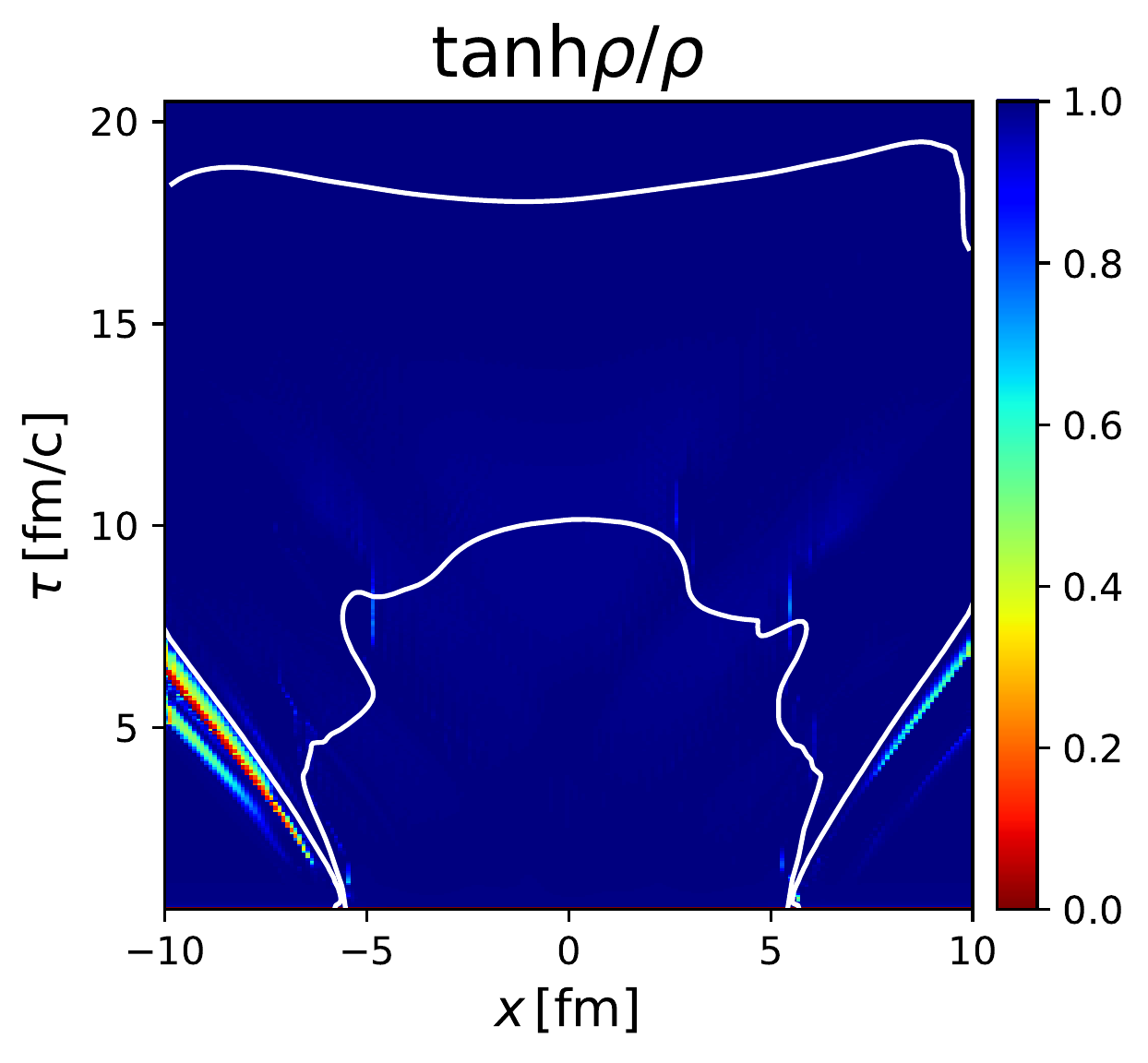}
  \vspace{-3mm}
  \caption{Proper time evolution of the quantity $\tanh\rho/\rho$ in the $x$-direction that controls the strength of the regularization of the $\pi^{\mu\nu}$. A value of $\ll 1$ strongly dampens the physical viscous effects while a value of $1$ corresponds to no regularization at all.}
  \label{regulationFig}
\end{figure*}

In Fig.~\ref{visCompFig} we show the evolution of the energy density in the transverse plane at $z=0$ at different points in proper time for ideal fluid dynamics (left column) and viscous fluid dynamics with only shear effects (middle column) and with shear and bulk effects (right column). We used the nonconformal QCD EoS defined in Sec.~\ref{sec:eos}, the parameterization of $\zeta/s$ in Eq.~(\ref{eq:zetas}), $\eta/s=0.2$, an initial temperature $T_0=0.6~\mathrm{GeV}$ at $\tau_0=0.5~\mathrm{fm/c}$, and $\Pi$ and $\pi^{\mu\nu}$ initialized to their respective Navier-Stokes values (where appropriate). Comparison of the left and middle column shows the smoothing effects that the shear viscosity has on the evolution of the energy density. The effects of the bulk viscosity are negligible until late times when the temperature approaches the critical temperature $T_c$ where $\zeta/s$ peaks.  Its effects are seen as a ring of high energy density in the last panel of the right column. 

\section{Validity of the effective theory}
\label{sec:validity}
As we already mentioned previously, the macroscopic effective description relies on the two types of small expansion parameters, namely the Knudsen number and inverse Reynolds number. The Knudsen number is defined as 
\begin{equation}
\mathrm{Kn}\equiv\frac{\ell_\mathrm{micro}}{L_\mathrm{macro}}\;,
\end{equation}
where the microscopic time scales in our problem are $\tau_\pi$ and $\tau_\Pi$, and the macroscopic time scales are gradients of the macroscopic variables (i.e. the thermodynamic variables that arise from the leading order interaction). In order to measure the strength of the deviation from leading order, the inverse Reynolds numbers are defined as
\begin{align}
  &\mathrm{R}^{-1}_\pi\equiv\frac{\sqrt{\pi^{\mu\nu}\pi_{\mu\nu}}}{\peq},
  \quad 
  \mathrm{R}^{-1}_\Pi\equiv\frac{|\Pi|}{\peq}.
  \label{inverseReynoldsNumbers}
  \end{align}
One of the assumptions we made to simplify Eq.~(\ref{relaxationEquations}) was to assume that $\mathrm{Kn}\sim\mathrm{R}^{-1}_{i}$ rather than treating them as independent dynamical quantities. Then the terms contained in ${\cal J}$ and ${\cal J}^{\mu\nu}$ in Eq.~(\ref{14_moment_terms}) are ${\cal O}(\mathrm{Kn}^2)$. In order for this to be a reasonable approximation they should be smaller than the ${\cal O}(\mathrm{Kn})$ Navier-Stokes terms in Eq.~(\ref{relaxationEquations}). To measure this, we define the second-order inverse Reynolds numbers
\begin{align}
  &(\mathrm{R}^{(2)}_\pi)^{-1}\equiv\frac{\sqrt{{\cal J}^{\mu\nu}{\cal J}_{\mu\nu}}}{2\eta\sqrt{\sigma^{\mu\nu}\sigma_{\mu\nu}}},
  \quad 
  (\mathrm{R}^{(2)}_\Pi)^{-1}\equiv\frac{|{\cal J}|}{\zeta|\theta|}.
    \label{inverseReynoldsNumbers2}
  \end{align}

In order to examine how small the above quantities are during a high-energy nuclear collision, we use the same initial conditions as Sec.~\ref{sec:results}. The proper time evolution of $\mathrm{Kn}_{\theta\pi}\equiv\tau_\pi\theta$ (left panel) and $\mathrm{Kn}_{\theta\Pi}\equiv\tau_\Pi\theta$ (right panel) are plotted in Fig.~\ref{KnudsenNumberFig} along the $x$-direction. Additionally, two white isothermal contours at $T_c=200$ MeV (inner contour) and the freezeout surface at $T_f=155$ MeV (outer contour) are plotted. The Knudsen numbers defined with other gradients of the macroscopic variables show the same overall behavior as $\mathrm{Kn}_{\theta\pi}\equiv\tau_{\pi}\theta$ so we do not show them here. When $\mathrm{Kn}\gtrsim 1$, the microscopic and macroscopic scales are of the same order and the viscous expansion will not converge. In these situations the fluid dynamic approximation in the form discussed here breaks down and is not the correct effective description.%
\footnote{It may still be possible to absorb some of the large terms into the leading-order dynamics and thus improve the validity of the hydrodynamic approach by extending the framework to viscous anisotropic hydrodynamics~\cite{Bazow:2013ifa,Molnar:2016vvu}. This will be explored in future work.}
In Fig.~\ref{KnudsenNumberFig} this happens only in the dilute regions of the plasma $|x|\gtrsim 5\,\mathrm{fm}$, or on the $T_c$ hypersurface for the Knudsen number $\mathrm{Kn}_{\theta\Pi}$ caused by the large value of $\tau_\Pi$. Fig.~\ref{inverseReynoldsNumberFig} shows the proper time evolution of $\mathrm{R}^{-1}_\pi$ and $\mathrm{R}^{-1}_\Pi$ (top panels)  and $(\mathrm{R}^{(2)}_\pi)^{-1}$ and $(\mathrm{R}^{(2)}_\Pi)^{-1}$ (bottom panels) along the $x$-direction. There are large regions in the $\tau{-}x$ space, particularly near the center of the plasma at $|x|\lesssim 5\,\mathrm{fm}$, where the second order inverse Reynolds numbers are not small. This casts doubts on the application of the 14-moment approximation that we used to simplify the fluid dynamic equations of motion. Figure \ref{regulationFig} shows the grid points in the transverse plane where $\pi^{\mu\nu}$ has been regulated according to Eq.~\ref{regPimunu}.  A value of $\ll 1$ strongly dampens the physical viscous effects while a value of 1 corresponds to no regularization at all. Here, only areas outside of the freeze-out surface are strongly regulated.

\section{Performance benchmarks}
\label{sec:performance}

\begin{figure*}[t!]
  \centering
  \vspace{3mm}
  \includegraphics[width=0.75\linewidth]{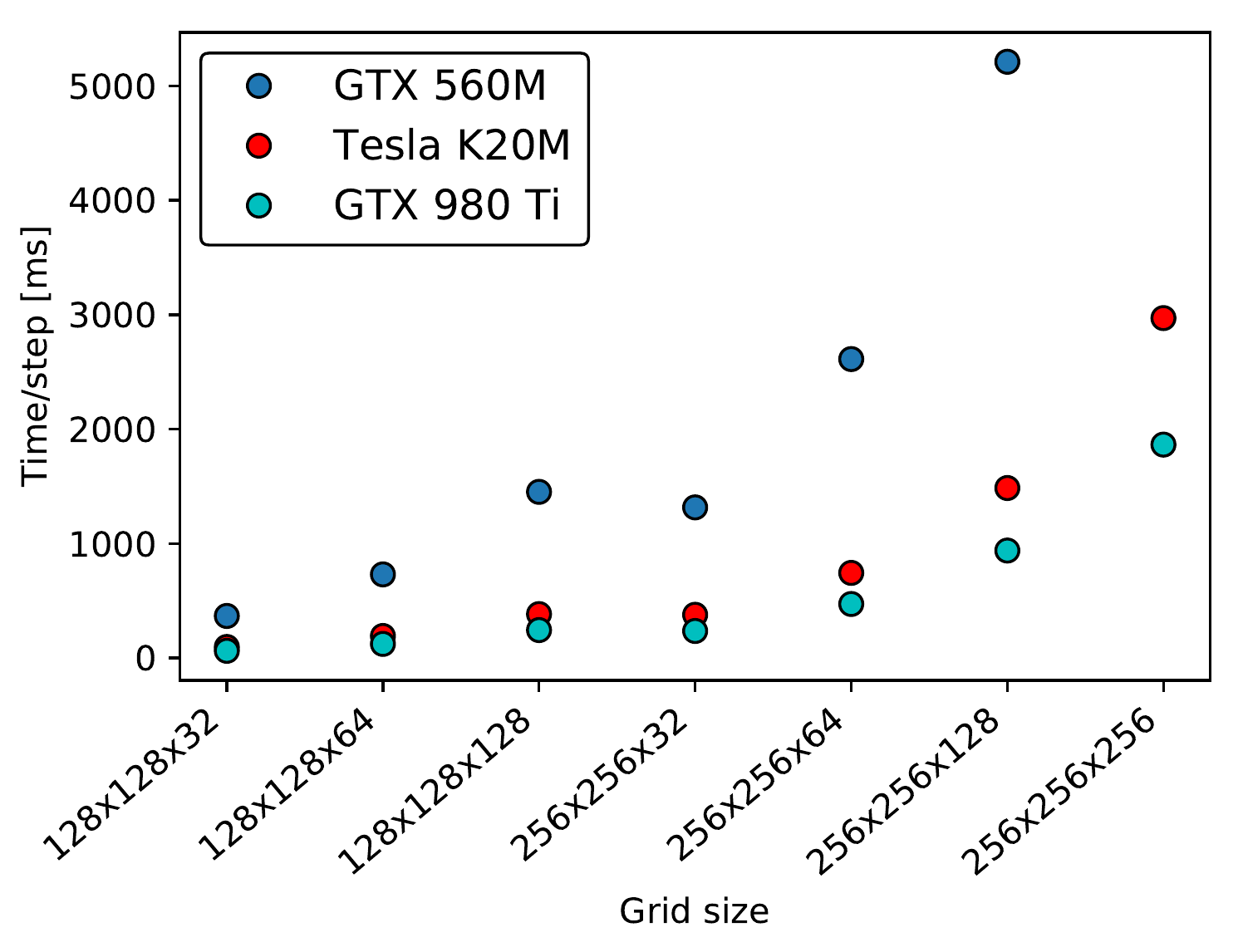} 
  \vspace{-3mm}
  \caption{Performance comparison of GPU-VH run on three different graphics cards [GTX 560M (blue circles), Tesla K20M (red circles), and GTX 980 Ti (cyan circles)] by measuring the computer time
it takes to complete one full RK step (described in Fig.~\ref{fig:twoStepRungeKutta}) averaged over 100 time steps at different spatial resolutions. In the current version of our code, we were unable to get the data point corresponding to the 256x256x256 grid size because it requires too much global memory for the GTX 560M graphics card.}
  \label{graphicsCardCompFig}
\end{figure*}

We now compare the speed up that can be seen from a single GPU over a single core CPU for (3+1)-dimensional fluid dynamics. For the CPU we use a separate implementation of GPU-VH (written in CUDA C) in order to get a highly optimized serial\footnote{A serial implementation runs the code on a single CPU core.} code using the C programming language. For the GPU implementation we ran GPU-VH on three different cards GTX 560M, Tesla K20M, and GTX 980 Ti whose technical specifications can be found in Table~\ref{cardInfo}. The simulation is the same as in Sec.~\ref{sec:vis}, for the case of non-zero shear viscosity and a non-vanishing bulk viscous pressure. To measure the performance we use the amount of computer time
it takes to complete one full Runge-Kutta time step (described in Fig.~\ref{fig:twoStepRungeKutta}), averaged over 100 time steps, at different spatial resolutions.
Table~\ref{980ti_compare_table} shows the speedup of GPU-VH run on the graphics card GeForce GTX 980 Ti relative to the CPU-VH code run on the host machine with a 2.6 GHz Intel Xeon CPU E5-2697 v3. Table~\ref{k20M_compare_table} compares GPU-VH against CPU-VH run on the graphics card Nvidia Tesla K20M and a 1.8 GHz Intel Xeon CPU E5-2630L v3, respectively. In both cases we see ${\cal O}(100)$ times speed up of GPU-VH over CPU-VH. In Fig.~\ref{graphicsCardCompFig} we plot the time to complete one full time integration step for three different graphics cards GTX 560M, Tesla K20M, and GTX 980 Ti. 

\begin{table*}[t!]
\small
\centering
  \scalebox{0.82}{
  \begin{tabular}{|c|c||c|c|c||c|c|c|}
    \hline
     \multicolumn{2}{|c||}{} 
     & \multicolumn{2}{|c||}{Clock speeds (MHz)} 
     &  \multicolumn{2}{|c|}{Memory Configuration} 
     & \multicolumn{2}{|c|}{Processing power (GFLOPS)}
     \\
    \hline
      Model 
      & \multicolumn{1}{|p{1.5cm}|}{\centering Processor \\ Cores }
      & Core 
      & Memory 
      & \multicolumn{1}{|p{1.25cm}|}{\centering Size \\ (GB)} 
      & \multicolumn{1}{|p{1.25cm}|}{\centering Bandwidth \\ (GB/s)}  
      & \multicolumn{1}{|p{1.25cm}|}{\centering Single \\ precision} 
      & \multicolumn{1}{|p{1.25cm}|}{\centering Double \\ precision}
      \\ 
    \hline
        \multicolumn{1}{|p{1.5cm}|}{\centering GeForce \\ GTX 560M} & 192 
        & 775  & 2500 
        & 3.076 & 60  
        & 595.2 & N/A
        \\
    \hline
        \multicolumn{1}{|p{1.5cm}|}{\centering GeForce \\ GTX 980 Ti} & 2816
        & 1000 & 7012  
        & 6.144 & 336  
        & 5632 & 176
        \\
    \hline
        \multicolumn{1}{|p{1.5cm}|}{\centering Tesla \\ K20M} & 2496    
        & 706  & 2600 
        & 5.120 & 208 
        & 3524 & 1175
         \\  
    \hline
\end{tabular}
}
\caption{General information on the three graphics cards used herein.  The information reported is based on the official Nvidia specifications.}\label{cardInfo}
\end{table*}
%
\begin{table}[t!]
\small
\centering
\begin{tabular}{|c|c|c|c|}
  \hline
  Number of grid points & C/CPU & CUDA/GPU & Speedup \\
  & (ms/step) & (ms/step) & \\
  \hline
  128 $\times$ 128 $\times$ 32 	& 7145.978  & 63.261 & 112.960 \\
  128 $\times$ 128 $\times$ 64 	& 13937.896 & 123.527 & 112.833 \\
  128 $\times$ 128 $\times$ 128 	& 30717.367 & 244.450 & 125.659 \\
  256 $\times$ 256 $\times$ 32 	& 25934.547 & 236.593 & 109.617 \\
  256 $\times$ 256 $\times$ 64 	& 57387.141 & 472.391 & 121.482  \\
  256 $\times$ 256 $\times$ 128 	& 129239.959 & 939.340 & 137.586 \\   
  256 $\times$ 256 $\times$ 256 	& 268448.459 & 1865.142 & 143.929 \\
  \hline
\end{tabular}
\caption{Performance results of the C/CPU and CUDA/GPU versions of our (3+1)-dimensional fluid 
dynamic codes CPU-VH and GPU-VH by measuring the computer time
it takes to complete one full RK step (described in Fig.~\ref{fig:twoStepRungeKutta}), averaged over 100 time steps, at different spatial resolutions. The GPU-VH code is run on the graphics card GeForce GTX 980 Ti and the CPU-VH code is run on the host machine with a 2.6 GHz Intel Xeon CPU E5-2697 v3.}\label{980ti_compare_table}
\end{table}
%
\begin{table}[t!]
\small
\centering
\begin{tabular}{|c|c|c|c|}
  \hline
  Number of grid points & C/CPU & CUDA/GPU & Speedup \\
  & (ms/step) & (ms/step) & \\
  \hline
  128 $\times$ 128 $\times$ 32 	& 7690.069  & 96.923 & 79.342 \\
  128 $\times$ 128 $\times$ 64 	& 16315.976 & 192.751 & 84.648 \\
  128 $\times$ 128 $\times$ 128 	& 38428.056 & 384.255 & 100.007 \\
  256 $\times$ 256 $\times$ 32 	& 30401.898 & 378.178 & 80.390 \\
  256 $\times$ 256 $\times$ 64 	& 72240.973 & 744.168 & 97.076  \\
  256 $\times$ 256 $\times$ 128 	& 144744.290 & 1485.703 & 97.423 \\   
  256 $\times$ 256 $\times$ 256 	& 322536.875 & 2970.727 & 108.572 \\
  \hline
\end{tabular}
\caption{Same as Table~\ref{980ti_compare_table}, but for the graphics card Nvidia Tesla K20M compared with a 1.8 GHz Intel Xeon CPU E5-2630L v3.}\label{k20M_compare_table}
\end{table}

It is difficult to say anything about how the scaling of the performance depends on the number of cores because memory bandwidth and amount of FLOPS also affect how fast a calculation can run. We can conclude, however, that for our purposes high end gaming cards would benefit us more (i.e.~the GTX 980 Ti) than cards geared towards high-accuracy mathematical calculations where double precision is needed (i.e.~the Tesla K20M). The difference in single/double precision FLOPS can be seen in Table~\ref{cardInfo}. It should be mentioned that all results shown in this paper were performed using single precision manipulations.

\section{Conclusions and Outlook}
\label{sec:conc}

In this paper we presented in detail how to implement relativistic fluid dynamics on GPUs using CUDA, and we validated the code GPU-VH against various test cases. In addition, although not new, we documented the regions of validity where we can expect the fluid dynamic approximation used in high-energy nuclear collisions to hold. For such simulations, we demonstrated a ${\cal O}(100)$ times speedup for GPU-VH against a highly optimized serial CPU implementation of our code. For a $256^3$ grid size (which is much larger than one would typically take) our code can simulate a single high-energy nuclear collision with 1000 time steps in $~30\,\mathrm{min}$. This means that we could simulate $1000$ random nuclear collisions on a small Beowulf cluster of 4 GPUs in approximately four days. Of course, there are other bottlenecks in the full dynamical evolution of a nuclear collision (of which fluid dynamics is only one part) that would benefit from being ported to GPUs. In order to precisely know all of the many parameters that enter these simulations (which exhibit complicated coupling to one another), it is necessary to use Gaussian Process emulators based on Bayesian statistics~\cite{Novak:2013bqa}. Full 3+1d simulations are computationally expensive, and it is infeasible at the moment to perform statistical parameter extraction without the aid of GPUs to accelerate other components in the chain as well including, for example, the late-stage hadronic evolution. 

\appendix

\section{(3+1)-dimensional equations in $(\tau,x,y,\eta_{s})$-coordinates}
\label{sec:milne_eqns}
Although the necessary tensor components in Cartesian and Milne coordinates needed herein for the r.h.s.~of the relaxation equations (\ref{relEqs}) are nicely listed in Refs.~\cite{Molnar:2009tx,Molnar:2014zha}, we choose to list them in this appendix for completeness.
The covariant derivative of the fluid velocity can be written in its irreducible form as,
\begin{equation}
u_{\nu;\mu}=u_{\mu}Du_{\nu}+\frac{1}{3}\Delta_{\mu\nu}\theta+\sigma_{\mu\nu}
+\omega_{\mu\nu}\;.
\end{equation}
For a covariant vector field $u_{\mu}$, we have that $Du_{\mu}\equiv du_{\mu}-\Gamma^{\beta}_{\mu\alpha}u^{\alpha}u^{\beta}$.
The expansion rate $\theta$, the shear stress tensor $\sigma^{\mu\nu}$, and vorticity tensor $\omega^{\mu\nu}$ are generally defined in any coordinate system as
\begin{align}
\theta & \equiv \nabla _{\mu }u^{\mu }=\partial _{\mu }u^{\mu }+\Gamma
_{\lambda \mu }^{\lambda }u^{\mu } \ ,
\end{align}
\begin{align}
\sigma^{\mu\nu} & \equiv \nabla^{\langle\mu}u^{\nu\rangle}=\frac{1}{2}\Delta^{\mu
\alpha }\Delta ^{\nu \beta }(u_{\alpha ;\beta }+u_{\beta ;\alpha })-\frac{%
\theta }{3}\Delta ^{\mu \nu }  \notag \\
& =\frac{1}{2}\left( 
\partial^{\mu}u^{\nu}-u^{\mu}du^{\nu}+\partial^{\nu}u^{\mu}-u^{\nu}du^{\mu}
 \right)
 \notag \\ 
& \hspace{5mm} +\frac{1}{2}\left(
\Delta^{\mu\alpha}u^{\beta}\Gamma^{\nu}_{\alpha\beta}
+\Delta^{\nu\alpha}u^{\beta}\Gamma^{\mu}_{\alpha\beta}
\right)
-\frac{\theta }{3}\Delta ^{\mu \nu }
,
\end{align}%
\begin{align}
\omega _{\hspace*{0.1cm}\nu }^{\mu }& \equiv \frac{1}{2}\Delta ^{\mu \alpha
}\Delta _{\hspace*{0.1cm}\nu }^{\beta }\left( u_{\alpha ;\beta }-u_{\beta
;\alpha }\right)  \notag \\
& =\frac{1}{2}\left[ g^{\mu \alpha }\left( \partial _{\nu }u_{\alpha
}-u_{\nu }du_{\alpha }\right) -g_{\nu }^{\beta }\left( \partial ^{\mu
}u_{\beta }-u^{\mu }du_{\beta }\right) \right] \,  \notag \\
& \hspace{5mm} +\frac{1}{2}\left( g^{\mu \alpha }u_{\nu }-g_{\hspace*{0.1cm}\nu }^{\alpha
}u^{\mu }\right) u^{\beta }\Gamma _{\alpha \beta }^{\lambda }u_{\lambda }.
\label{vorticity}
\end{align}
In Milne coordinates, the necessary convective derivative of the fluid velocity components are:
\begin{align}
du^{\tau}&\equiv du_{\tau}=u_{\tau}\partial_{\tau}u_{\tau}+u_{x}\partial_{x}u_{\tau}
+u_{y}\partial_{y}u_{\tau}+u_{\eta}\partial_{\eta}u_{\tau}\;,
\\
du^{x}&\equiv-du_{x}=u_{\tau}\partial_{\tau}u_{x}+u_{x}\partial_{x}u_{x}
+u_{y}\partial_{y}u_{x}+u_{\eta}\partial_{\eta}u_{x}\;, 
\\
du^{y}&\equiv-du_{y}=u_{\tau}\partial_{\tau}u_{y}+u_{x}\partial_{x}u_{y}
+u_{y}\partial_{y}u_{y}+u_{\eta}\partial_{\eta}u_{y}\;,
\\
du^{\eta}&\equiv-\tau^{-2}du_{\eta}-\tau^{-1}u_{\tau}u_{\eta}=
u_{\tau}\partial_{\tau}u_{\eta}+u_{x}\partial_{x}u_{\eta}
+u_{y}\partial_{y}u_{\eta}+u_{\eta}\partial_{\eta}u_{\eta}\;.
\end{align}
For the Milne coordinate system, the $x$ and $y$ components are not affected by the transformation. The proper time and convective derivatives of $u_{\mu}$ are then formally the same: $Du_{x}=du_{x}$ and $Du_{y}=du_{y}$. The $\tau$ and $\eta$ components are 
\begin{align}
Du_{\tau}&= du_{\tau}+\tau u^{2}_{\eta}\;,
\\
Du_{\eta}&=du_{\eta}
=-\tau^{2}du^{\eta}-2\tau u_{\tau}u_{\eta}\;.
\end{align}
The expansion rate is
\begin{equation}
\theta=u_{\tau}/\tau+\partial_{\tau}u_{\tau}
+\partial_{x}u_{x}+\partial_{y}u_{y}+\partial_{\eta}u_{\eta}\;.
\end{equation}
The various shear tensor components are
{\small
\bea \nonumber 
\sigma^{\tau \tau}
&=& - \tau u_{\tau} u^2_\eta + \partial_\tau u_\tau
- u_\tau Du_\tau
+ \left(u_\tau^2 - 1\right)\frac{\theta}{3} \, ,\\ \nonumber
\sigma^{\tau x}
&=& - \frac{\tau u^2_\eta u_x}{2}
+ \frac{1}{2} \left[ \partial_\tau (u_x)
- \partial_x \gamma \right] 
- \frac{1}{2} \left[u_\tau D (u_x)
+ u_x Du_\tau \right]
+ u_\tau u_x \frac{\theta}{3} \, ,\\ \nonumber
\sigma^{\tau y}
&=& - \frac{\tau u^2_\eta u_y}{2}
+ \frac{1}{2}\left[\partial_\tau (u_y)
- \partial_y u_\tau \right] 
- \frac{1}{2}\left[u_\tau D (u_y)
+ u_y Du_\tau \right]
+ u_\tau u_y \frac{\theta}{3} \, ,\\ \nonumber
\sigma^{\tau \eta}
&=& - \frac{u_\eta}{2\tau} \left(2u_\tau^2
+ \tau^2 u^2_\eta \right)
+ \frac{1}{2}\left[\partial_\tau (u_\eta)
- \frac{1}{\tau^2} \partial_\eta u_\tau \right] 
- \frac{1}{2}\left[\gamma D (u_\eta)
+ u_\eta Du_\tau \right]
+ u_\tau u_\eta \frac{\theta}{3} \, ,\\ \nonumber
\sigma^{\eta \eta}
&=& -\frac{u_\tau}{\tau^3} \left(1
+ 2 u^2_\eta \tau^2 \right)
- \frac{1}{\tau^2}\partial_\eta (u_\eta) 
- u_\eta D (u_\eta) + (\frac{1}{\tau^2}
+ u^2_\eta) \frac{\theta}{3} \, ,\\ \nonumber
\sigma^{x\eta}
&=& -\frac{u_\tau u_x u_\eta}{\tau}
- \frac{1}{2} \left[ \partial_x (u_\eta)
+ \frac{1}{\tau^2}\partial_\eta (u_x) \right] 
- \frac{1}{2} \left[ u_x D (u_\eta)
+ u_\eta D (u_x) \right]
+ u_x u_\eta \frac{\theta}{3} \, , \\ \nonumber
\sigma^{y\eta}
&=& -\frac{u_\tau u_y u_\eta}{\tau}
- \frac{1}{2} \left[ \partial_y (u_\eta)
+ \frac{1}{\tau^2}\partial_\eta (u_y) \right]
- \frac{1}{2} \left[ u_y D (u_\eta)
+ u_\eta D (u_y) \right]
+ u_y u_\eta \frac{\theta}{3} \, , \\ \nonumber
\sigma^{xx}
&=& - \partial_x (u_x)
- u_x D (u_x)
+ (1 + u^2_x) \frac{\theta}{3} \, ,  \\
\sigma^{yy}
&=& - \partial_y (u_y)
- u_y D (u_y) 
+ (1 + u^2_y) \frac{\theta}{3} \, ,  \\
\sigma^{xy}
&=& -\frac{1}{2} \left[ \partial_x (u_y)
+ \partial_y (u_x) \right]  
- \frac{1}{2} \left[ u_x D (u_y)
+ u_y D (u_x) \right] 
 +
u_x u_y \frac{\theta}{3}
\, .
\eea
}
The non-vanishing components of the vorticity are given as
\begin{align}
\omega _{\left. {}\right. x}^{\tau }& \equiv \omega _{\left. {}\right. \tau
}^{x}=\frac{1}{2}\left[ \partial _{\tau } u_{x}
+\partial _{x}u_{\tau} \right]   
 +\frac{1}{2}\left[ u_{x}du_{\tau} -u_{\tau} d u_{x} %
\right] +\frac{1}{2}\tau u_{\eta }^{2}u_{x}\ , 
\\
\omega _{\left. {}\right. y}^{\tau }& \equiv \omega _{\left. {}\right. \tau
}^{y}=\frac{1}{2}\left[ \partial _{\tau }u_{y}
+\partial _{y}u_\tau \right]  
 +\frac{1}{2}\left[ u_{y}du_{\tau} -u_{\tau} du_{y} %
\right] +\frac{1}{2}\tau u_{\eta }^{2}u_{y}\ , 
\\
\omega _{\left. {}\right. \eta }^{\tau }& \equiv \tau ^{2}\omega _{\left.
{}\right. \tau }^{\eta }=\frac{1}{2}\left[ \partial _{\tau }\left( \tau
^{2}u_{\eta }\right) +\partial _{\eta }u_{\tau} \right]   
 +\frac{1}{2}\left[ \tau ^{2}u_{\eta }du_{\tau} -u_{\tau} d\left( \tau
^{2}u_{\eta }\right) \right] \ +\frac{1}{2}\tau ^{3}u_{\eta }^{3}\ ,
\\
\omega _{\left. {}\right. y}^{x}& \equiv -\omega _{\left. {}\right. x}^{y}=%
\frac{1}{2}\left[ \partial _{y}u_{x} -\partial
_{x}u_{y} \right]   
+\frac{1}{2}\left[ u_{y}du_{x} -
u_{x}du_{y} \right] \ , 
\\
\omega _{\left. {}\right. \eta }^{x}& \equiv -\tau ^{2}\omega _{\left.
{}\right. x}^{\eta }=\frac{1}{2}\left[ \partial _{\eta } u_{x} -\partial _{x}\left( \tau ^{2}u_{\eta }\right) \right]  
 +\frac{1}{2}\left[ \tau ^{2}u_{\eta }du_{x}
-u_{x}d\left( \tau ^{2}u_{\eta }\right) \right] \ , 
\\
\omega _{\left. {}\right. \eta }^{y}& \equiv -\tau ^{2}\omega _{\left.
{}\right. y}^{\eta }=\frac{1}{2}\left[ \partial _{\eta }u_{y} -\partial _{y}\left( \tau ^{2}u_{\eta }\right) \right]  
 +\frac{1}{2}\left[ \tau ^{2}u_{\eta }du_{y}
-u_{y}d\left( \tau ^{2}u_{\eta }\right) \right] \ .
\end{align}
The $I^{\mu\nu}_{1}$ terms are
\begin{align}
I_{1}^{xx}& =2u_{x}\left( \pi ^{\tau x}Du_{\tau }+\pi ^{xx}Du_{x}+\pi
^{yx}Du_{y}+\pi ^{\eta x}Du_{\eta }\right) \ , \\
I_{1}^{yy}& =2u_{y}\left( \pi ^{\tau y}Du_{\tau }+\pi ^{xy}Du_{x}+\pi
^{yy}Du_{y}+\pi ^{\eta y}Du_{\eta }\right) \ , 
\\
I_{1}^{xy}& =\left( \pi ^{\tau x}u_{y}+\pi ^{\tau
y}u_{x}\right) Du_{\tau }+\left( \pi ^{xx}u_{y}+\pi ^{xy}u_{x}\right)
Du_{x} 
\notag \\ & \hspace{5mm} 
+\left( \pi ^{yx}u_{y}+\pi ^{yy}u_{x}\right) Du_{y}+\left( \pi
^{\eta x}u_{y}+\pi ^{\eta y}u_{x}\right) Du_{\eta } \ ,
\\
%
I_{1}^{x\eta }& =\left( \pi ^{\tau x}u_{\eta }+\pi ^{\tau \eta
}u_{x}\right) Du_{\tau }+\left( \pi ^{xx}u_{\eta }+\pi ^{x\eta }u_{x}\right)
Du_{x}
\notag \\ & \hspace{5mm} 
+\left( \pi ^{yx}u_{\eta }+\pi ^{y\eta }u_{x}\right) Du_{y}+\left(
\pi ^{\eta x}u_{\eta }+\pi ^{\eta \eta }u_{x}\right) Du_{\eta } \ , 
\\
I_{1}^{y\eta }& =\left( \pi ^{\tau y}u_{\eta }+\pi ^{\tau \eta
}u_{y}\right) Du_{\tau }+\left( \pi ^{xy}u_{\eta }+\pi ^{x\eta }u_{y}\right)
Du_{x}
\notag \\ & \hspace{5mm} 
+\left( \pi ^{yy}u_{\eta }+\pi ^{y\eta }u_{y}\right) Du_{y}+\left(
\pi ^{\eta y}u_{\eta }+\pi ^{\eta \eta }u_{y}\right) Du_{\eta } \ .
\end{align}
The $I^{\mu\nu}_3$ terms are
\begin{align}
I_{3}^{xx}& =2\left( \pi ^{x\tau }\omega _{\left. {}\right. \tau }^{x}+\pi
^{xy}\omega _{\left. {}\right. y}^{x}+\pi ^{x\eta }\omega _{\left. {}\right.
\eta }^{x}\right) , 
\\
I_{3}^{yy}& =2\left( \pi ^{y\tau }\omega _{\left. {}\right. \tau }^{y}+\pi
^{yx}\omega _{\left. {}\right. x}^{y}+\pi ^{y\eta }\omega _{\left. {}\right.
\eta }^{y}\right) , 
\\
I_{3}^{xy}& =\pi ^{x\tau }\omega _{\left. {}\right. \tau }^{y}+\pi ^{y\tau
}\omega _{\left. {}\right. \tau }^{x}+\pi ^{xx}\omega _{\left. {}\right.
x}^{y}  
 +\pi ^{yy}\omega _{\left. {}\right. y}^{x}+\pi ^{x\eta }\omega _{\left.
{}\right. \eta }^{y}+\pi ^{y\eta }\omega _{\left. {}\right. \eta }^{x}\ , 
\\
I_{3}^{x\eta }& =\pi ^{x\tau }\omega _{\left. {}\right. \tau }^{\eta }+\pi
^{\eta \tau }\omega _{\left. {}\right. \tau }^{x}+\pi ^{xx}\omega _{\left.
{}\right. x}^{\eta }  
 +\pi ^{xy}\omega _{\left. {}\right. y}^{\eta }+\pi ^{\eta y}\omega
_{\left. {}\right. y}^{x}+\pi ^{\eta \eta }\omega _{\left. {}\right. \eta
}^{x}\ , 
\\
I_{3}^{y\eta }& =\pi ^{y\tau }\omega _{\left. {}\right. \tau }^{\eta }+\pi
^{\eta \tau }\omega _{\left. {}\right. \tau }^{y}+\pi ^{yx}\omega _{\left.
{}\right. x}^{\eta }  
 +\pi ^{\eta x}\omega _{\left. {}\right. x}^{y}+\pi ^{yy}\omega _{\left.
{}\right. y}^{\eta }+\pi ^{\eta \eta }\omega _{\left. {}\right. \eta }^{y}\ ,
\end{align}
The $I^{\mu\nu}_4$ terms are
\begin{align}
I_{4}^{xx}& =\left( \pi ^{x\tau }\sigma ^{x\tau }-\pi ^{xx}\sigma ^{xx}-\pi
^{xy}\sigma ^{xy}-\tau ^{2}\pi ^{x\eta }\sigma ^{x\eta }\right) 
+\frac{1}{3}\left( 1+u_{x}^{2}\right) \pi _{\beta }^{\alpha
}\sigma _{\alpha }^{\beta }\ , 
\\
I_{4}^{yy}& =\pi ^{y\tau }\sigma ^{y\tau }-\pi ^{yx}\sigma ^{yx}-\pi
^{yy}\sigma ^{yy}-\tau ^{2}\pi ^{y\eta }\sigma ^{y\eta }  
+\frac{1}{3}\left( 1+u_{y}^{2}\right) \pi _{\beta }^{\alpha
}\sigma _{\alpha }^{\beta }\ , 
\\
I_{4}^{xy}& =\frac{1}{2}\left( \pi ^{x\tau }\sigma ^{y\tau }+\pi ^{y\tau
}\sigma ^{x\tau }\right) -\frac{1}{2}\left( \pi ^{xx}\sigma ^{yx}+\pi
^{yx}\sigma ^{xx}\right)  \notag \\
& \hspace{5mm} -\frac{1}{2}\left( \pi ^{xy}\sigma ^{yy}+\pi ^{yy}\sigma ^{xy}\right) -%
\frac{\tau ^{2}}{2}\left( \pi ^{x\eta }\sigma ^{y\eta }+\pi ^{y\eta }\sigma
^{x\eta }\right)  \notag \\
& \hspace{5mm} +\frac{1}{3}u_{x}u_{y} \pi _{\beta }^{\alpha
}\sigma _{\alpha }^{\beta }\ , \\
I_{4}^{x\eta }& =\frac{1}{2}\left( \pi ^{x\tau }\sigma ^{\eta \tau }+\pi
^{\eta \tau }\sigma ^{x\tau }\right) -\frac{1}{2}\left( \pi ^{xx}\sigma
^{\eta x}+\pi ^{\eta x}\sigma ^{xx}\right)  \notag \\
& \hspace{5mm} -\frac{1}{2}\left( \pi ^{xy}\sigma ^{\eta y}+\pi ^{\eta y}\sigma
^{xy}\right) -\frac{\tau ^{2}}{2}\left( \pi ^{x\eta }\sigma ^{\eta \eta
}+\pi ^{\eta \eta }\sigma ^{x\eta }\right)  \notag \\
& \hspace{5mm} +\frac{1}{3}u_{x}u_{\eta } \pi _{\beta }^{\alpha
}\sigma _{\alpha }^{\beta }\ , \\
I_{4}^{y\eta }& =\frac{1}{2}\left( \pi ^{y\tau }\sigma ^{\eta \tau }+\pi
^{\eta \tau }\sigma ^{y\tau }\right) -\frac{1}{2}\left( \pi ^{yx}\sigma
^{\eta x}+\pi ^{\eta x}\sigma ^{yx}\right)  \notag \\
& \hspace{5mm} -\frac{1}{2}\left( \pi ^{yy}\sigma ^{\eta y}+\pi ^{\eta y}\sigma
^{yy}\right) -\frac{\tau ^{2}}{2}\left( \pi ^{y\eta }\sigma ^{\eta \eta
}+\pi ^{\eta \eta }\sigma ^{y\eta }\right)  \notag \\
& \hspace{5mm} +\frac{1}{3}u_{y}u_{\eta } \pi _{\beta }^{\alpha
}\sigma _{\alpha }^{\beta }\ .
\end{align}
%
\section{Fluid dynamic equations in conservative form}
\label{sec:cons_form}
In this appendix we show how both the conservation laws (\ref{dtT00}) and (\ref{dtT03}) and the relaxation equations (\ref{relEqs_Pi}) and (\ref{relEqs}) can be written in the same flux-conserving form
\begin{equation}
\frac{\partial q}{\partial t}+\frac{\partial (v^{i} q)}{\partial x^i}=
J[q](t,x)
\equiv S[q](t,x)+G[q](t,x)\;.
\label{B1}
\end{equation}
Comparing this form with Eqs.~(\ref{dtT00})-(\ref{dtT03}) we read off
{\small
\begin{align}
S^{\tau}_{c}&\equiv 
-\partial _{x}\left( v_{x}\p-v_{x}\pi ^{\tau \tau }+\pi ^{\tau x}\right)
-\partial _{y}\left( v_{y}\p-v_{y}\pi ^{\tau \tau }+\pi ^{\tau y}\right) 
 -\partial _{\eta }\left( v_{\eta }\p-v_{\eta }\pi ^{\tau \tau }+\pi ^{\tau
\eta }\right) \,, 
 \\
S^{x}_{c}&\equiv 
-\partial _{x}\left( \p-v_{x}\pi ^{\tau x}+\pi ^{xx}\right) -\partial
_{y}\left( -v_{y}\pi ^{\tau x}+\pi ^{xy}\right)  
 -\partial _{y}\left( -v_{\eta }\pi ^{\tau x}+\pi ^{x\eta }\right) \,, 
  \\
S^{y}_{c}&\equiv
 -\partial _{x}\left( -v_{x}\pi ^{\tau y}+\pi ^{xy}\right) -\partial
_{y}\left( \p-v_{y}\pi ^{\tau y}+\pi ^{yy}\right) 
 -\partial _{\eta }\left( -v_{\eta }\pi ^{\tau y}+\pi ^{y\eta }\right) \,, 
  \\
S^{\eta}_{c}&\equiv 
-\partial _{x}\left( -v_{x}\pi ^{\tau \eta }+\pi ^{x\eta }\right)
-\partial _{y}\left( -v_{y}\pi ^{\tau \eta }+\pi ^{y\eta }\right)  
 -\partial _{\eta }\left( \frac{\p}{\tau ^{2}}-v_{\eta }\pi ^{\tau \eta
}+\pi ^{\eta \eta }\right) \,
\end{align}
}
as well as
\begin{align*}
G^{\tau}_{c}\equiv -\frac{1}{\tau}(T^{\tau\tau}+\tau^{2}T^{\eta\eta})\,,\quad 
G^{x}_{c}\equiv -\frac{1}{\tau}T^{\tau x} \,,\quad
G^{y}_{c}\equiv -\frac{1}{\tau}T^{\tau y} \,,\quad 
G^{\eta}_{c}\equiv -\frac{3}{\tau}T^{\tau \eta}\;.\quad
\end{align*}
The relaxation equations (\ref{relEqs_Pi}) and (\ref{relEqs}) are a subclass of a more general class of equations describing a dissipative system, namely the convective-diffusion equation 
\begin{equation}
\frac{\partial q}{\partial t}+v^{i}\frac{\partial q}{\partial x^i}=
H[q](t,x)\;.
\end{equation}
These equations can be recast into the form (\ref{B1}) by introducing an additional source term on the right hand side:
\begin{equation}
\frac{\partial q}{\partial t}+\frac{\partial}{\partial x^{i}}(v^{i}q)=
q\frac{\partial v^{i}}{\partial x^{i}}+H[q](t,x)+G[q](x,t)\;.
\label{relaxConsForm}
\end{equation}
Written in conservative form (\ref{relaxConsForm}), and splitting the right hand side $J[q]\equiv q\partial_{i}v_{i}+H[q]$ again into a source terms $S$ and a geometric term $G$, as in (\ref{B1}), the relaxation equations~(\ref{relEqs_Pi}) and (\ref{relEqs}) [with the convective derivative $d\equiv u_{\tau}\partial_{\tau}+u_{x}\partial_{x}
+u_{y}\partial_{y}+u_{\eta}\partial_{\eta}$] become
\begin{align}
\partial_{\tau}\Pi 
&+\partial_{x}(v_{x}\Pi)
+\partial_{y}(v_{y}\Pi)
+\partial_{\eta}(v_{\eta}\Pi) 
=S_{\Pi}\;,
\\
\partial_{\tau}\pi^{\mu\nu} 
&+\partial_{x}(v_{x}\pi^{\mu\nu})
+\partial_{y}(v_{y}\pi^{\mu\nu})
+\partial_{\eta}(v_{\eta}\pi^{\mu\nu}) 
=S^{\mu\nu}_{\pi}+G^{\mu\nu}_{\pi}\;,
\end{align}
where (with $\partial_{i}v_{i}\equiv\partial_{x}v_{x}+\partial_{y}v_{y}+\partial_{\eta}v_{\eta}$)
\begin{align}
S_{\Pi}&\equiv\frac{1}{u_\tau}\left(-\frac{\zeta}{\tau_\Pi}\theta-\frac{\Pi}{\tau_\Pi}-I_{\Pi}\right)
+\Pi\partial_{i}v_{i} \, ,
\\
S^{\mu\nu}_{\pi}&\equiv \frac{1}{u_\tau}\left(2\frac{\eta}{\tau_\pi}\sigma^{\mu\nu}-\frac{\pi^{\mu\nu}}{\tau_\pi}-I^{\mu\nu}\right)
+\pi^{\mu\nu}\partial_{i}v_{i}
\;.
\end{align}
The only non-zero components of $G^{\mu\nu}_{\pi}$ are
\begin{align}
G^{\tau\tau}_{\pi}&\equiv
-2\tau v_\eta \pi^{\tau\eta} \;,
\nonumber \\
G^{\tau x}_{\pi}&\equiv
-2\tau v_\eta \pi^{x\eta}\;,
\nonumber \\
G^{\tau y}_{\pi}&\equiv
-2\tau v_\eta \pi^{y\eta} \;,
\nonumber \\
G^{\tau \eta}_{\pi}&\equiv
-\tau v_\eta \pi^{\eta\eta}-\frac{1}{\tau}(\pi^{\tau\eta}+v_\eta\pi^{\tau\tau}) \;,
\nonumber \\
G^{x \eta}_{\pi}&\equiv
-\frac{1}{\tau}(\pi^{x\eta}+v_\eta\pi^{\tau x}) \;,
\nonumber \\
G^{y \eta}_{\pi}&\equiv
-\frac{1}{\tau}(\pi^{y\eta}+v_\eta\pi^{\tau y}) \;,
\nonumber \\
G^{\eta \eta}_{\pi}&\equiv
-\frac{2}{\tau}(\pi^{\eta\eta}+v_\eta\pi^{\tau \eta}) \;.
\end{align}
Finally, this yields all dynamical equations in the form (\ref{B1}), with
\begin{eqnarray}
&&\hat{S}_{c}\equiv
\begin{pmatrix}
S^{\tau}_{c} & S^{x}_{c} & S^{y}_{c} & S^{\eta}_{c}
\end{pmatrix}^T \, \nonumber
\\
&&\vec{S}_{\pi}\equiv 
 \begin{pmatrix}
  S^{\tau \tau}_{\pi} & S^{\tau x}_{\pi} & S^{\tau y}_{\pi}
   & S^{\tau \eta}_{\pi} &
  S^{xx}_{\pi} & S^{xy}_{\pi} & S^{x\eta}_{\pi} &
  S^{yy}_{\pi} & S^{y\eta}_{\pi} & S^{\eta\eta}_{\pi} & S_{\Pi}
 \end{pmatrix}^T \, \nonumber
\\ 
&&\hat{G}_{c}\equiv 
 \begin{pmatrix}
  G^{\tau}_{c} & 
  G^{x}_{c} & 
  G^{y}_{c} & 
  G^{\eta}_{c}
 \end{pmatrix}^T \,,\nonumber
 \\ 
&&\vec{G}_{\pi}\equiv 
 \begin{pmatrix}
  G^{\tau \tau}_{\pi} & G^{\tau x}_{\pi} & G^{\tau y}_{\pi}
   & G^{\tau \eta}_{\pi} &
  0 & 0 & G^{x\eta}_{\pi} &
  0 & G^{y\eta}_{\pi} & G^{\eta\eta}_{\pi} & 0
 \end{pmatrix}^T \, .
 \label{B13}
\end{eqnarray}

\begin{center}
\rule{6cm}{0.4pt}
\end{center}


\section*{Acknowledgments}
This work was supported by the U.S. Department of Energy, Office of Science, Office of Nuclear Physics under Award No. \rm{DE-SC0004286} as well as by the National Science Foundation, Division for Advanced Cyberinfrastructure, in the framework of the JETSCAPE Collaboration under Award No. 1550223. M. Strickland was supported by the U.S. Department of Energy, Office of Science, Office of Nuclear Physics under Award No. DE-SC0013470.

\bibliographystyle{elsarticle-num}

\bibliography{gpu-vh}

\end{document}